\documentclass[aps,prb,floatfix,superscriptaddress,hyperref=pdftex]{revtex4-2}

\usepackage{amssymb}
\usepackage{hyperref}
\usepackage{graphicx}
\usepackage{placeins}
\usepackage{booktabs}
\usepackage{svg}

\usepackage{dsfont}
\usepackage[utf8]{inputenc}
\usepackage[normalem]{ulem}

\usepackage{upgreek}
\usepackage{siunitx}
\usepackage{physics}
\usepackage{soul}
\usepackage{xcolor}
\usepackage{tabularx,multirow,array}
\usepackage{siunitx}
\usepackage{chemmacros}
\usepackage{cancel}

\usepackage{bbm}

\begin{document}


\title{Spectroscopic Visualization of Hard Quasi-1D Superconductivity Induced in Nanowires Deposited on a Quasi-2D Indium film}


\author{Ambikesh Gupta}
\affiliation{Department of Condensed Matter Physics, Weizmann Institute of Science, Rehovot 7610001, Israel}

\author{Pranab Kumar Nag}
\affiliation{Department of Condensed Matter Physics, Weizmann Institute of Science, Rehovot 7610001, Israel}

\author{Shai Kiriati}
\affiliation{Department of Condensed Matter Physics, Weizmann Institute of Science, Rehovot 7610001, Israel}

\author{Samuel D. Escribano}
\affiliation{Department of Condensed Matter Physics, Weizmann Institute of Science, Rehovot 7610001, Israel}

\author{Man Suk Song}
\affiliation{Department of Condensed Matter Physics, Weizmann Institute of Science, Rehovot 7610001, Israel}

\author{Hadas Shtrikman}
\affiliation{Department of Condensed Matter Physics, Weizmann Institute of Science, Rehovot 7610001, Israel}

\author{Yuval Oreg}
\affiliation{Department of Condensed Matter Physics, Weizmann Institute of Science, Rehovot 7610001, Israel}

\author{Nurit Avraham}
\email{nurit.avraham@weizmann.ac.il}
\affiliation{Department of Condensed Matter Physics, Weizmann Institute of Science, Rehovot 7610001, Israel}

\author{Haim Beidenkopf}
\email{haim.beidenkopf@weizmann.ac.il}
\affiliation{Department of Condensed Matter Physics, Weizmann Institute of Science, Rehovot 7610001, Israel}

\begin{abstract}
Following significant progress in the visualization and characterization of hybrid superconducting-semiconducting systems, greatly propelled by reports of Majorana zero modes in nanowire devices, considerable attention has been devoted to investigating the electronic structure at the buried superconducting-semiconducting interface and the nature of the induced superconducting correlations. The properties of that interface and the structure of the electronic wave functions that occupy it determine the functionality and the topological nature of the induced superconducting state. Here, we introduce a novel
hybrid platform for proximity-inducing superconductivity in InAs$_{0.6}$Sb$_{0.4}$ nanowires, leveraging a unique architecture and material combination.  By dispersing these nanowires over a superconducting Indium film we exploit Indium's high critical temperature of 3.7~K and the anticipated high spin-orbit and Zeeman couplings of InAs$_{0.6}$Sb$_{0.4}$. This design preserves the pristine top facet of the nanowires, making it highly compatible with scanning tunneling spectroscopy. Using this architecture we demonstrate that the mechanical contact supports Cooper-pair transparency as high as 90\%, comparable with epitaxial interfaces. The anisotropic angular response to an applied magnetic field shows the quasi-two-dimensional nature of the parent superconductivity in the Indium film and the quasi-one-dimensional nature of the induced superconductivity in the nanowires. Our platform offers robust and advantageous foundations for studying the emergence of topological superconductivity and the interplay of superconductivity and magnetism using atomic-scale spectroscopic tools.
\end{abstract}

\date{\today}
\maketitle

\section{Introduction}

The superconducting proximity effect remarkably enables to induce superconducting correlations from an intrinsic superconductor (SupC) into an otherwise normal metal through Andreev tunneling of Cooper pairs~\cite{deutcher1969proximity,tinkham2004introduction}. Inducing superconductivity into a semiconductor (SemC) instead of a  metal offers tunability of the electronic properties of the hybrid system such as control of the doping level of the SemC~\cite{cohen1964existence, burkard2020superconductor, van1988superconductor}. Dimensionality also plays an important role both for the SemC and for the SupC. The typical low electronic mass in SemCs results in a large Bohr radius that leads to size quantization in narrow constrictions imposed either by the physical dimension of a nano-scale device as a nanowire (NW)~\cite{lu2006semiconductor} or by gate-confinement through electrostatic depletion~\cite{PhysRevB.106.L241301}. This further enables the incorporation of quantum dots and tunnel barriers and their integration into complex circuits. The low-dimensionality of the SupC renormalizes the superconducting penetration depth~\cite{zavaritskii1952properties,tinkham2004introduction} and increases its robustness to threading of magnetic flux through the narrow facet~\cite{toxen1961size}. In certain SupCs even the critical transition temperature is found to increase with decreasing thickness, such as in the commonly studied \ch{Al}~\cite{khukhareva1963superconducting} and in \ch{In}~\cite{vogel1967superconductivity} that we investigate here. Accordingly, the geometry and architecture of the hybrid system lends a powerful design tool for optimizing its properties. 

Hybrid SemC-SupC devices provide an exciting bedrock for a range of superconducting electronic and spintronic technologies with low power consumption~\cite{linder2015superconducting,martinez2020interfacial,eschrig2015spin,nadeem2023superconducting}. These hybrids have attracted vast scientific attention also as a platform for fault tolerant quantum computation~\cite{KITAEV20032,nayak2008non} by proximity inducing topological superconductivity into SemC NWs with high spin-orbit coupling and a high $g$-factor~\cite{oreg2010helical,Lutchyn2010Aug, leijnse2012introduction} as InAs, InSb~\cite{Lutchyn2018May} and PbTe~\cite{gao2024hard}. Achieving this requires fine tuning of the chemical potential across the device to within an overlap of the superconducting gap and the Zeeman gapped Kramer degeneracy of a quantized sub-band. Topological superconductivity thus harnesses many of the advantages hybrid SemC-SupC devices offer.

Yet, the incorporation of SemCs also introduces several challenges that must be adequately addressed to facilitate scientific and technological progress. Due to their low carrier concentration, SemCs are sensitive to charge disorder that cannot be efficiently screened~\cite{queisser1998defects}. This increased susceptibility often leads to the formation of charge accumulation or depletion layers on boundaries of the SemC~\cite{zhang2012band} that may promote or limit the superconducting proximity effect. Indeed, initial studies could not establish a hard induced gap in SemC NWs from ex-situ deposited SupC electrodes~\cite{mourik2012signatures,das2012zero,deng2012anomalous,PhysRevLett.110.186803}. This challenge was surmounted by growing epitaxial SupC layers over the SemC which nowadays is considered essential for inducing a hard gap~\cite{gül2017hard,chang2015hard,gao2024hard}. Because of poor wetting of the SemC surface by the SupC metal achieving continuous coverage typically requires cryogenic deposition and subsequent introduction of a rigid capping layer~\cite{kang2017robust,geng2024epitaxial}. However, the strong coupling between the SemC and the SupC metal can lead to the metalization of the SemC NW resulting in suppression of its desired properties such as spin-orbit and $g$-factor, and limited ability to gate-tune its doping~\cite{cole2015effects}. Therefore, optimized balance is needed to mitigate these counteracting effects.

 In this study, we introduce an innovative SemC-SupC architecture optimized for atomic-scale scanning tunneling microscopy (STM) and spectroscopic studies of induced superconductivity within SemC NWs. Unlike common setups in which the NWs are partially or fully deposited by a SupC shell, our approach involves direct deposition of bare SemC NWs onto a SupC thin film. This configuration offers unrestricted access to the pristine facets of the NWs, providing the capability of visualizing the spatial profile of electronic states with high resolution. We surprisingly find that by simple non-epitaxial mechanical coupling of this architecture the NW exhibits robust induced superconductivity.

We demonstrate this with \ch{InAs_{1-x}Sb_x} with \ch{x} $\approx0.4$ (InAsSb) NWs deposited on top of a thin film of \ch{In}, both possessing favourable material properties. The malleable nature of \ch{In} allows slight sagging of the NWs into the \ch{In} substrate, promoting the electronic coupling between the two materials that enhances the induced superconductivity in the NW. In addition, dipping the metallic (\ch{PtIr}) STM tip into the \ch{In} substrate enables to induce superconductivity in the tip, thereby offering a platform to measure the state of the NWs with a SupC tip. Apart from enhancing the energy resolution of the measurement by the coherence peaks of the SupC tip~\cite{pan1998vacuum}, this allows creating SupC-insulator-SupC junctions between the SupC tip and sample that could even support superconducting Josephson spectroscopy for probing coherence properties. In addition to this, \ch{In} is a standard $s$-wave SC, with a high $T_{\rm c}=3.7$~K and large superconducting gap $\Delta\simeq0.6$~meV relative to commonly studied Al. Parity conservation has also been demonstrated in \ch{In} islands grown over \ch{InAs} NWs~\cite{bjergfelt2021superconductivity}. Furthermore, the \ch{InAsSb} SemC NW exhibits high spin-orbit coupling and a significantly larger $g$-factor compared to other binary III-V compound SemCs~\cite{PhysRevLett.117.076403, PhysRevMaterials.2.044202}. We reproduce the result with theoretical simulation that enables us to predict that while we do not obtain topological superconductivity, a reduced \ch{In} film thickness could achieve it.

\section{\label{sec:SC-Nanowire}Results}

We perform spectroscopic STM measurements of the proximitized quasi-1D \ch{InAsSb} NWs deposited on superconducting \ch{In} substrate. Ternary InAsSb NWs with a typical diameter of about $100$~nm were grown by \ch{Au}-assisted molecular beam epitaxy on reclining InAs stems that are grown on (001) InAs substrate~\cite{ercolani2012growth} (see Appendix~\ref{sm-fab} for further details). The InAs stem has a wurtzite crystal structure and a round morphology while \ch{InAsSb} has a zinc-blende structure and a well-faceted hexagonal morphology. For STM measurements, the NWs were mechanically harvested in situ immediately after their growth and deposited on a $\sim 50$~nm thick \ch{In} film prior to their transfer to the STM chamber. The entire harvest and transfer process of the NWs was carried out under ultrahigh vacuum conditions, in a home-built vacuum suitcase~\cite{reiner2017hot}. Tunneling spectroscopy was performed by measuring the tunneling differential conductance, dI/dV, using a lock-in detection scheme. The effective electron temperature was extracted from the tunneling spectrum on the superconducting \ch{In} substrate by fitting it to the Dyne's formula~\cite{dynes1978direct}. The tunneling spectra were obtained through two distinct configurations: either in a normal-insulator-SupC setup using metallic STM Pt/Ir tips as tunneling probes, or in an SupC-Insulator-SupC configuration. In the latter case, we immersed the Pt/Ir tips into the \ch{In} substrate, resulting in the attachment of an In cluster at the tip's apex rendering it superconducting. A topographic STM image of an \ch{InAsSb} NW partially submerged in the \ch{In} substrate is shown in Fig.~\ref{Fig1}A. The extruded profile of the NW gets convoluted with the mesoscopic shape of the tip apex distorting the global topographic image of the NW by creating replications by secondary tips (clearly seen in the upper right panel of Fig.~\ref{Fig1}A). An atomically resolved topographic image taken on a top facet of a NW (bottom-right panel), reveals an ordered crystalline structure. Yet, the InAsSb NWs have shown a larger degree of disorder relative to InAs NWs~\cite{reiner2017hot} possibly due to their ternary composition.


\subsection{Induced superconductivity in \texorpdfstring{I\MakeLowercase{n}A\MakeLowercase{s}S\MakeLowercase{b}}{InAsSb} Nanowires}

We explore the superconducting state induced in the NW due to its contact with the underlying \ch{In} substrate. Tunneling dI/dV spectra measured at $350$~mK with a normal metallic tip both on the \ch{In} substrate and on the InAsSb NW top facet are shown in Fig.~\ref{Fig1}B (red and blue lines, respectively). In both we find a clear gap structure around zero bias flanked by coherence peaks. The gap we image over the NW is slightly narrower than that we measure on the \ch{In} film. 
To determine the parent and induced energy gaps, we fit the dI/dV spectra to the Dynes formula~\cite{dynes1978direct} 

\begin{eqnarray}
\rho_{\rm t}(E)=\rho_{0} \left|\mathrm{Re}\left\{\frac{E-i\Gamma_{\rm t}}{\sqrt{(E-i\Gamma_{\rm t})^2-\Delta^2}}\right\}\right|,
\label{Eq:dynes}
\end{eqnarray}
which phenomenologically encodes a finite quasiparticle lifetime, $\hbar / \Gamma$, ~\cite{PhysRevB.94.144508} to the conventional Bardeen-Cooper-Schrieffer (BCS) density of states of a $s$-wave superconductor (dashed black lines in Fig.~\ref{Fig1}B).
By gathering statistics from many spectra measured across multiple \ch{In} and 6 NW locations we extract average energy gaps of $\Delta_{\rm NW}=0.548 \pm 0.116$~meV on the top NW facets, compared to $\Delta_{\rm In}=0.623 \pm 0.059$~meV on the \ch{In} substrate. We thus obtain an average induced gap which is 88\% of the parent superconducting gap when measured 100 nm away from the physical SupC-NW interface. We stress that this is achieved simply by mechanical coupling between the NWs and substrate. We attribute the variance in gap size measured on \ch{In} to inhomogeneity of the \ch{In} film thickness. The stronger variability in gap size of the NWs stems from both substrate inhomogeneity and variations in NW-substrate coupling. Nevertheless, over each individual NW we find that the gapped spectrum remains remarkably consistent across the extent of the upper facet as visualized in the linecut of Fig.~\ref{Fig1}C. This underscores the robust and uniform nature of the induced superconductivity in the NW in face of atomic-scale disorder.
The temperature dependence of the \ch{In} superconducting gap is shown in  Fig.~\ref{Fig1}D. The gap disappears at $T_{\rm c}\simeq3.7$~K, in agreement with previous reports on bulk \ch{In} ~\cite{Matthias1963Jan, Eisenstein1954Jul}.

The slight suppression of the induced gap in the NW relative to the parent \ch{In} gap in the substrate reflects the intricate characteristics of the NW-substrate heterostructure. This includes the coupling between the NW and substrate, and the work function difference between both materials, which leads in turn to a bend bending at the NW's interface. To quantify the strength of these effects and gain further insight about the impact of the various parameters of the heterostructure on the proximity effect in the NW, we compare the experimental results with numerical simulations. We model the NW-substrate heterostructure with a Hamiltonian~\cite{antipov2018effects, Thesis_sam} that describes the electrons inside the nanostructure and treats on equal footing the NW and the substrate through its spatial dependence
\begin{eqnarray}
H =\left[\vec{k}\frac{\hbar^{2}}{2m^{*}(\vec{r})}\vec{k}-E_{\mathrm{F}}(\vec{r})+e\phi(\vec{r})\right]\sigma_{0}\tau_{z} \nonumber \\ 
+\frac{1}{2}\mu_{B}g(\vec{r})\vec{B}\cdot\vec{\sigma}\tau_{z} +\frac{1}{2}[\vec{\alpha}(\vec{r})\cdot(\vec{\sigma}\times\vec{k})+(\vec{\sigma}\times\vec{k})\cdot\vec{\alpha} (\vec{r})]\tau_{z} \nonumber \\ 
+\Delta(\vec{r},\vec{B})\sigma_{y}\tau_{y}. \; \;
\end{eqnarray}
The first three terms correspond to the kinetic energy, where $m^{*}$ is the effective mass; $E_{\rm F}$ the band-bottom of the conduction band with respect to the Fermi level; and $\phi(\vec{r})$ the electrostatic potential. We compute $\phi(\vec{r})$ self-consistently in the Thomas-Fermi approximation, considering the experimental geometry and the band bending towards the NW facets. The next term in the Hamiltonian is the Zeeman field, in both the NW and SC, where $g(\vec{r})$ is the $g$-factor and $\vec{B}$ the magnetic field. The following term is the spin-orbit interaction with a coupling parameter $\vec{\alpha}$ that is nonzero only within the NW. Both, the spin-orbit coupling and the $g$-factor of the ternary \ch{InAsSb} NW, are obtained by interpolating between the parameters of the binaries \ch{InAs} and \ch{InSb}. The last term in the Hamiltonian is the superconducting pairing, with $\Delta$ the superconducting pairing amplitude, which is only present inherently in the SupC, and depends on $B$ due to Zeeman and orbital effects. For comparison with the experimental results we use the measured $\Delta(B)$ as a phenomenological  parameter. While most parameters for this Hamiltonian are extracted from more fundamental theories, the band-bending and the interface transparency can be extracted by adjusting them in such a way that the theoretical dI/dV resembles the experimental one (see Appendix~\ref{sm-theory} for further details).

 A comparison between the measured and the best fitted calculated density of states is shown in Fig.~\ref{Fig1}E,F, respectively. Similarly to the measured spectra, the density of states is calculated on the \ch{In} substrate (red line) and in the middle of the top facet of the NW (blue line). By fitting the spectra in both locations, we obtain two different superconducting gaps, $\Delta=0.62$~meV for the parent gap in the \ch{In} substrate, and $\Delta=0.52$~meV for the induced gap in the NW, closely aligned with the experimental values. To obtain such a good agreement, we have tuned the transparency of the substrate-NW interface, $\kappa$, and the band offset across it. We find $\kappa\backsimeq0.83$, where $\kappa=1$ corresponds to a fully transparent interface and $\kappa=0$ to a fully opaque one. We thus observe that the interface of the mechanically paired NW and substrate is highly transparent allowing for the proximity effect to take place. For the band-bending at the interface we obtain $0.1$~eV in our simulations. This value closely aligns with the band-bending observed on the bare facets of InAs NWs~\cite{Antipov2018EffectsNanowires}. In its presence, the wavefunction of the electrons localizes in proximity to the facets in a ring-like profile. This spatial distribution is advantageous both for inducing superconductivity and for probing the electronic states on the top facet in the STM.

Our simulations also give a plausible explanation for the increased width of the coherence peaks in the NW relative to what we find in the \ch{In} substrate. The band-bending at the NW-substrate interface pulls the electron's wavefunction close to the interface, where the proximity effect is enhanced. However, higher-energy subbands possess larger kinetic energy, causing their wavefunctions to spread more across the NW's section and diminishing their induced superconductivity. Consequently, each subband exhibits a distinct superconducting gap that, when convoluted with temperature, collectively resembles a single broadened coherence peak. The observed decoherence process in the NW is therefore a possible consequence of the existence of subgap states with different superconducting pairing correlations.

\subsection{Effects of magnetic field on the \ch{In} substrate and superconducting tip}
To further explore the superconducting proximity effect in the NW, we examine its response to the application of a magnetic field in various orientations relative to the substrate and the NW. We first characterize the magnetic field dependence of the \ch{In} substrate. The evolution of the dI/dV tunneling spectrum between a normal tip and a superconducting substrate under a magnetic field perpendicular to the substrate, $H_\perp$, is shown in Fig.~\ref{Fig2}A. As expected, the superconducting gap is suppressed with increasing $H_{\perp}$, until it disappears completely at a critical magnetic field of $H_{\rm c,\perp}=300$~Oe, in agreement with the \ch{In} bulk critical field. In the direction parallel to the film, $H_\parallel$, the critical field is nonetheless enhanced to $H_{\rm c,\parallel}=1250$~Oe, see Fig.~\ref{Fig2}C, which is about 4 times higher than $H_{\perp}$. This enhancement is due to finite-size effects of the superconductor~\cite{Tinkham1996}, since the Meissner effect is suppressed in low-dimensional structures where orbital effects are significantly smaller.

Next, we examine the evolution when tunneling from a superconducting tip while increasing the magnetic field both in the out-of-plane and in-plane directions relative to the substrate, shown in Fig.~\ref{Fig2}B,D, respectively. In both we observe a kink that occurs at the substrate's critical field of either orientation with respect to the substrate. This suggests that due to mesoscopic effects the quasi-0D SupC dot at the apex of the STM tip has a higher critical field than that of the quasi-2D substrate. To distinguish the evolution of the tip gap from the sample gap, we first treat the spectrum in Fig.~\ref{Fig2}D as if it is of a normal-insulator-SupC junction and extract the value of the effective superconducting gap using Dyne's equation, shown in Fig.~\ref{Fig2}E. At the kink the superconductivity in the substrate dies out and the gap that continues across it is that of the tip alone. We thus fit a square-root magnetic field dependence to that high-field single gap regime (dotted line). By extrapolating it to zero field we extract the value of the tip gap, $\Delta_{\rm tip}$. 

Now we can isolate the substrate gap, $\Delta_{\rm In}$ contribution by plotting the deviation of the combined gap from the fitted tip gap, shown in Fig.~\ref{Fig2}F. Surprisingly, following this rather simple procedure we reproducibly find a gap on the sample side which is about half of its value measured with a normal tip. We attribute this suppression to current fluctuations generated by the magnet controller (even at zero field, see Fig.~\ref{sm-Bcontrol}) that translate to magnetic field fluctuations. These suppress the superconducting gap of the quasi-2D \ch{In} film but hardly the quasi-0D tip, in accordance with our observations.
Still, since this suppression is consistent on both the \ch{In} substrate and the NW we use it to obtain an accurate estimation of the relative gaps on both.


\subsection{Evolution of proximity-induced superconductivity in the \ch{InAsSb} NWs under magnetic field} \label{NW}

We thus proceed to apply the same analysis for the induced superconductivity in the InAsSb NWs. The dI/dV on the top facet of the NW in the presence of increasing in-plane magnetic field is displayed in Fig.~\ref{Fig3}A and B  with applied field perpendicular to the NW axis, $H_{\rm \perp NW}$, and parallel to the NW, $H_{\rm \parallel NW}$ respectively. Like in the quasi-2D \ch{In} film in Fig.~\ref{Fig2}B, the plots demonstrate two critical magnetic fields in the in-plane direction, corresponding to the different critical magnetic fields for the superconductivity induced in the quasi-1D NW and that of the quasi-0D STM tip. The discrete magnetic field spectra in Fig.~\ref{Fig3}A show how a hard gap at 0 T turns softer gradually and vanishes as we increase the strength of $H_{\rm \perp NW}$. To extract the finer value of $\Delta_{\rm NW}$, we look at the continuous evolution of the spectra as a function of $H_{\rm \parallel NW}$. While the perpendicular field measurements exhibit similar features to the ones of the substrate, featuring a critical magnetic field of $\sim300$~Oe, the measurements parallel to the NW show a much larger critical magnetic field. This can be better appreciated in Fig.~\ref{Fig3}C and D, where we follow the same procedure we demonstrated on the \ch{In} film to extract the induced gap in the NW. We obtain from the fitting a critical magnetic field for this NW of $\sim700$~Oe. This smaller critical magnetic field is due to the large Zeeman splitting in the NW, which possesses a large $g$-factor, $g\simeq110$, that contributes significantly to the closing of the gap. 

Intriguingly, on one of the NWs we have examined the induced superconducting gap was found to decrease over the repeated measurements, as seen in Fig.~\ref{fig:lose_NW}. The deterioration of the induced gap started with a mild touch of the STM tip over the NW that had left a local topographic mark. Afterwards with consecutive field sweeps the double gap feature below the kink gradually decreased until eventually the kink disappeared completely. The tip gap was unaffected in this process. It is remarkable that the reduced coupling between the NW and the \ch{In} substrate that suppresses tunneling of Cooper pairs still allows tunneling of electrons that facilitate the STM measurement. Eventually, it had become completely opaque to tunneling of Cooper pairs as superconductivity could no longer be induced. It also provides compelling evidence that the gap we measure when tunneling to the top facet of the NW represents the local electronic correlations in the NW itself, rather than a remote tunneling interface between the NW bottom facet and the substrate. The \ch{In} substrate clearly remains superconducting throughout all measurements and therefore the lower tunneling junction would have remained gapped as long as tunneling is supported. This signifies that the gapped spectra that we measure represent the electronic state at the vicinity of the tunneling junction with the STM tip at the top facet of the NW.

Lastly, we examine the angular dependence of the in-plane magnetic field response of the superconductivity induced in quasi-1D NWs by the quasi-2D \ch{In} film. We measure the dI/dV spectrum on top of a NW as we rotate the magnetic field within the plane of the \ch{In} film with a fixed magnitude of $H_{\parallel}=1000$~Oe slightly below the film's and this NW's in-plane critical field, shown in Fig.~\ref{fig4}A. A $\pi$-periodically modulated pattern is observed in the evolution of the superconducting gap of the NW in which the coherence peaks shift in energy in a periodic fashion. Upon fitting the individual spectra with Dynes formula, the evolving superconducting gap can be extracted as is shown in white dots in Fig.~\ref{fig4}A. In addition to this, the modulation of the inflection point, $\mathrm{max\left\{d^{2}I/dV^{2}\right\}}$, shown with black dots, is clearly apparent with an expected period of $\pi$. The periodicity is aligned with the NW orientation in topography. The gap is maximal when the field is aligned along the NW and minimal when the field is normal to the NW. A control experiment on the adjacent \ch{In} substrate displayed in Fig.~\ref{fig4}B finds no such periodicity. We thus conclude that unlike the parent superconducting state in the quasi-2D film, the superconducting state induced in the NW inherits its 1D nature.

The variation in critical magnetic fields with orientation is attributed to orbital effects in the nanowire. The spin-orbit coupling in the NW renormalizes the effective $g$-factor depending on the area threaded by the magnetic field (see Appendix~\ref{Sec:theo_topo} for more details). Our simulations of this configuration shown in Fig.~\ref{fig4}C and D, obtain similar quantitative results. We thus estimate the difference in spin-orbit coupling strength between the directions parallel and normal to the NW to be $\Delta \alpha \approx$20 meV$\cdot$nm. This further serves as a proof of the existence of subgap states in the NW and the presence of spin-orbit coupling in the NW as the $g$-factor renormalization due to orbital effects is a consequence of the spin-orbit coupling.

Encouraged by the substantial agreement between theory and experiment, we make predictions about how to obtain a topological superconducting phase in this device (more details in Appendix~\ref{Sec:theo_topo}). We propose to use the same \ch{InAs_{0.6}Sb_{0.4}} NWs, with diameters ranging from $120$ to $150$~nm deposited on top of a $10$~nm thick \ch{In} substrate (as opposed to the 50 nm thick \ch{In} films we have used here). Our analysis indicates that this specific configuration is poised to exhibit topological superconductivity. Furthermore, we outline the anticipated outcomes of a potential dI/dV measurement in such a setup: a discernible gap closing and reopening at the end of the NW, together with a robust zero-energy mode exhibiting exponential decay into the NW. Our work unveils a pioneering approach for investigating induced superconductivity in low-dimensional systems, offering a unique spatial resolution of subgap states. Our proposed configuration showcases promise for achieving topological superconductivity, opening new avenues for exploration and potential applications in quantum information processing, and advancing our understanding of exotic quantum states in low-dimensional systems.


\subsection*{Data Availability Statement}
All data used in the analysis is available to any researcher for purposes of reproducing or extending the analysis upon request to the corresponding authors.


%


\newpage{}

\newpage{}

\begin{figure}
\begin{centering}
\includegraphics[width=0.95\columnwidth]{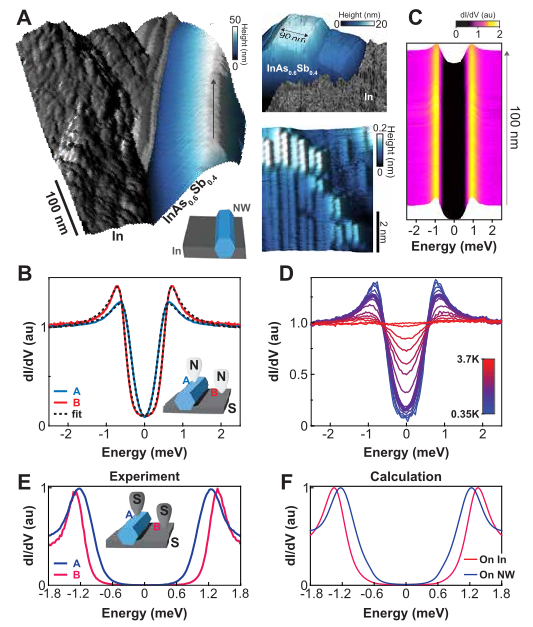}
\par\end{centering}
\caption{\textbf{\label{Fig1} Induced superconductivity in \ch{InAsSb} NWs. (a)} STM topographic image of an \ch{InAsSb} NW deposited on an \ch{In} substrate. The lower inset shows a schematic drawing of the experimental system: The NW (hexagonal, blue) is placed on top of the \ch{In} substrate (dark grey), and measured from the top by the STM tip. Upper-right panel shows the topography of another nanowire with flat top facet and a width of $90$~nm. Lower-right panel shows an atomic scale topography of the top facet of a NW revealing the surface reconstruction of the NW's facet. \textbf{(b)}  Characteristic dI/dV spectra, at $T=350$~mK, showing the induced superconducting gap on the top facet of the NW (blue) versus the parent superconducting gap measured on the \ch{In} substrate (red) fitted with a $s$-wave BCS spectrum (dashed lines). Both are measured  with a metallic tip.\textbf{(c)} Waterfall plot of dI/dV measurement at $T=350$~mK along the arrow in \textbf{(a)} using a superconducting tip.   \textbf{(d)} dI/dV spectra measured with a metallic tip on the \ch{In} substrate for different temperatures. \textbf{(e)} Characteristic spectra measured on \ch{In} (red) and NW (blue) using a superconducting tip. \textbf{(f)} Simulation results for spectra on \ch{In} (red) and NW (blue) convoluted with the typical DOS of a superconducting tip.}
\end{figure}

\newpage{}

\begin{figure}
\begin{centering}
\includegraphics[width=0.95\columnwidth]{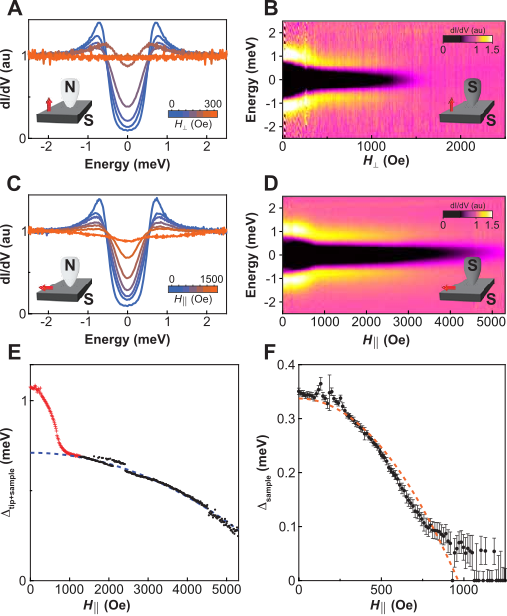}
\par\end{centering}
\caption{\textbf{\label{Fig2} Evolution of superconductivity in an \ch{In} substrate as a function of magnetic field.} \textbf{(a)} Discrete dI/dV curves on the \ch{In} substrate for different values of $H_\perp$ (blue to orange curves) using a metallic tip. \textbf{(b)} Color plot of dI/dV as a function of ramping $H_\perp$ using a superconducting tip. Notice the kink in the peak position at $H_\perp = 300$~Oe when coherence peaks vanish in (a). \textbf{(c, d)} Same as (a, b) but for a parallel magnetic field $H_\parallel$.  {\textbf{(e)} Extracted superconducting gap $\Delta$ of (d) using N-I-S Dynes fit. The blue dashed line corresponds to a Dynes fitting of the black dots. \textbf{(f)} $\Delta_{\rm In}$ extracted by fitting the red part of (e) with S-I-S Dynes equation with $\Delta_{\rm Tip} = 0.5$~meV.}}
\end{figure}

\newpage{}

\begin{figure}
\begin{centering}
\includegraphics[width=0.95\columnwidth]{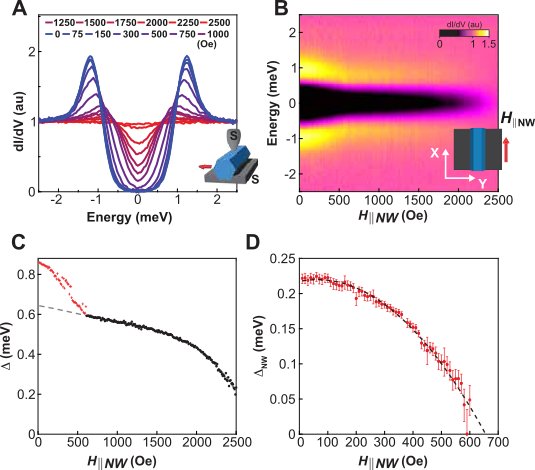}
\par\end{centering}
\caption{\textbf{\label{Fig3} Evolution of proximity induced superconductivity in the NW as a function of magnetic field. (a)} Discrete dI/dV curves on the NW for different values of $H_{\perp NW}$ (blue to red curves) using a superconducting tip. \textbf{(b)} Color plot of the dI/dV on the NW as a function of ramping $H_{\parallel NW}$ using a superconducting tip. \textbf{(c)} Extracted superconducting gap $\Delta$ of (b) using N-I-S Dynes fit. \textbf{(f)} $\Delta_{\rm NW}$ extracted by fitting the red part of (c) with S-I-S Dynes equation with $\Delta_{\rm Tip} = 0.5$~meV.}
\end{figure}

\newpage{}

\begin{figure}
\begin{centering}
\includegraphics[width=0.98\columnwidth]{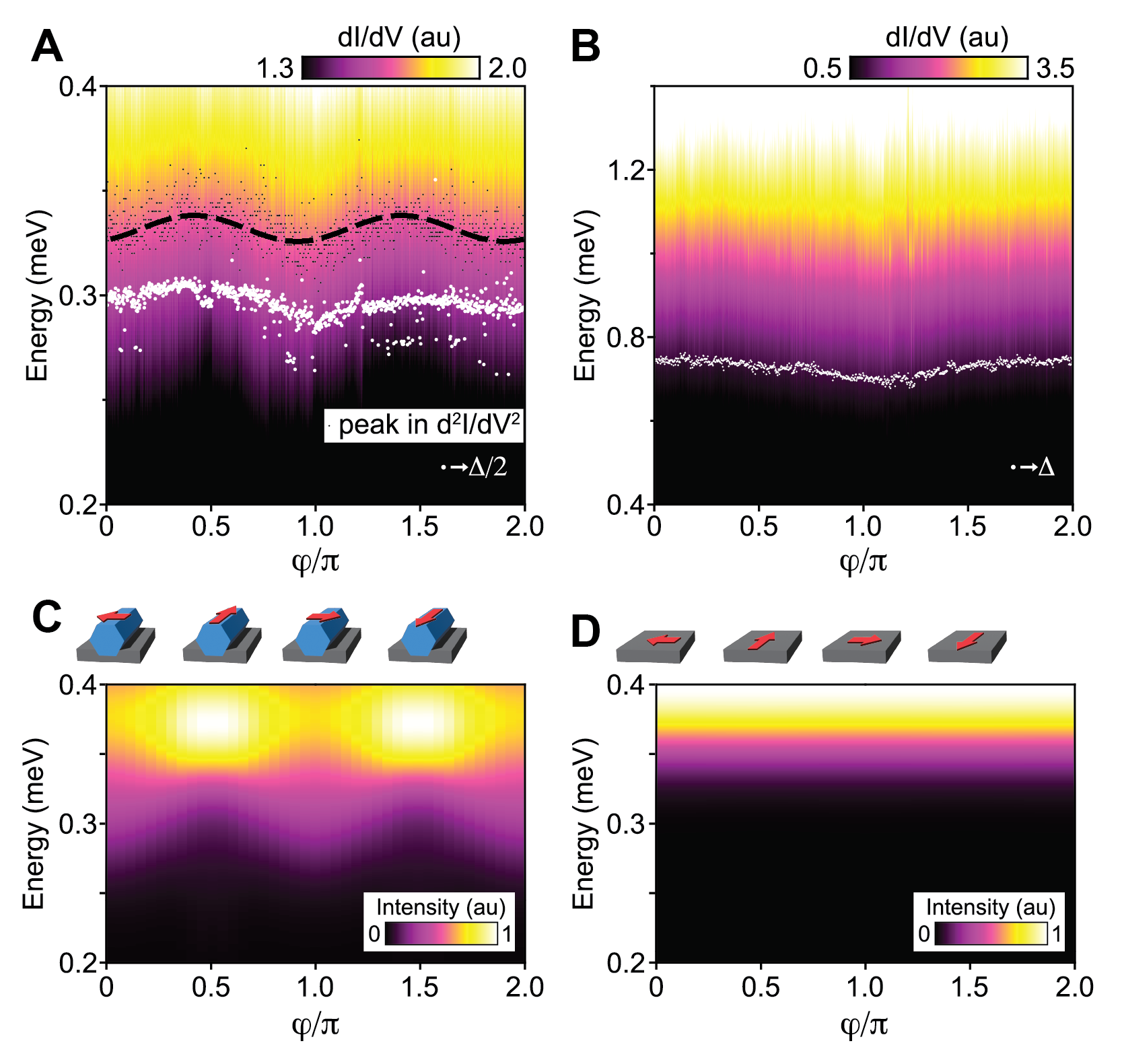}
\par\end{centering}
\caption{\textbf{\label{fig4} Proximity-induced superconductivity in the NW as a function of $H_{\parallel}$ orientation, i.e. in the plane of the substrate. (a)} Color plot of dI/dV on the \ch{NW} as a function of angular rotation of $H_{\parallel} = 1000$~Oe using a superconducting tip. Notice the modulation of the curve with increasing $\phi$. The white dots correspond to $\Delta/2$ where $\Delta$ is extracted from Dynes fit as described in Fig.~\ref{Fig2}. The black dots correspond to the peaks in the d$^2$I/dV$^2$ curves. The black dashed line is the sinusoidal fit to these peaks that emphasizes the periodic modulation. \textbf{(b)} Same as (a) but directly on the In substrate. Here the white dots correspond to $\Delta$ extracted from Dynes fit. \textbf{(c), (d)} Numerical simulations of the NW-substrate heterostructure, showing the LDOS as a function of angular rotation of $H_{\parallel}$, on (c) the top-facet of the NW or (d) the top facet of the In substrate}.
\end{figure}

\newpage{}

\clearpage{}

\section{Sample fabrication process} \label{sm-fab}
Reclining InAsSb nanowires (NWs) were grown by Au-assisted vapor liquid solid (VLS) molecular beam epitaxy (MBE) in a Riber 32 system with vacuum in the mid $10^{-11}$~Torr. An ultrathin ($<1$~nm) layer of Au was evaporated in-situ on the (001) InAs at $\sim100$~°C right after oxide blow-off in a separate chamber attached to the MBE growth chamber. For NWs growth on the (001) surface the substrate was first heated to $\sim600$~°C  under As overpressure, where the gold droplets form, then gradually cooled to the growth temperature $\sim400$~°C. Midway between the two temperatures the In shutter ($\sim5\cdot10^{-7}$) was opened for InAs stem growth. During this cool down process, the (001) surface initially becomes covered with craters comprised of two opposite (111)B facets, which facilitate the nucleation of typically rounded NWs that grow in two opposite (111) directions, parallel to one of the (011) directions. InAs NWs growth in the (111) direction is maintained for an hour forming the stem~\cite{Kang2013Nov, Kang2018Jul}. For axial growth of the InAsSb NWs on the InAs stems the Sb shutter was opened, and the substrate temperature ramped to $\sim450$~°C at a rate of $\sim10$~°C per minute right away. The growth of InAsSb continues for another hour. The temperature (and flux) of Sb and As were $405$~°C ($5.5\cdot10^{-11}$~Torr) and $185$~°C ($2.8\cdot10^{-11}$~Torr), respectively. A one-hour pause is used for adjusting the fluxes between the InAs stem and InAsSb segment growth. The InAsSb NWs were characterized by field emission scanning electron microscopy (FE-SEM, Zeiss Supra-55, 3~kV, working distance $\sim4$~mm), transmission electron microscopy (TEM, Thermo Fisher Scientific Talos F200X, 200 kV). EDS data were obtained with a SuperX G2 four-segment SDD detector with a probe semi-convergence angle of 30~mrad and a beam current of approximately 200~pA. The EDS hyperspectral data was quantified using the Velox software (Thermo Fisher Scientific Electron Microscopy Solutions, Hillsboro, USA), by background subtraction and spectrum deconvolution.

\begin{figure}[h]
\begin{centering}
\includegraphics[width=1\columnwidth]{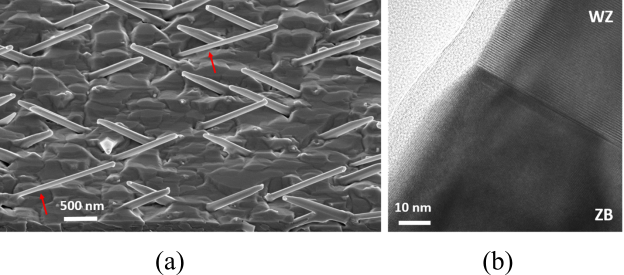}
\end{centering}
\caption{{\bf Nanowires after the growing process.} {\bf (a)} Bird's-view SEM image of as-grown \ch{InAs_{1-x}Sb_{x}} (x = 0.3$\sim$0.4) NWs. The red-colored arrows indicate the transition from InAs stem to InAsSb. {\bf (b)} TEM image of the interface between wurtzite InAs and zincblende InAsSb, showing pure crystallinity.}
\label{fig:NW_image} 
\end{figure}

The low density of reclining InAsSb NWs seen in Fig.~\ref{fig:NW_image}(a) was controlled for successful harvest and transfer of NWs into the STM via a dedicated pumped suitcase attached to the MBE system~\cite{reiner2017hot}. As indicated by the red-colored arrows, the interface between the InAs stem and the InAsSb nanowire can be distinguished by a slight diameter difference. The TEM image in Fig.~\ref{fig:NW_image}(b) shows the clear transition from stacking faults free wurtzite structure InAs to the high crystal quality zincblende InAsSb. 

In the experimental device measured in STM, \ch{InAs_{0.6}Sb_{0.4}} NWs were deposited on top of a $\sim50$~nm thick thin film substrate made of \ch{In} by the process of stamping in high vacuum conditions. In this stamping process, the nanowires break from their stems and are transferred to the In substrate. As a result of this transfer, the nanowire distribution on top of the thin film can be uneven. Additionally, the scanning range of the STM tip is smaller than the sample size, meaning some of the NWs end up in a region inaccessible to the tip.

\begin{figure}
\begin{centering}
\begin{tabular}{cc}
      \includegraphics[width=0.49\columnwidth]{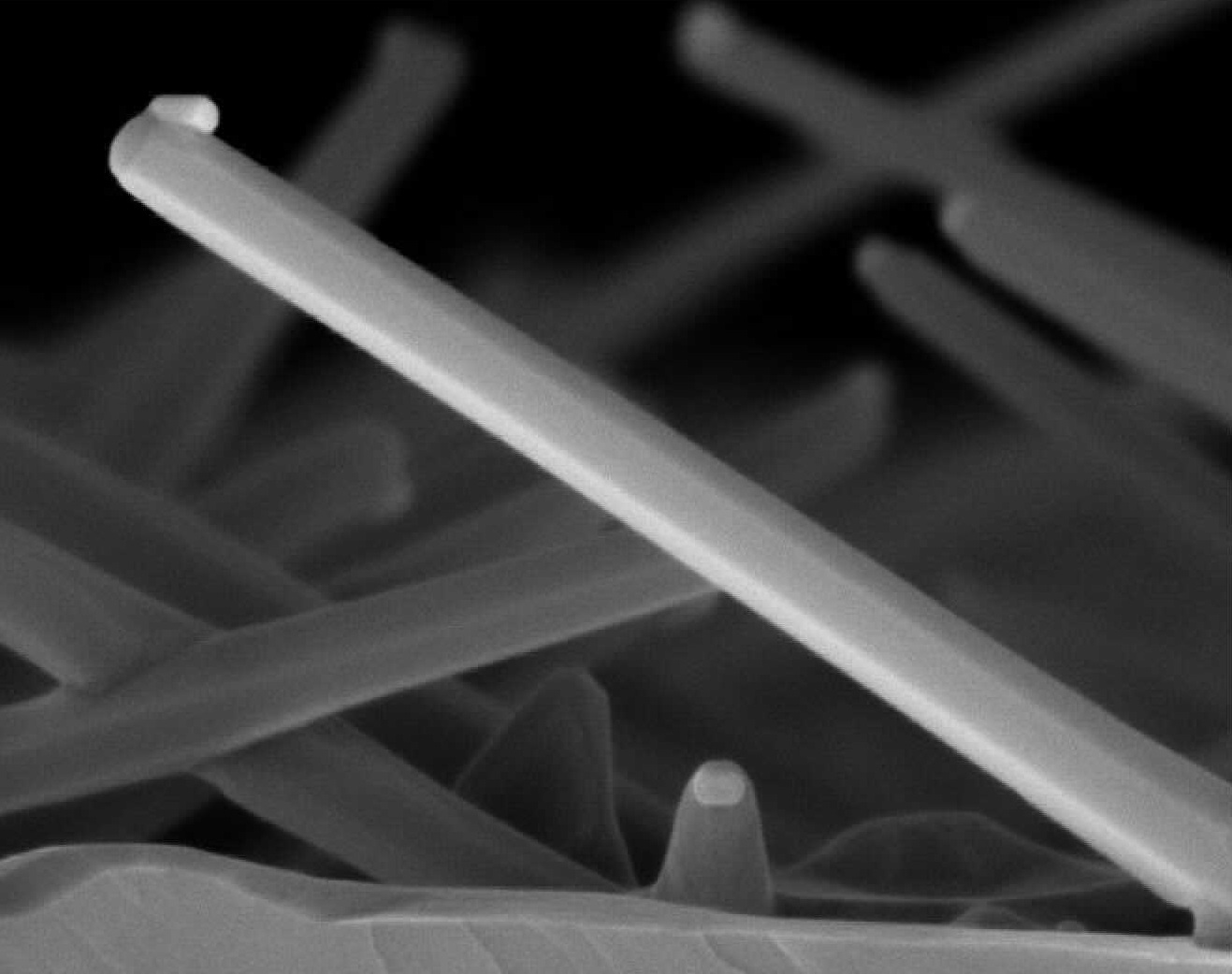} & \includegraphics[width=0.49\columnwidth]{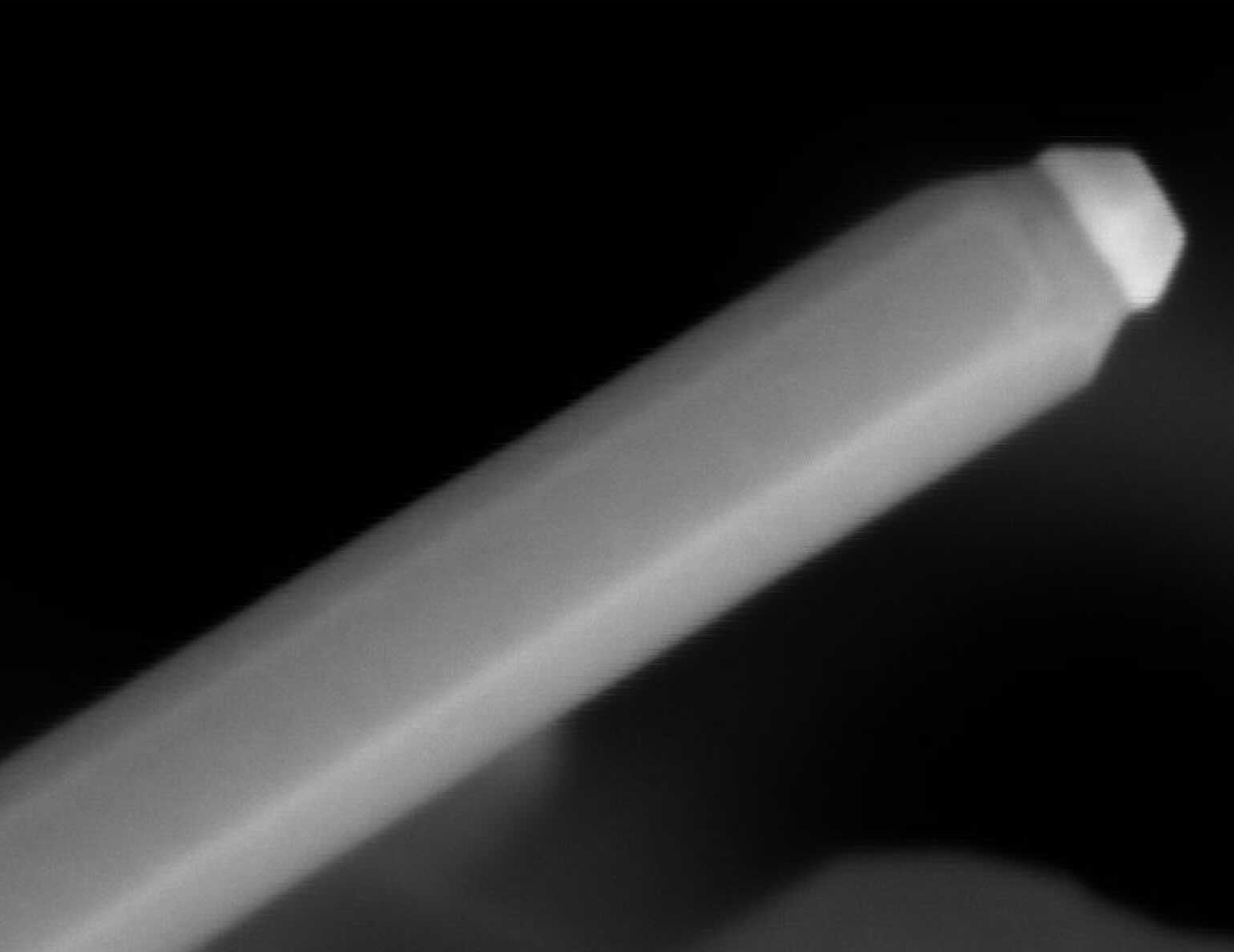}\\
      (a) & (b)\\
      \includegraphics[width=0.49\columnwidth]{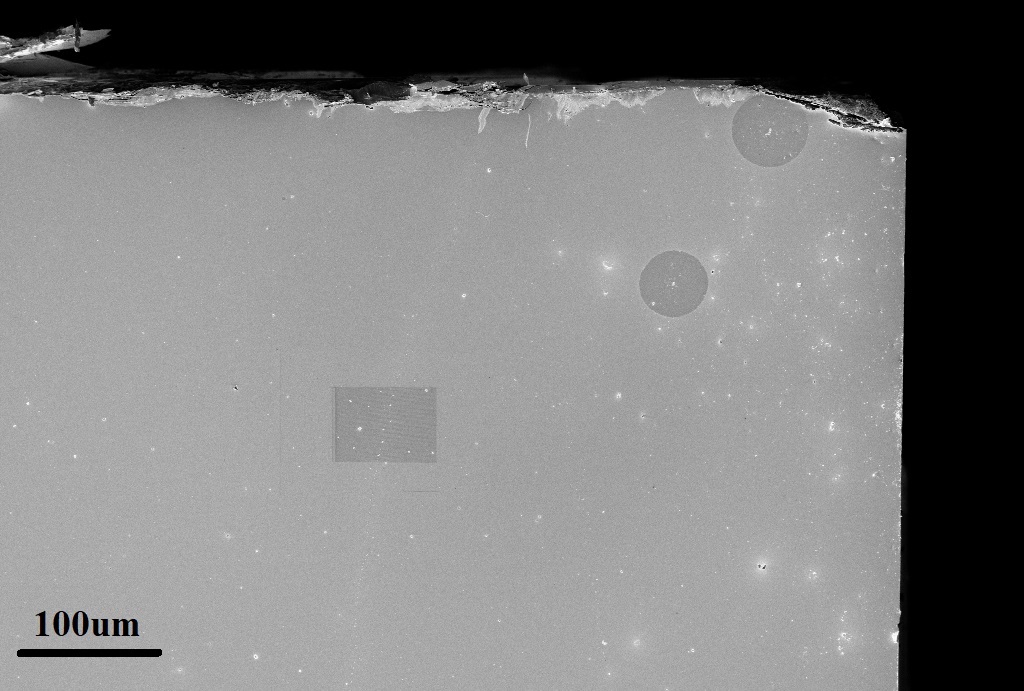} & \includegraphics[width=0.49\columnwidth]{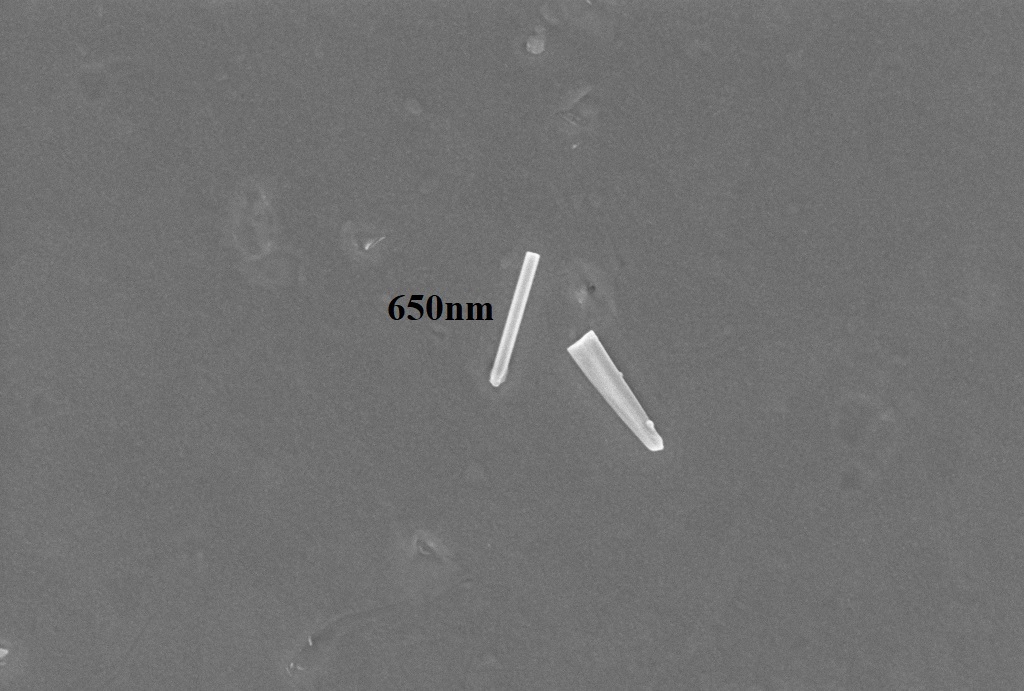}\\
      (c) & (d)\\
    \end{tabular}
\end{centering}
\caption{\textbf{SEM images of the sample before and after stamping. (a,b)} SEM images of nanowires on growth system, before stamping. \textbf{(c)} SEM image of the sample after stamping. Nanowire distribution is sparse. \textbf{(d)} SEM image of nanowire pieces on the sample, after stamping.}\label{fig:SEM}
\end{figure}

SEM images of the nanowires before stamping show many long wires with the expected hexagonal shape, grown at an angle to allow better stamping [Fig.~\ref{fig:SEM}(a,b)]. Post-experiment SEM images of our sample show nanowires concentrated in patches, most of them broken into small pieces [Fig.~\ref{fig:SEM}(c,d)].

\section{Additional Measurements}
\subsection{Temperature dependence of the superconducting gap in a S-I-S configuration}
\begin{figure}
\begin{centering}
\includegraphics[width=0.7\columnwidth]{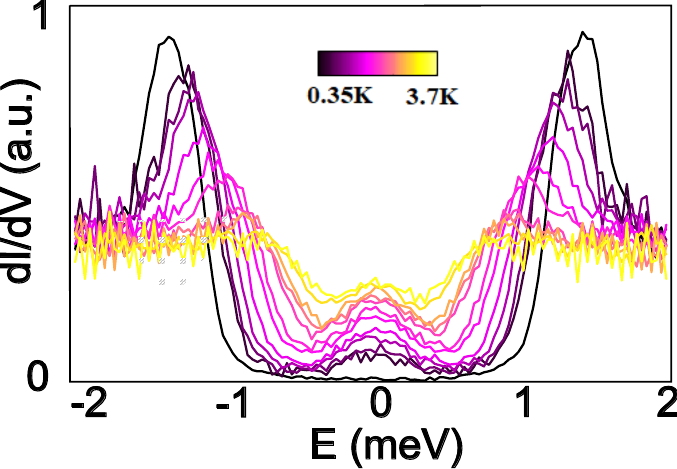}
\end{centering}
\caption{\textbf{Evolution of the dI/dV with respect to temperature in a S-I-S configuration.} dI/dV measured on \ch{In} using a SC tip for increasing temperatures, from $T=0.35$~K (black) to $T=3.7$~K (yellow).}
\label{fig:temp_SC_tip} 
\end{figure}

While measuring the DOS of a sample using a STM, a typical dI/dV measurement provides actually a convolution between the DOS of the sample and the tip (in the tunneling regime). Hence, when using a superconducting tip, the interpretation of the dI/dV measurement is no longer trivial: theoretically, one should observe two coherence peaks at $E=\pm(\Delta_{\rm sample}+\Delta_{\rm tip})$ and another two at $E=\pm\left|\Delta_{\rm sample}-\Delta_{\rm tip}\right|$. The later, nonetheless, should be strongly suppressed at ultra-low temperatures. To illustrate this phenomenon, we show in Fig.~\ref{fig:temp_SC_tip} a typical measurement of the \ch{In} substrate using a SC tip. Each curve is taken at a different temperature, shown in the colorbar. As expected, we observe two main coherence peaks at $E\simeq\pm2\Delta_{\rm In}=\pm1.2$~meV which decrease with temperature as a result of thermally-induced decoherence effects, completely disappearing at the superconducting critical temperature $T_{\rm c}=3.7$~K. In addition, there is a small broadened peak at $E\simeq 0$, clearly visible at higher temperatures as the Fermi function starts deviating substantially from the step function at $T\rightarrow0$. This peaks is the result of the small difference between the superconducting gaps of the sample and the tip. We stress that this zero-bias peak does not correspond to an Andreev bound state but it is rather an artifact of the convolution with temperature of both the individual DOS.

\subsection{Changes in the nanowire-substrate coupling} \label{sm-coupling}
\begin{figure}
\begin{centering}
\includegraphics[width=0.8\columnwidth]{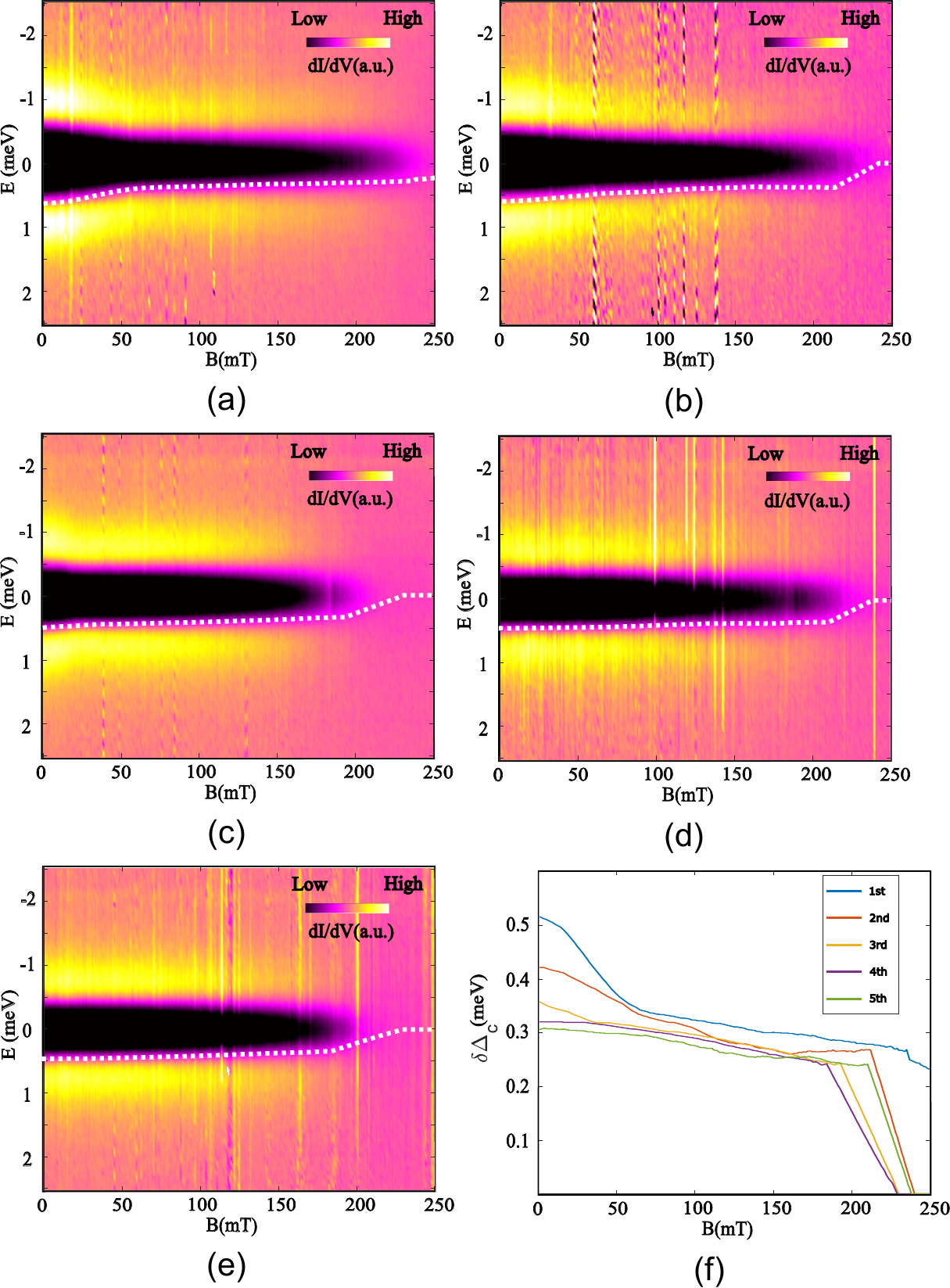}
\par\end{centering}
\caption{\textbf{\label{fig:lose_NW} Subsequent dI/dV measurements on the superconducting NW under increasing $B_{in-plane}$ using a superconducting tip.} {\bf (a-e} Consecutive measurements of the dI/dV vs $B_{in-plane}$ taken at the same place on the NW. Notice how the kink in (a) that gradually disappears over (b-e). The dotted white line in the plots corresponds to the maximum of the first derivative of the dI/dV at each $B_{in-plane}$. These lines are plotted all together in \textbf{(f)} for comparison.}
\end{figure}

Due to differences in contact surface and the degree of penetration of the different NWs and the amorphous In substrate, the mechanical coupling and therefore the hybridization between the superconducting In and the NW may vary. In this appendix we show that it is actually possible to experimentally change the hybridization by physically/electrically disturbing the NW using the STM tip, altering its hybridization with the surface. Such a process is demonstrated in Fig.~\ref{fig:lose_NW} where we take the dI/dV measurement as a function of increasing magnetic field 5 times in a row. All the measurements in the figure were taken at the same place on the NW using a SC tip. In the false color plot Fig.~\ref{fig:lose_NW}(a), as we increase the magnetic field from $0$~T, we see a kink that it is better appreciated in the dotted white line, which draws the limits of the superconducting gap. This kink occurs at the magnetic field at which the NW loses its superconductivity. For magnetic fields higher than this kink, superconductivity can only arise from the SC tip. We observe in the subsequent measurements (see next plots) that the kink shifts to lower magnetic fields, completely disappearing In the 5th [see Fig.~\ref{fig:lose_NW}(e)]. To better appreciate this phenomenon, we represent in Fig.~\ref{fig:lose_NW}(f) all the white dotted lines of previous plot.

We attribute these changes in the induced superconductivity to the changes in mechanical coupling between the NW with the superconducting substrate. During the course of the measurement, the electrical interaction between the tip and the NW can lead to the physical movement of the NW. The changes in the topographic profile (not shown) corroborate this interpretation.

\begin{figure}
\begin{centering}
\begin{tabular}{cc}
\includegraphics[width=0.3\columnwidth]{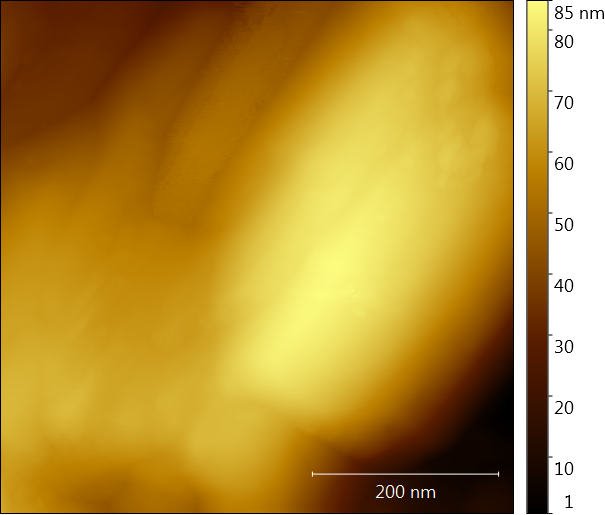} &
\includegraphics[width=0.3\columnwidth]{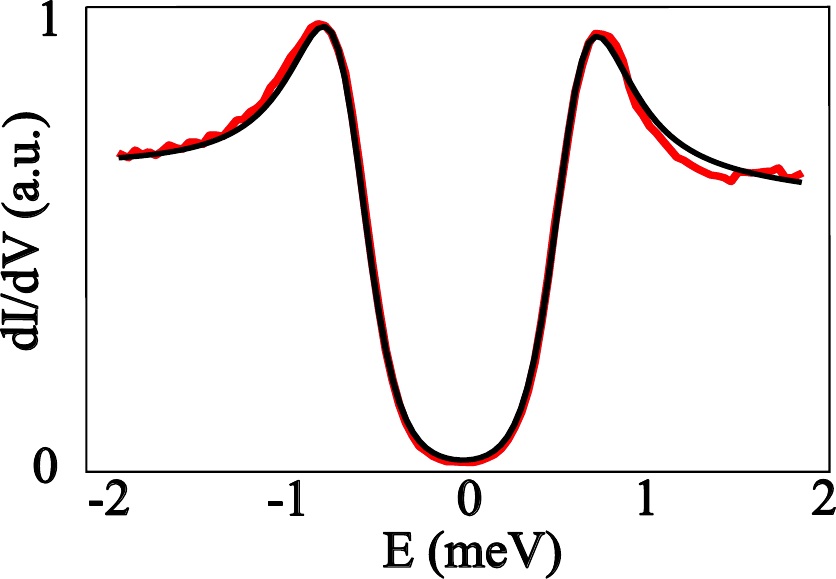}\\
      (a) & (b)\\
 \includegraphics[width=0.3\columnwidth]{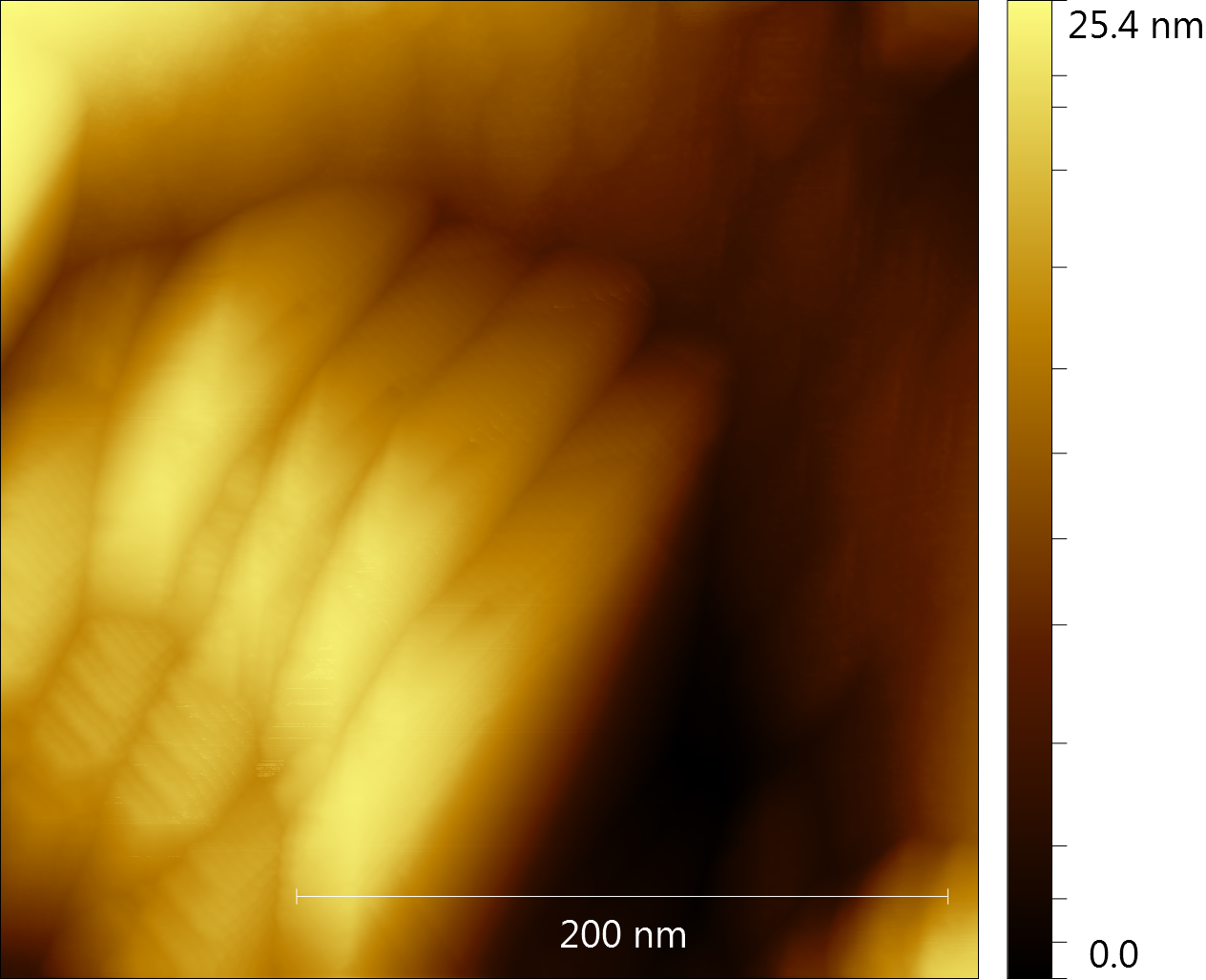} &
\includegraphics[width=0.3\columnwidth]{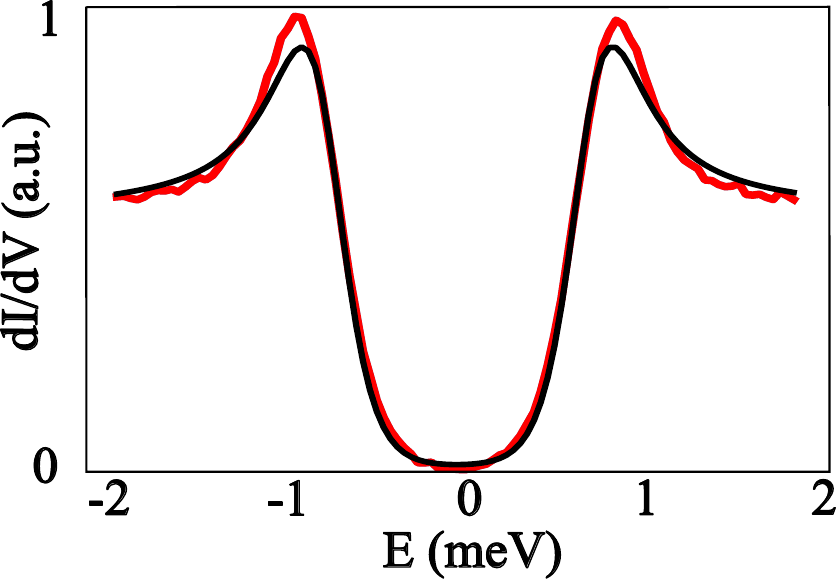}\\
      (c) & (d)\\
      \includegraphics[width=0.3\columnwidth]{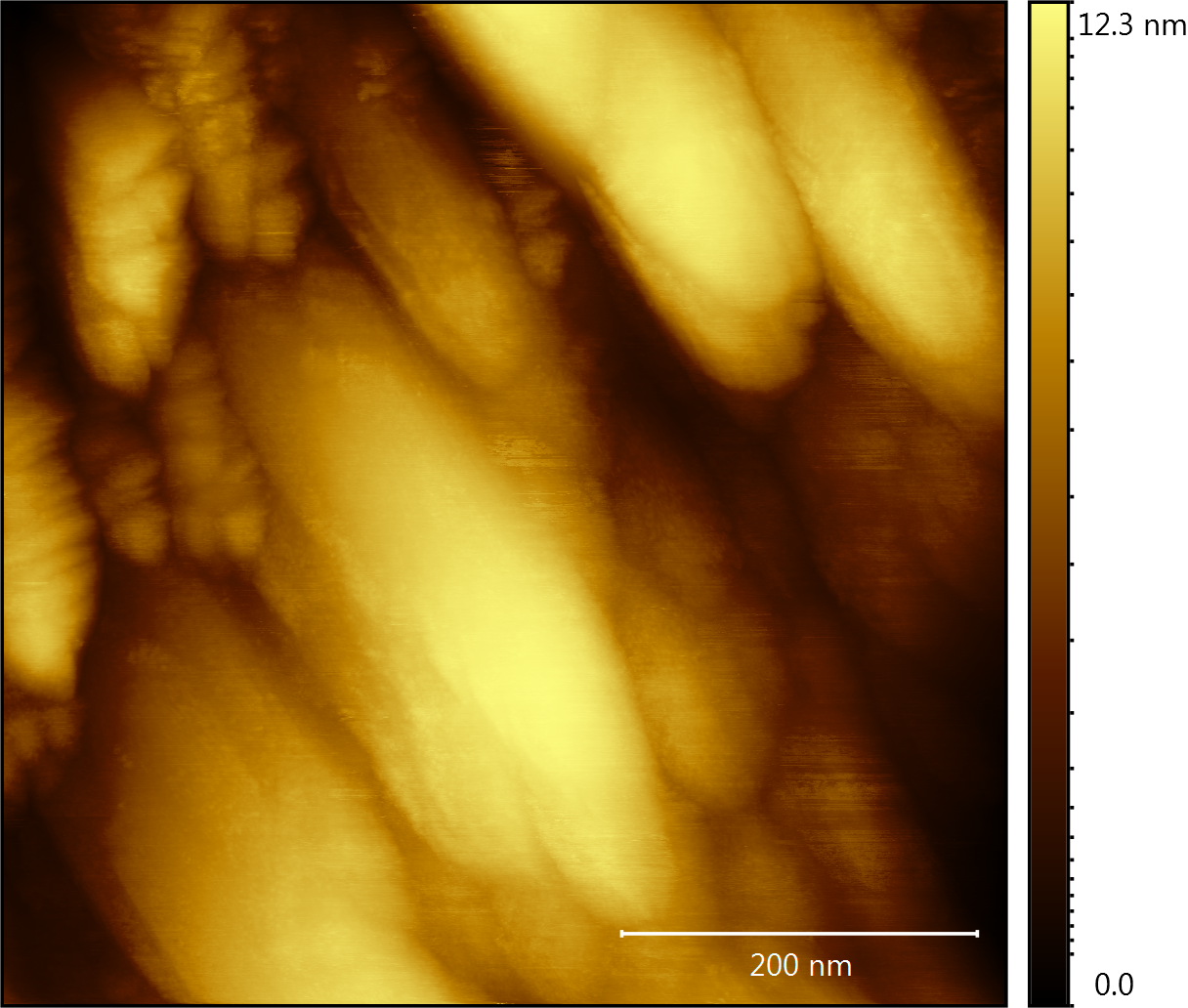} &
\includegraphics[width=0.3\columnwidth]{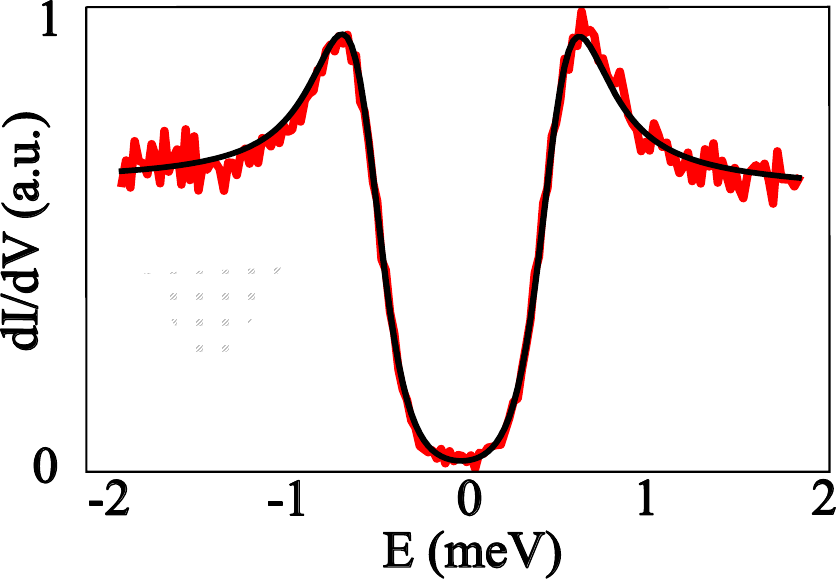}\\
    (e) & (f)\\
 \includegraphics[width=0.3\columnwidth]{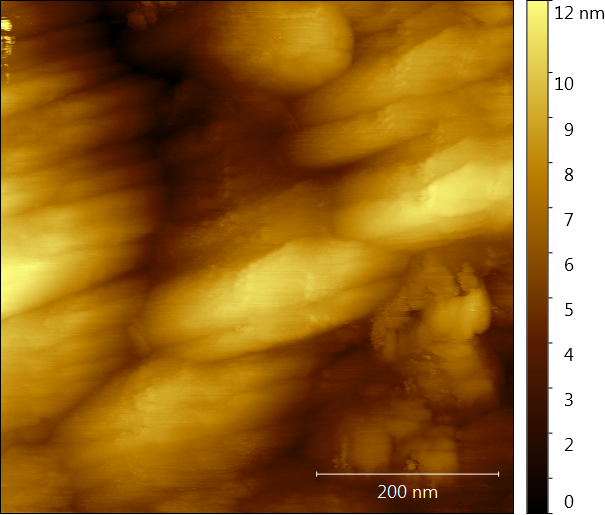} &
\includegraphics[width=0.3\columnwidth]{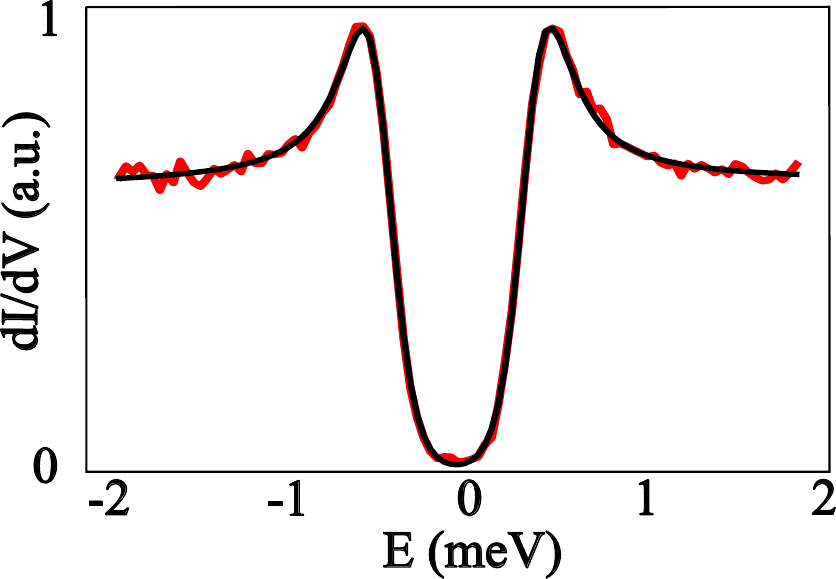}\\
    (g) & (h)\\
    \end{tabular}

\end{centering}
\caption{\textbf{\label{fig:different_NWs}} {\bf Comparison of the dI/dV among different NWs.} Topographies of 4 different NWs encountered during the experiment {\bf (a,c,e,g)} with characteristic single gap spectrum from each {\bf (b,d,f,h)}. We observe that our sample includes NWs of different length, thickness and quality. And moreover, each of them are coupled differently to the In substrate, exhibiting different superconducting gaps as a result ($0.65$~meV, $0.76$~meV, $0.56$~meV, and  $0.55$~meV, correspondingly).}
\end{figure}

We have also observed small changes in the coupling between different, untouched NWs. In Fig.~\ref{fig:different_NWs} we show four NWs measured during the experiment that in addition can be compared to the one appearing in the main text. Each shows the characteristic DOS measured with a normal metallic tip. For every NW, the gap is consistent along different locations on the wire, but is different among different NWs. This agree with our interpretation that different NWs come to have a distinct coupling strength to the underlying SC substrate, resulting in variance of the induced SC gap size.

\subsection{Dependence of the superconducting gap on the magnetic field direction}

\begin{figure}
\begin{centering}
\includegraphics[width=0.9\columnwidth]{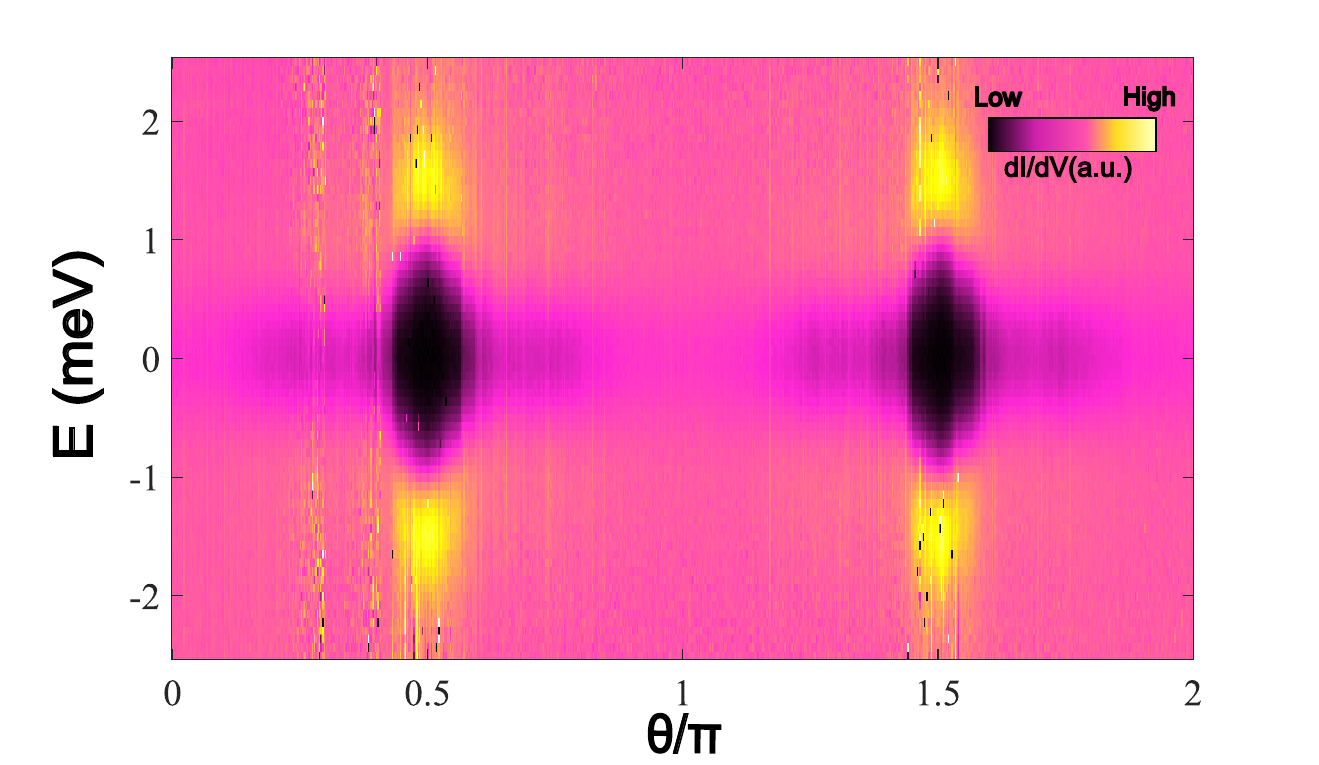}
\par\end{centering}
\caption{\textbf{\label{fig:In_YZ} Evolution of the dI/dV with the in/out-of-plane magnetic field orientation.} dI/dV vs the magnetic field orientation. The angle $\theta=\left\{\pi/2,3\pi/2\right\}$ corresponds to a parallel magnetic field to the \ch{In} substrate while $\theta=\left\{0,\pi\right\}$ is perpendicular to it.}
\end{figure}

In order to complete the characterization of our nanodevice, we show in Fig.~\ref{fig:In_YZ} a typical measurement of the dI/dV changing the orientation of the magnetic field in the Y/Z plane, i.e., in/out-of-plane, at $B=0.1$~T. In this configuration, the angle $\theta=\left\{\pi/2,3\pi/2\right\}$ corresponds to a parallel magnetic field to the \ch{In} substrate while $\theta=\left\{0,\pi\right\}$ is perpendicular to it. We observe that there is only a superconducting gap when the magnetic field is parallel to the substrate, while it completely disappears when it is perpendicular. This phenomenon is due to the suppression of the Meissner effect in low-dimensional heterostructures: when the magnetic field is parallel to the substrate, orbital effects are much weaker and thus the Meissner effect is diminished. Interestingly, Fig.~\ref{fig:In_YZ} shows a tiny residual superconducting gap after the main one is gone. As explained before, this gap corresponds to the one of tip, whose dimensionality is smaller than the substrate itself. However, the fact that it does not exhibit superconductivity at $\theta=\left\{\pi/2,3\pi/2\right\}$ implies that is not purely 0D, but has some shape instead that provides different critical magnetic fields for different directions.

\subsection{Effect of Current Noise due to Magnet Controller on the SC spectrum}

\begin{figure}
\begin{centering}
\includegraphics[width=0.9\columnwidth]{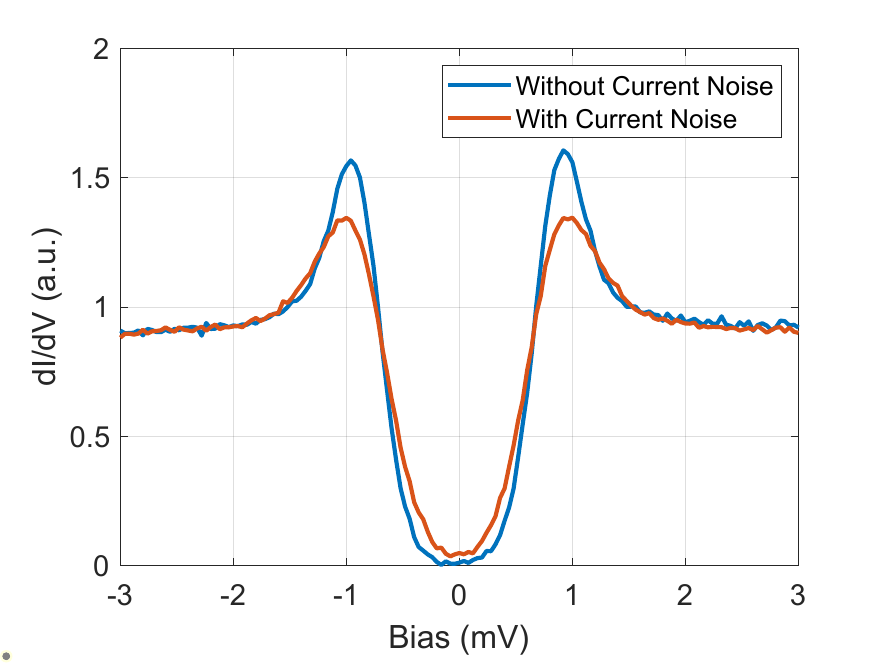}
\par\end{centering}
\caption{\textbf{\label{fig:mageff} dI/dV spectrum on top of the NW with and without Current Noise.} The SC gap and the coherence peaks get affected due to the noise from Magnet Control Power Supply. The two spectrum were taken on the same location when the magnet controller was off(blue) and when the magnet controller was on(orange).}
\label{sm-Bcontrol}
\end{figure}

The underestimation of the superconducting gap in Fig. \ref{Fig2} and Fig. \ref{Fig3} while fitting the dI/dV spectrum with S-I-S Dynes formula can be attributed to the current fluctuations generated by the magnet controller. In Fig. \ref{fig:mageff}, we demonstrate this effect where in the absence of the current noise arising from the magnet controller, the coherence peaks are sharper and we observe a hard gap(blue). Upon turning on the magnet controller(still 0T B field), the spectrum smears, the coherence peaks become less sharp and the gap turns softer(orange). For continuous field measurement, we were not able to eliminate this noise, but since the effect is the same for the NW and the Indium substrate, we still use SIS Dynes fit to obtain an accurate estimation of the relative gaps
on both.

\section{Fitting equations}

The differential conductance dI/dV in the tunneling regime, including temperature effects, is given by
\begin{eqnarray}
    \frac{\mathrm{dI}}{\mathrm{dV}}(\vec{r},V)&\propto&\int\left(\rho(\vec{r},E)\frac{\partial\rho_{\mathrm{t}}(E+eV)}{\partial V}\left[f(E)-f(E+eV)\right]\right.\nonumber \\
    &&\left.-\rho(\vec{r},E)\rho_{\mathrm{t}}(E+eV)\frac{\partial f(E+eV)}{\partial V}\right)\mathrm{dE}.
    \label{Eq:dIdV with T}
\end{eqnarray}
where $\rho(\vec{r},E)$ is the density of states of the sample at the point $\vec{r}$ where the STM tip is located, and $\rho_{t}(E)$ is the density of states of the tip, which in its simplest form can be written as
\begin{eqnarray}
\rho_{\mathrm{t}}(E)=\left|\Re\left\{ \frac{\rho_{0}\left(E-i\Gamma_{\mathrm{t}}\right)}{\sqrt{\left(E-i\Gamma_{\mathrm{t}}\right)^{2}-\Delta^{2}}}\right\} \right|
\end{eqnarray}
being $\Gamma_{t}$ the Dynes parameter of the tip and $\rho_{0}$ the normal DOS of the tip at its Fermi level. The function $f(E)$ in the above equation is the Fermi-Dirac distribution for a given temperature $T$. We fit our dI/dV measurements with Eq.~\eqref{Eq:dIdV with T} using $\Delta$, $\rho_0$ and $\Gamma_{\rm t}$ as free parameters. For the magnetic field dependent measurements, we moreover assume that $\Delta_{t}$ depends on $B$ as
\begin{equation}
    \Delta_{t}\left(B\right)=\Delta_{t}^{0}\left(1-\frac{B}{B_{t}^{c}}\right)^{1/2},
\end{equation}
where $B_{\rm c}$ is the critical magnetic field of the sample.

\begin{figure}
\begin{centering}
\includegraphics[width=0.9\columnwidth]{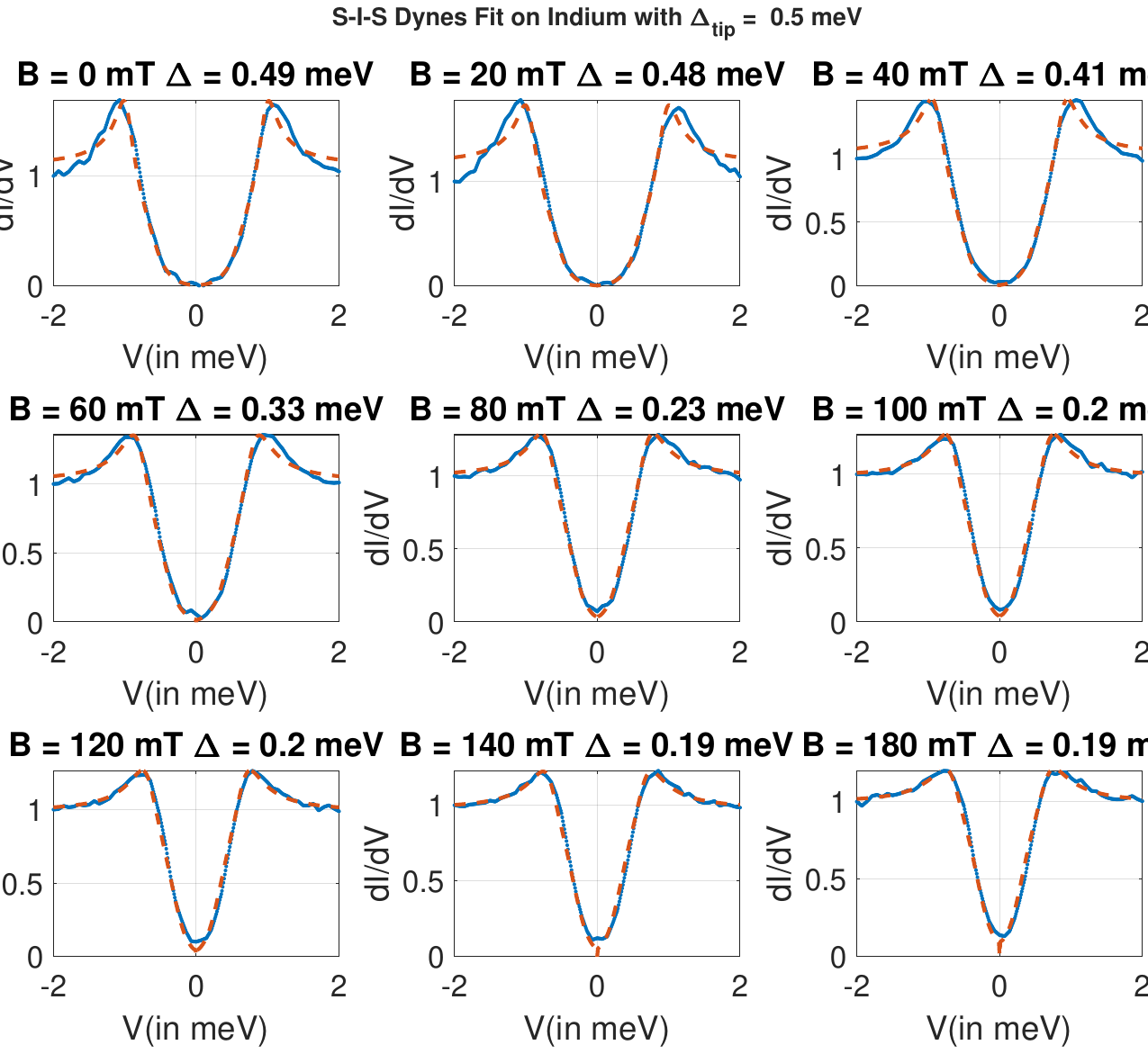}
\par\end{centering}
\caption{\textbf{\label{fig:SIS_In} S-I-S Dynes fit overlaid on dI/dV spectra on Indium substrate for a series of in-plane magnetic field with $\Delta_{t} = 0.5~\mathrm{meV}$.} }
\end{figure}

\begin{figure}
\begin{centering}
\includegraphics[width=0.9\columnwidth]{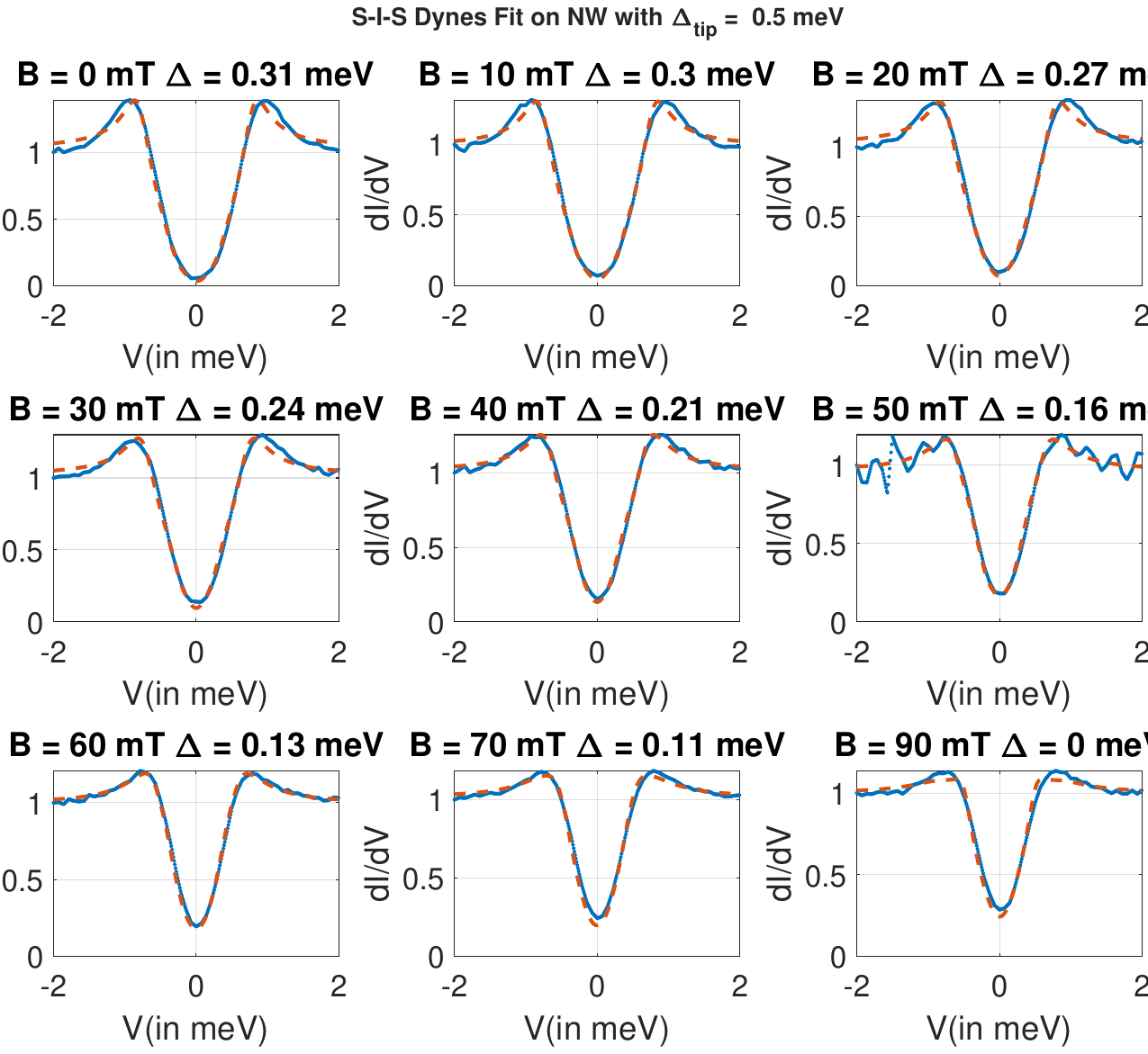}
\par\end{centering}
\caption{\textbf{\label{fig:SIS_NW} S-I-S Dynes fit overlaid on dI/dV spectra on NW for a series of in-plane magnetic field with $\Delta_{t} = 0.5~\mathrm{meV}$.}}
\end{figure}

\section{Theoretical methods} \label{sm-theory}

Our model describes the experimental setup illustrated in Fig.~\ref{Fig1}(a): an \ch{InAs_{0.6}Sb_{0.4}} NW with hexagonal cross-section that is deposited on top of an \ch{In} substrate. Below some critical temperature, the substrate becomes a superconductor and may induce superconductivity in the wire by proximity effect. Following Ref.~\cite{Thesis_sam}, we describe the NW-substrate heterostructure through the Bogoliubov-de-Gennes Hamiltonian
\begin{eqnarray}
H &=& \left[ \vec{k}  \frac{\hbar^2}{2m^*(\vec{r})} \vec{k} - E_{\rm F}(\vec{r}) + e\phi(\vec{r})\right]\sigma_0\tau_z \nonumber \\ 
 &+& \frac{1}{2} \left[\vec{\alpha}(\vec{r})\cdot \left(\vec{\sigma}\times\vec{k}\right) + \left(\vec{\sigma}\times\vec{k}\right)\cdot \vec{\alpha}(\vec{r}) \right]\tau_z \nonumber \\
&+&\frac{1}{2}\mu_Bg(\vec{r})\vec{B}\cdot\vec{\sigma}\tau_z+ \Delta(\vec{r},\vec{B})\sigma_y\tau_y,
\label{Eq:H}
\end{eqnarray}
written in the Nambu basis $\Psi=(\psi_{\uparrow},\psi_{\downarrow},\psi_{\uparrow}^\dagger,\psi_{\downarrow}^\dagger)$. The first three terms correspond to the kinetic energy, being $m^*$ the effective mass; band-bottom of the conduction band with respect to the Fermi level $E_{\rm F}$; and the electrostatic potential $e\phi(\vec{r})$, respectively. The following term is the spin-orbit (SO) interaction, where $\vec{\alpha}$ is the SO coupling. The next term is the Zeeman field, being $g(\vec{r})$ the $g$-factor and $\vec{B}$ the magnetic field. And the last term is the superconducting pairing, with $\Delta$ the superconducting pairing amplitude. This Hamiltonian applies for both, the NW and the substrate. This is why all the parameters are spatial dependent, as they take different values for different materials.

\begin{figure}
    \centering
    \includegraphics[width=0.95\columnwidth]{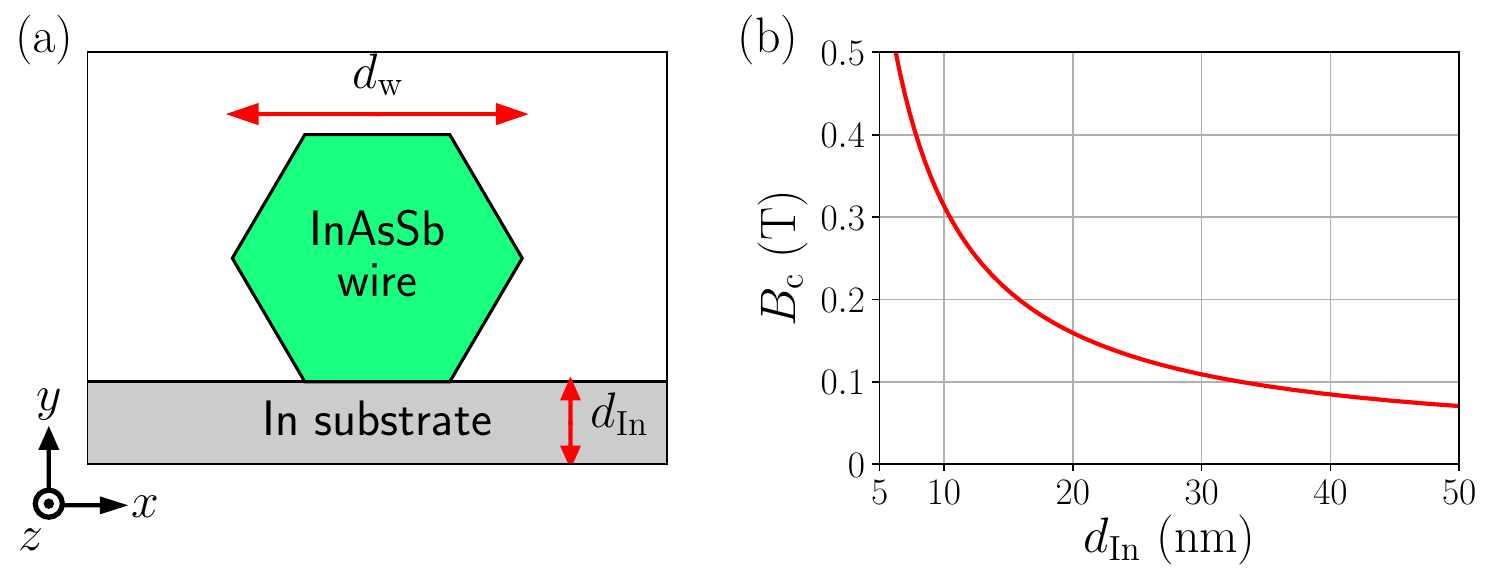}
    \caption{\textbf{Sketch of the theoretical model.} {\bf (a)} Sketch of the model: an hexagonal cross-section InAsSb nanowire (green) of width $d_{\rm w}$ is deposited on top of an In substrate (grey) of thickness $d_{\rm In}$. {\bf (b)} Critical magnetic field of the superconducting phase of thin-film In as a function of its thickness.}
    \label{Fig1-theo}
\end{figure}

Let us make a few remarks regarding this Hamiltonian. Firstly, we assume the Hamiltonian is translational invariant along the wire's direction, which lies on the $z$-direction. Hence, $k_z$ is a good quantum number that is conserved, and $\vec{r}=(x,y)$. Secondly, orbital magnetic effects are included through the minimal substitution $\vec{k}\rightarrow \vec{k}-(e/\hbar)\vec{A}$, where $\vec{A}$ is the magnetic potential vector. We use the symmetric gauge to describe this potential vector, that is $\vec{A}=-\frac{1}{2}\vec{r}\times\vec{B}=\frac{1}{2}(-yB_z,xB_z,yB_x-xB_y)$, being $(x,y)=(0,0)$ the center of the wire. In the next section we will discuss which new terms introduce this orbital effects. Thirdly, we introduce some (random chemical potential) disorder in the substrate to model the roughness of the substrate. 

Notice we do not include a winding of the superconducting phase in the Hamiltonian as this effect is expected to be small since just one facet of the NW is in contact with the SC. We nevertheless take into account the magnetic field dependence of the SC gap through~\cite{tinkham:book04}
\begin{equation}
    \Delta(B)=\Delta_0\sqrt{1-\left(\frac{B}{B_{\rm c}}\right)^2},
\end{equation}
where $B_{\rm c}$ is the critical magnetic field. This critical magnetic field depends on the thickness of the thin film substrate $d_{\rm In}$ (and thus, on the magnetic field orientation with respect to the substrate plane). More particularly, using London theory, one finds~\cite{tinkham:book04}
\begin{equation}
    B_{\rm c}=\frac{B_{\rm c}^{\rm (bulk)}}{\sqrt{1-\frac{2\lambda}{d_{\rm In}}\tanh{\left(\frac{d_{\rm In}}{2\lambda}\right)}}},
    \label{Eq:Bc}
\end{equation}
where $\lambda$ is the effective penetration length, that we take\footnote{In principle, the penetration depth also depends on the thickness of the SupC in thin films as $\lambda\simeq\lambda^{\rm (bulk)}\coth{\left(d/2\lambda_0\right)}$ (see general discussion on \ch{In} films in Refs.~\cite{Toxen:PR61, Toxen:PR62, Chaudhari:PR66, tinkham:book04}). However, this equation provides values for $\lambda$ orders of magnitude larger than $30$~nm, meaning that it may work for a dirty SC. For a clean one, if $\lambda\ll l_e$ (being $l_e$ the electron mean free path), the equation that holds is $\lambda\simeq 0.64\lambda^{\rm (bulk)}\sqrt{\xi_0/l_e}$ that we approximate to $30$~nm.} $\lambda\simeq30$~nm. In Fig.~\ref{Fig1-theo}(b), we show this critical magnetic field as a function of the In thickness $d_{\rm In}$.

In addition to our description of the nanostructure, we compute the electrostatic potential $\phi(\vec{r})$ by solving the Poisson equation in the Thomas-Fermi approximation
\begin{eqnarray}
\vec{\nabla}\left(\epsilon(\vec{r})\vec{\nabla}\phi(\vec{r})\right)=\rho_{\rm wire}^{\rm (TF)}(\vec{r})+\rho_{\rm acc}(\vec{r}),
\label{Eq:Poisson}
\end{eqnarray}
where $\epsilon(\vec{r})$ is the dielectric permittivity across the system, $\rho_{\rm wire}^{\rm (TF)}(\vec{r})$ accounts for the charge density inside the wire, and $\rho_{\rm acc}(\vec{r})$ allows to describe the charge accumulation layer that it is typically present at the uncapped facets of these wires. We compute this equation self-consistently, as the charge density of the wire depends in turn on the electrostatic potential $\phi(\vec{r})$. We impose a boundary condition in the substrate $V_{\rm In}$, which corresponds to the band bending between InAsSb and In as a result of the work-function difference between both materials. We try with different values for $V_{\rm In}$ so that we obtain similar DOS than the one observed experimentally.

A comprehensive list with the values of all the parameters used in our simulations can be found in Table~\ref{tab:table_params}. To obtain the parameters for the values of \ch{InAs_{0.6}Sb_{0.4}}, we simply interpolate between the ones of \ch{InAs} and \ch{InSb}, using a non-linear interpolation, as shown in Refs.~\cite{Webster:JOP15, Suchalkin:IOP16, Manago:AIP21}.

\begin{table}
\centering
\caption{\label{tab:table_params}%
 Material parameters used in our simulations. The spin-orbit coefficient $\alpha$ is computed through the Kane parameter $P$, and semiconductor gaps $\Delta_{\rm g}$ and $\Delta_{\rm soff}$, as explained in Refs.~\cite{Escribano2020Jul, Thesis_sam}. The effective mass $m_{\rm eff}$, the $g$-factor and the semiconductor gaps are found using a non-linear interpolation, as shown in Refs.~\cite{Webster:JOP15, Suchalkin:IOP16, Manago:AIP21}. The parent superconducting gap of \ch{In} is inferred from experiments. The rest of \ch{In} parameters correspond to the ones of a normal metal.}
\setlength\tabcolsep{0pt}
\begin{tabular*}{0.48\textwidth}{@{\extracolsep{\fill} } lcc }
\hline\hline
\textrm{}&
\textrm{\ch{InAs_{0.6}Sb_{0.4}}}&
\textrm{In}\\
\hline\hline
$m_{\rm eff}$ [$m_0$] & 0.0082  & 1\\
$E_{\rm F}$ [eV] & 0  & -5 \\
$g$ & 110  & -2 \\
$\Delta$ [meV]  & 0  & 0.6 \\
$P$ [eV$\cdot$nm]  & 1  & - \\
$\Delta_{\rm g}$ [eV]  & 0.10  & - \\
$\Delta_{\rm soff}$ [eV]  & 0.20 & - \\
$\epsilon_{r}$  & 15.5  & $\infty$ \\ \hline\hline
\end{tabular*}
\end{table}

The Schr\"odinger-Poisson equation is solved self-consistently using FDM and FEM with the routines implemented in Ref.~\cite{MajoranaNanowiresQSP_v1}. We diagonalize the Hamiltonian for different momentum $k_z$ so that we obtain the different subbands $E^{(j)}(k_z)$ and their related wavefunction $\Psi^{(j)}(\vec{r},k_z)$. From there, we can compute the local density of states $\rho(\vec{r},E)$ in any point of the system at a given temperature $T$
\begin{eqnarray}
\rho(\vec{r},E)=\sum_{n} \int_{k_z} \dd k_z  \frac{1}{k_BT\sqrt{2\pi}}\nonumber \\
\cdot \exp{-\frac{1}{2}\left(\frac{E-E^{(n)}(k_z)}{k_BT}\right)^2} \left| \Psi^{(n)}(\vec{r},k_z)\right|^2.
\end{eqnarray}
The differential conductance measured by the STM tip at any point $\vec{r}$ at small bias voltage $V$ can be computed using linear response theory as
\begin{eqnarray}
\mathrm{I}(\vec{r},V)\propto\int \rho(\vec{r},E)\rho_{\rm t}(E+eV) \mathrm{dE} \nonumber \\ 
\cdot \left[f(E)-f(E+eV)\right],
\end{eqnarray}
\begin{eqnarray}
\mathrm{\frac{dI}{dV}}(\vec{r},V)&\propto& \int \left(\rho(\vec{r},E)\frac{\partial \rho_{\rm t}(E+eV)}{\partial V}\left[f(E)-f(E+eV)\right]\right.\nonumber \\
&&\left.- \rho(\vec{r},E)\rho_{\rm t}(E+eV) \frac{\partial f(E+eV)}{\partial V} \right)\mathrm{dE},
\end{eqnarray}
where $\rho(\vec{r},E)$ is the density of states of the sample at the point $\vec{r}$ where the STM tip is located (at the $T\rightarrow 0$ limit), and $\rho_{\rm t}(E)$ is the density of states of the tip, that we model as
\begin{eqnarray}
\rho_{\rm t}(E)=\left|\mathrm{Re}\left\{\frac{\rho_{0}(E-i\Gamma_{\rm t})}{\sqrt{(E-i\Gamma_{\rm t})^2-\Delta^2}}\right\}\right|,
\label{Eq:DOS_tip}
\end{eqnarray}
being $\Gamma_{\rm t}$ the Dynes parameter of the tip and $\rho_0$ the normal DOS of the tip at its Fermi level. The function $f(E)$ is the Fermi-Dirac distribution for a given temperature $T$.

\section{Comparison theory-experiment}

\subsection{Superconducting gap in the substrate vs in the wire}
We compute the dI/dV on the top of the wire and on the substrate (we observe no variation regarding the exact point of the substrate/wire top facet where the dI/dV is computed) and we compare it with the experimental measurements. We find that there are two parameters that dramatically control the dI/dV in the wire in our simulations: the band-bending between the wire and substrate $V_{\rm In}$, and the transparency $\kappa$ of the coupling between the wire and the substrate. This last parameter is not (completely) naturally present in our model. It can be understood as a barrier between the wire and substrate as a result of some mechanism that suppress the tunneling between both. This can be due to the lattice mismatch, that the wire is deposited and it is not epitaxially growth (and thus the wire is at some distance from the substrate), strain, impurities, or any mix of them. Only the fact that the effective mass is different between both materials, and, thus, there is an abrupt change at the interface, is already included in our simulations. To further introduce any other possible mechanism, we artificially change the hopping at the interface by a factor of $\kappa$. Therefore, $\kappa=1$ means that no other mechanism is included, and $\kappa=0$ means that wire and substrate are completely decoupled because of them. 

Another parameter that modifies the dI/dV is the doping of the wire (the value of $E_{\rm F}$ in the wire), although to a lower extend. However, to reduce the number of free parameters, we take an educated value for this doping, assuming that the Fermi level lies close to the bottom of the conduction band (as the experimental dI/dV in the normal state suggests, not shown) and we study how the dI/dV changes with the other two. 

\begin{figure}
    \centering
    \includegraphics[width=0.99\columnwidth]{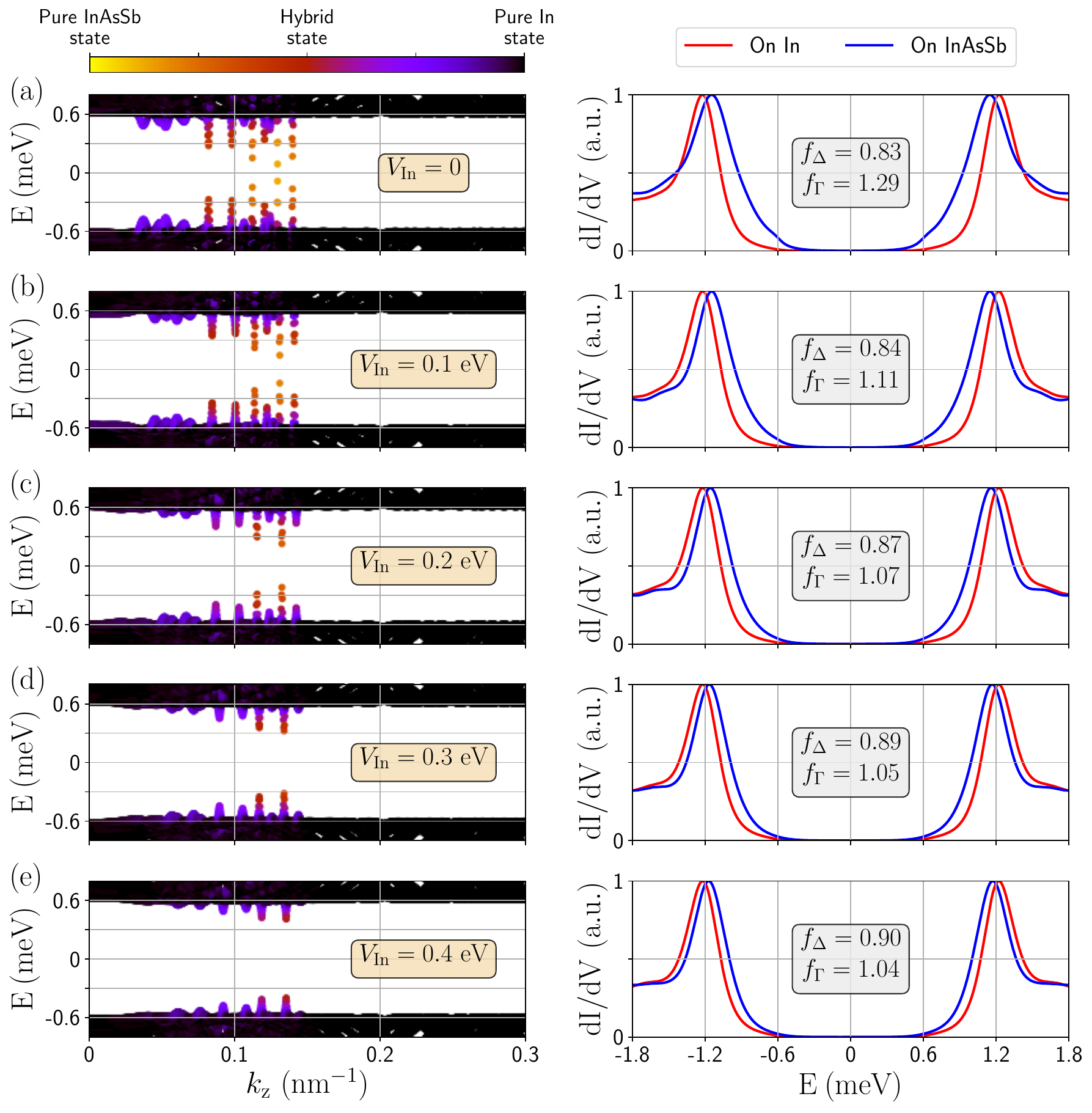}
    \caption{\textbf{Spectrum and conductance for different band-bendings.} (Left) Energy spectrum vs momentum for the In-InAsSb hybrid system. With colors we represent the weight of each state on the different parts of the heterostructure, from a pure state in the InAsSb wire (yellow) to a pure state in the In substrate (black). Any color in between is a mixed state. (Right) Differential conductance from the top facet of the InAsSb wire to the bottom of the substrate (blue) or from the top of the In substrate to its bottom part (red). The STM tip is superconducting and made of In as well, like in the experiments. Different rows (a-e) correspond to different simulations using different values of the band-bending between the In and the InAsSb $V_{\rm In}$.}
    \label{Fig2-theo}
\end{figure}

We start studying the results for different band-bendings $V_{\rm In}$. In Fig.~\ref{Fig2-theo}, we show the energy spectrum (left) and corresponding dI/dV (right) at the top facet of the wire (blue) and at the substrate (red), for different $V_{\rm In}$ (rows). To provide more insight, we fit each dI/dV with the theoretical curves to obtain the effective (fitted) superconducting gap and Dynes parameters in the wire ($\Delta_{\rm w}$ and $\Gamma_{\rm w}$) and substrate ($\Delta_{\rm s}$ and $\Gamma_{\rm s}$). We show the relations
\begin{equation}
    f_\Delta\equiv\frac{\Delta_{\rm w}}{\Delta_{\rm s}}, \ \ \ \ \ \ f_\Gamma\equiv\frac{\Gamma_{\rm w}}{\Gamma_{\rm s}},
\end{equation}
in the dI/dV plots. For low $V_{\rm In}$ [see Fig.~\ref{Fig2-theo}(a,b)], several subgap states, weakly coupled to the substrate, emerge in the spectrum. These states are mainly distributed all across the cross-section of the wire (not shown), so that the hybridization with the SC is small and, thus, the proximity effect is poorly induced. This shows up in the dI/dV as a soft gap below the parent SC gap. In this case, effectively the system behaves as a double-gap superconductor, one is hard (and close to the parent gap) and the second is soft. This is reflected as a large Dynes parameter (as compared to the substrate). When increasing the band-bending $V_{\rm In}$ [see Fig.~\ref{Fig2-theo}(d,e)], the subgap states are pushed towards the parent gap, as the wavefunction localizes closer to the substrate (not shown), enhancing the hybridization with the SC. The dI/dV in the wire only exhibits one hard gap in this case, although with a slightly smaller gap than the parent gap. This in turn provide a very similar Dynes parameter between the wire and substrate. Looking more closely into the evolution of $f_{\Delta}$ and $f_{\Gamma}$, we conclude that $V_{\rm In}$ rules the Dynes parameter while it changes only slightly the induced gap (actually the coherence peak almost does not change). 

\begin{figure}
    \centering
    \includegraphics[width=0.99\columnwidth]{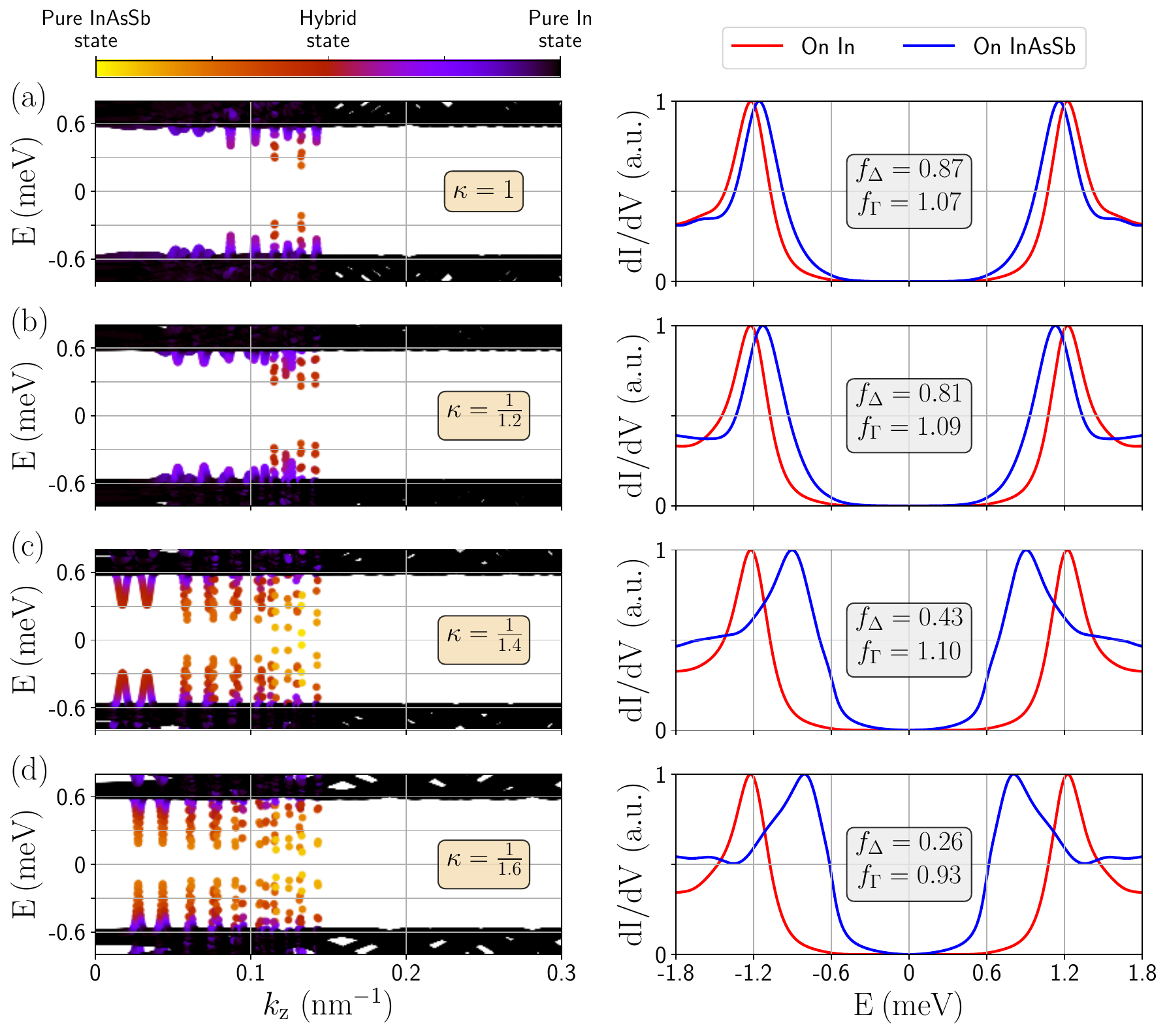}
    \caption{\textbf{Spectrum and conductance for different transparencies between wire and substrate.} Same as in Fig.~\ref{Fig2-theo} but changing the coupling between wire and substrate by a factor of $\kappa$.}
    \label{Fig3-theo}
\end{figure}

In Fig.~\ref{Fig3-theo}, we show the same simulations but changing the transparency $\kappa$ between the wire and substrate (different rows). Again, $\kappa=1$ implies that the interface is perfectly transparent, while $\kappa=0$ means that wire and substrate are completely decoupled. In Fig.~\ref{Fig3-theo} one can appreciate that as one decreases the transparency $\kappa$ [(a) to (d)], the induced gap is dramatically suppressed: there are more subgap states, the coherence peak in the dI/dV of the wire is displaced to zero, and the ratio between the gaps goes to zero. Since all the subgap states decouple to the substrate almost with the same strength, the Dynes parameter remains more or less the same (because the coherence peak is roughly well-defined as all the subbands exhibit the same gap) while the ratio of gaps is the one that dramatically changes. Hence, we conclude that the transparency $\kappa$ rules the induced gap.

We can thus estimate the values of $V_{\rm In}$ and $\kappa$ by comparing to the ratios $f_\Delta$ and $f_\Gamma$ between the experiments and our theoretical simulations. We find that $V_{\rm In}=0.1$~eV and $\kappa = 1/1.2\simeq0.83$ fit the best the experiments, as shown in the main text. We find the ratios do not dramatically change with a small change of $V_{\rm In}$ and $\kappa$, but fine tuning makes no sense as there are other small details not included in this picture. 

On the one hand, the band-bending $V_{\rm In}=0.1$~eV that we obtain is not very large. It is of the order of the one created by the accumulation layer that is typically at the uncovered facets of the wire. That means that the wavefunction tends to localize across all the facets of the wire. This is somehow beneficial for transport measurements, as the electron can go from the top facet of the wire (where the STM tip probes) directly to the bottom one (close to the substrate). 

On the other hand, the transparency $\kappa\simeq 0.83$ is high, pointing to a good proximity effect and hybridization between both materials. Actually one can estimate the distance between the wire and the substrate from the transparency parameter $\kappa$. Assuming that the tunneling is exponentially suppressed with the distance, we use the ansatz
\begin{equation}
    \kappa=\exp(1-\frac{d}{a_{\rm InAsSb}})
\label{Eq:kappa}
\end{equation}
where $a_{\rm InAsSb}$ is the lattice constant of InAsSb and $d$ the distance between the bottom facet (atoms) of the wire and the upper facet (atoms) of the substrate. For  $\kappa\simeq 0.83$ we obtain $d\simeq 1.18 a_{\rm InAsSb}$, which is a reasonable number.

\subsection{Loss of the induced superconductivity after several measurements}

\begin{figure}
    \centering
    \includegraphics[width=0.95\columnwidth]{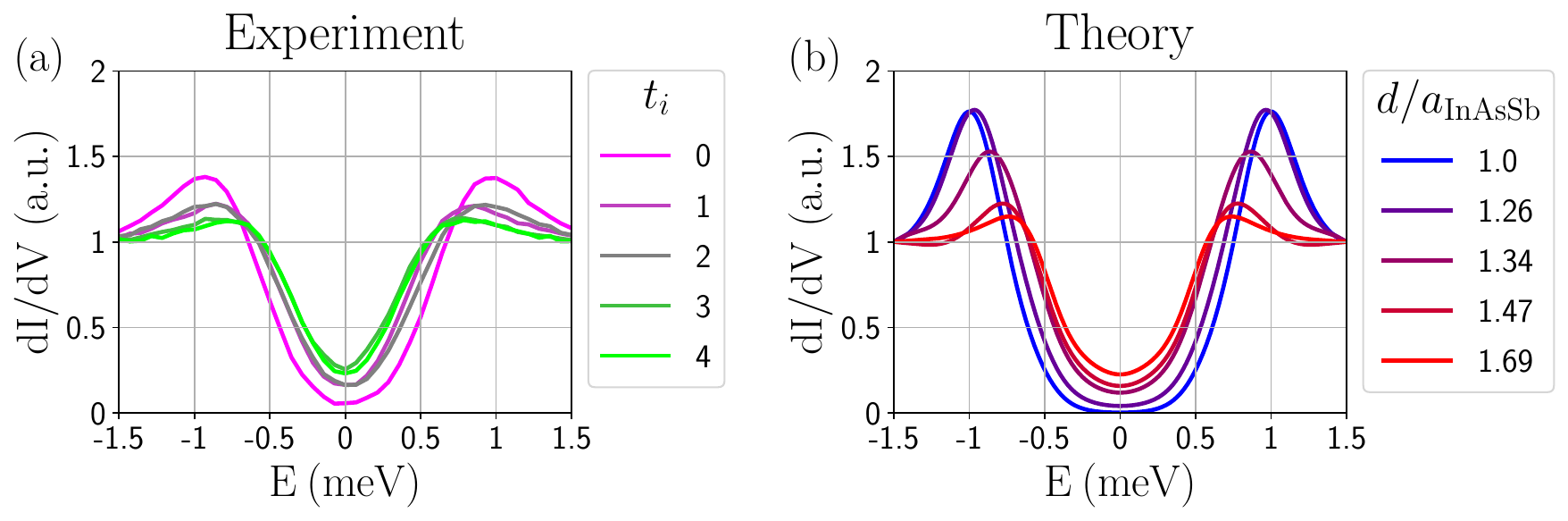}
    \caption{\textbf{Loss of the induced superconductivity.} {\bf (a)} dI/dV through the wire measured in the experiments at $B=0$, Different colors represent measurements at different (consecutive) times. {\bf (b)} Simulated dI/dV through the wire at $B=0$ for different transparencies $\kappa$, or alternatively, different distances between the wire and substrate $d$ [see Eq.~\eqref{Eq:kappa}].}
    \label{Fig4-2-theo}
\end{figure}

In the experiments, it has been observed sometimes that the induced gap in the wire is lost after several measurements on it. In Fig.~\ref{Fig4-2-theo}(a) we show one experimental example of the dI/dV measured on the wire at $B=0$. Different colors represent different measurements at different times. One can observe how the gap is reduced with time, going from $\sim 2\Delta_0$ to $\sim \Delta_0$. This means that, after several measurements, the superconductivity induced in the wire is lost, so that the only remaining gap is the one of the tip. Measurements on the substrate do not exhibit the same feature (not shown).

This evolution of the induced gap with time reassembles to the theoretical dI/dV of Fig.~\ref{Fig3-theo} for different transparencies. One can imagine that, after several measurements, the STM tip can either move the wire, increasing the distance with the substrate, or introduce impurities (or deformations) in it. Any of them reduces the value of $\kappa$, suppressing the hybridization with between the wire and substrate and, therefore, the induced superconductivity in the wire. In Fig.~\ref{Fig4-2-theo}(b) we show some the simulations of the dI/dV for different $\kappa$, or alternatively, different distances between the wire and substrate $d$ [see Eq.~\eqref{Eq:kappa}]. Both, experiment and theory, follow a very similar qualitative and quantitative behavior.

\subsection{Angle dependence of the gap} \label{Sec:ang}
Another proof that there are actually subgap states in the wire is that the gap in the wire changes with the magnetic field orientation (in-plane). In the substrate, the superconducting gap should not change while changing the orientation of the magnetic field. But in the wire, if there are states, the induced gap may change as a result of the anisotropy of the magnetic field. Even if the intrinsic $g$-factor of the wire is isotropic, as it happens in this zinc-blende III-V compound semiconductors, the orbital effects in the wire depends on the magnetic field orientation. To understand this, let us write the Hamiltonian of Eq.~\eqref{Eq:H} with the orbital effects explicitly, i.e., substituting $\vec{k}\rightarrow \vec{k}-\frac{e}{\hbar}\vec{A}$, so that we obtain
\begin{eqnarray}
H = \left[ \vec{k}  \frac{\hbar^2}{2m^*(\vec{r})} \vec{k} -\frac{1}{2}(e\hbar)\left(\vec{k}  \frac{1}{m^*(\vec{r})} \vec{A} +\vec{A}  \frac{1}{m^*(\vec{r})} \vec{k}\right) \right. \nonumber \\
\left.- \mu_{\rm eff}(\vec{r},B)\right]\sigma_0\tau_z \nonumber \\
+ \frac{1}{2} \left[\vec{\alpha}(\vec{r})\cdot \left(\vec{\sigma}\times\vec{k}\right) + \left(\vec{\sigma}\times\vec{k}\right)\cdot \vec{\alpha}(\vec{r}) \right]\tau_z \nonumber \\
+\frac{1}{2}\mu_B\mathbf{g_{\rm eff}}(\vec{r})\vec{B}\cdot\vec{\sigma}\tau_z + \Delta(\vec{r},\vec{B})\sigma_y\tau_y.\; \;\;\;\;
\label{Eq:H2}
\end{eqnarray}
Note that the Hamiltonian is the same than Eq.~\eqref{Eq:H} except that there is an additional term linear in $k$ (second term), and the chemical potential and $g$-factor are renormalised to
\begin{equation}
    \mu_{\rm eff}(\vec{r},B)=E_{\rm F}(\vec{r}) - e\phi(\vec{r}) - \frac{e^2 |\vec{A}|^2}{2m^*(\vec{r})},
\end{equation}
and 
\begin{equation}
    \mathbf{g_{\rm eff}}(\vec{r},\vec{B})=g(\vec{r})\mathbf{1}+\mathbf{g_{\rm orb}}(\vec{r},\vec{B}),
\end{equation}
respectively. Here $\mathbf{1}$ is the identity and $\mathbf{g_{\rm orb}}$ is a tensor that, for the symmetric gauge, can be written as
\begin{widetext}
\begin{equation}
\mathbf{g_{\rm orb}}(\vec{r},\vec{B})=\frac{2m_e}{\hbar^2}\begin{pmatrix}
\alpha_y(\vec{r}) y+\alpha_z(\vec{r}) z & -\alpha_x(\vec{r}) y & -\alpha_x(\vec{r}) z\\
-\alpha_y(\vec{r}) x & \alpha_x(\vec{r}) x+\alpha_z(\vec{r}) z & -\alpha_y(\vec{r}) z \\
-\alpha_z(\vec{r}) x & -\alpha_z(\vec{r}) y & \alpha_y(\vec{r}) y +\alpha_x(\vec{r}) x \\\end{pmatrix}.
\end{equation}
\end{widetext}
Fixing the modulus of the magnetic field at $|\vec{B}|^2=B_0$, one can check that only the Zeeman term $H_Z$ in the wire changes when changing the orientation of the magnetic field $\theta$ (in-plane) with respect to the wire's direction. More particularly, for a nanowire (i.e., with $\alpha_z=0$ and $z=0$), we have
\begin{equation}
    H_Z(\theta=0)= \frac{1}{2}\mu_B \left[g_0+\frac{2m_e}{\hbar^2}(\alpha_y(\vec{r}) y +\alpha_x(\vec{r}) x)\right]B_0\sigma_z,
\end{equation}
and
\begin{eqnarray}
    H_Z(\theta=\pi/2)= \frac{1}{2}\mu_B \left[g_0+\frac{2m_e}{\hbar^2}(\alpha_y(\vec{r}) y)\right]B_0\sigma_x \nonumber \\
    - \cancelto{\sim 0}{\frac{1}{2}\mu_B \frac{2m_e}{\hbar^2}\alpha_y(\vec{r}) x} B_0\sigma_y,
\end{eqnarray}
Notice that the Zeeman field is smaller at $\theta=\pi/2$ as the term proportional to $\alpha_x(\vec{r})x$ is missing. We can provide an estimation of this term. Taking $B_0=0.1$~T, $x\sim 30$~nm and $\alpha_x\simeq 20$~meV$\cdot$nm, we obtain $\Delta H_Z\sim 0.05$~meV.

\section{Theoretical predictions about the topological phase}
\label{Sec:theo_topo}

Given the strong agreement observed between theory and experiments, we can now employ our theoretical modeling to make predictions. The platform studied here boasts unique features that facilitate the study of Andreev bound states, particularly in the context of $p$-wave superconductors, owing to the spatial resolution of the subgap states enabled by the STM. A $p$-wave superconducting NW can be engineered when a semiconductor NW with robust spin-orbit coupling is proximitized by a superconductor~\cite{oreg2010helical, Lutchyn2010Aug}, as in the nanostructure studied here. In the presence of a magnetic field applied along the NW's axis, the NW is anticipated to undergo a topological phase transition for a sufficiently strong magnetic field ($B\ge B_{\rm topo}$), where the gap closes and reopens again. Consequently, after the reopening of the gap, two zero-energy states emerge, bound at the edges of the NW with an exponential decay toward the NW's center. These are the so-called Majorana zero modes (MZMs).

Over the past couple of decades, numerous experiments have sought to create and manipulate these states. However, definitive proof of their existence has remained elusive, primarily due to the reliance on conventional transport measurements that can only access to the spectrum at one end of the wire, or, at best, at a few fixed points along it. Such approaches have proven insufficient to probe the spatial profile of these states and distinguish them from other topologically-trivial counterparts. Hence, our platform stands out as a powerful tool to both create and provide definitive proof of the existence of MZMs.

In principle, our platform gathers all the ingredients to host MZMs: a large $g$-factor and spin-orbit coupling in the wire, together with a large proximitized superconducting gap. In particular, to obtain a topological phase, the $g$-factor of the wire must be large enough so that the Zeeman field can close the superconducting gap at $k_z=0$. The magnetic field for which the gap is closed is called topological critical magnetic field $B_{\rm c}^{\rm (t)}$ and it is given by 
\begin{equation}
    B_{\rm c}^{\rm (t)}=\frac{\Delta_{\rm ind}}{\frac{1}{2}\mu_B g_{\rm InAsSb}},
    \label{Eq:top_critical_B}
\end{equation}
where $\Delta_{\rm ind}$ is the gap induced in the wire, and $g$ is the $g$-factor of the wire. Notice that $g$ depends on the \ch{Sb} concentration, so does $B_{\rm c}^{\rm (t)}$. Nonetheless, there is another constraint for the system to develop a topological phase: the topological critical magnetic field must be smaller than the critical magnetic field of the superconductor $B_{\rm c}$. Otherwise, the magnetic field closes the gap of the superconductor (and destroys superconductivity) before the gap is closed in the wire and the topological state is achieved. 

\begin{figure}
    \centering
    \includegraphics[width=0.99\columnwidth]{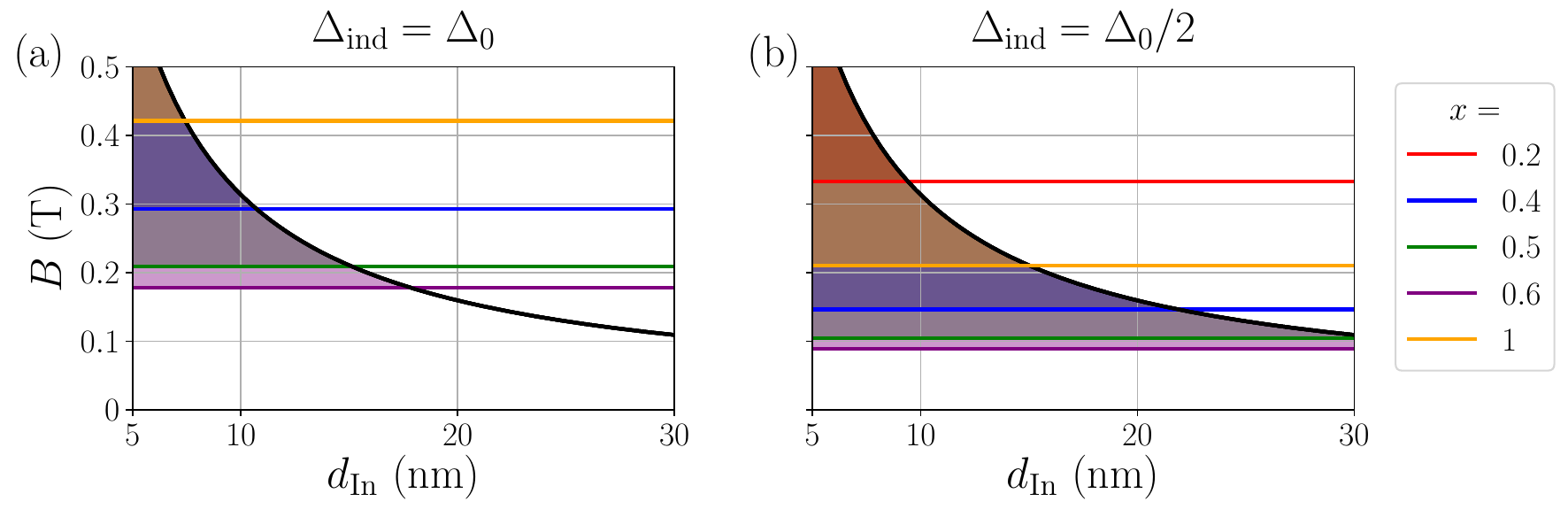}
    \caption{\textbf{Optimization of material parameters.} Critical magnetic field (black line) of the SC as a function of the In substrate thickness. With colors, we also show the topological critical magnetic field for the InAs$_{1-x}$Sb$_x$ wire. Different colors correspond to different concentrations of Sb, which modifies the $g$-factor [see Eq.~\eqref{Eq:top_critical_B}]. In {\bf (a)} we assume that the induced gap in the wire $\Delta_{\rm ind}$ is the same than the parent gap $\Delta_0$, while in {\bf (b)} we assume it is the half. Notice that topological states may only be found in the colored areas. Hence, for each concentration, there is a maximum \ch{In} thickness for which the wire may support topological superconductivity. }
    \label{Fig6-theo}
\end{figure}

To understand this interplay and to know which experimental parameters are the best to enhance the topological phase, we plot in Fig.~\ref{Fig6-theo} the different critical magnetic fields. The black line shows the critical magnetic field of the superconductor as a function of the \ch{In} substrate thickness [Eq.~\eqref{Eq:Bc}], and the colored lines correspond to the topological critical magnetic field for different Sb concentrations [Eq.~\eqref{Eq:top_critical_B}] assuming that $\Delta_{\rm ind}$ (a) is the parent gap $\Delta_0$ or (b) the half. Only when the topological critical magnetic field is larger than the critical magnetic field of the SC, the wire may support topological superconductivity. Colored areas precisely marked these regions in magnetic and \ch{In} thickness parameter space. Hence, for different \ch{Sb} concentrations, there is a maximum value of the In thickness for which topological superconductivity can be found. Notice that a concentration of $x=0.4$ will provide (roughly) the most extended topological phase, as this crystal present the larger $g$-factor. For this crystal, the thickness of In must be around $10$ to $20$~nm,far from the $30$ to $50$~nm used in the experimental setup. This is why subgap states have not been observed in our current devices. We hypothesize that for a thickness of $10$~nm, which is experimentally feasible, zero energy states can be found in this platform for an extended region of magnetic fields.

\begin{figure*}
    \centering
    \includegraphics[width=0.98\columnwidth]{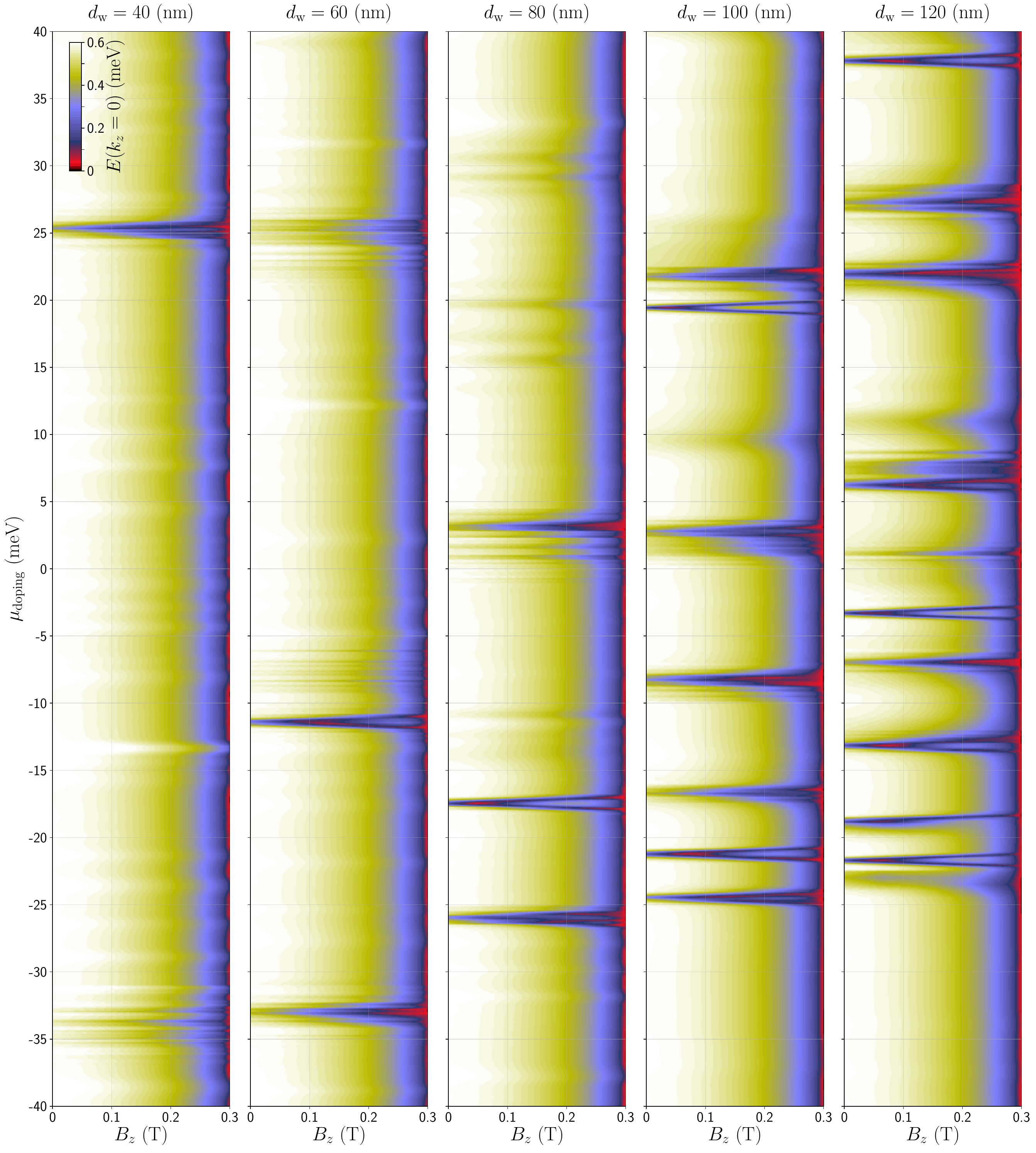}
    \caption{\textbf{Topological phase diagram.} Topological phase diagrams (energy spectrum at $k=0$ vs magnetic field along the wire $B_z$ and the intrinsic doping of the wire $\mu_{\rm doping}$) for different wire's diameters (different columns). }
    \label{Fig7-theo}
\end{figure*}

With this knowledge, we can now study the phase diagram for a $10$~nm thick In substrate and an InAs$_{0.6}$Sb$_{0.4}$ wire. In Fig.~\ref{Fig7-theo} we show the lowest-energy spectrum at $k_z=0$ vs the magnetic field along the wire's axis $B_z$ and the doping of the wire, for different wire diameters (different columns). The system becomes topological for any $\mu$ and $B$ value inside those parabolic regions delimited by a zero-energy crossing. Each parabolic region corresponds to a different subband with develops a topological phase. In STM experiments, unfortunately, is not possible to easily tune the doping of the wire $\mu_{\rm doping}$. However, as one can see in Fig.~\ref{Fig7-theo}, the larger the wire's diameter, the higher are the chances to find a topological state.

\begin{figure}
\begin{centering}
\includegraphics[width=0.98\columnwidth]{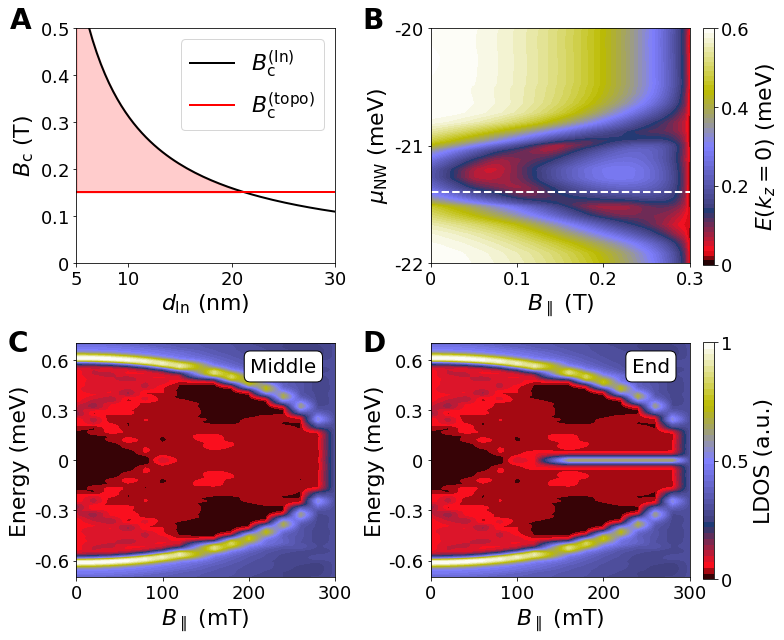}
\par\end{centering}
\caption{\textbf{\label{Fig5} Theoretical calculations for Identifying Topological Phase Transitions in \ch{InAsSb} NWs. (a)} Critical magnetic field as function of the In thickness (black) and critical magnetic field of the topological phase transition of the hybrid NW (red). The red area marks the values of $d_{\rm In}$ and $B$ for which it is possible to find topological states. \textbf{(b)} Energy at $k_z=0$ as a function of magnetic field $B$ and intrinsic doping $\mu_{\rm NW}$ for $d_{\rm In}=10$~nm. The zero energy crossing marks a topological phase transition.  \textbf{(c,d)} LDOS on top of the wire as a function of magnetic field at the middle of the NW (c) or at the end (d), for the intrinsic doping shown with a dotted white line in (b). We set $T=300$~mK in (c,b).}
\end{figure}

We now focus on the second subband of a $100$~nm thick NW deposited on top of a $10$~nm thick \ch{In} substrate. The lowest-energy spectrum is shown in Fig.~\ref{Fig5}(b), which is a zoom-in of Fig.~\ref{Fig7-theo}(d), and exhibits the typical parabolic boundary where a subgap state crosses zero-energy. For the fixed chemical marked with a white line in (b), we compute and show in (c) and (d) the LDOS in the middle of the NW and at the end, respectively. As expected, the LDOS at the end exhibits a zero-energy mode after the gap closing that is not present in the middle.

One limitation of this nanodevice, nonetheless, lies in the inability to adjust the chemical potential within the NW using a back gate, owing to the constraints imposed by the STM technique. Given that different NWs possess distinct intrinsic dopings resulting from the growth process, MZMs may be naturally present in the NW or not. Consequently, we rely on statistic to detect MZMs across a sample of multiple NWs. One way to overcome this is to grow tapered NWs, whose diameter changes along their length. In such NWs the level spacing between the quantized subbands would vary along the NW. In the current setup, in order to enhance the likelihood of encountering a NW with the desired doping, we propose the use of thick NWs, where the subband splitting is minimal, thus increasing the probability of observing a topological phase, as shown in Fig.~\ref{Fig7-theo}.


\begin{thebibliography}{59}%
\makeatletter
\providecommand \@ifxundefined [1]{%
 \@ifx{#1\undefined}
}%
\providecommand \@ifnum [1]{%
 \ifnum #1\expandafter \@firstoftwo
 \else \expandafter \@secondoftwo
 \fi
}%
\providecommand \@ifx [1]{%
 \ifx #1\expandafter \@firstoftwo
 \else \expandafter \@secondoftwo
 \fi
}%
\providecommand \natexlab [1]{#1}%
\providecommand \enquote  [1]{``#1''}%
\providecommand \bibnamefont  [1]{#1}%
\providecommand \bibfnamefont [1]{#1}%
\providecommand \citenamefont [1]{#1}%
\providecommand \href@noop [0]{\@secondoftwo}%
\providecommand \href [0]{\begingroup \@sanitize@url \@href}%
\providecommand \@href[1]{\@@startlink{#1}\@@href}%
\providecommand \@@href[1]{\endgroup#1\@@endlink}%
\providecommand \@sanitize@url [0]{\catcode `\\12\catcode `\$12\catcode
  `\&12\catcode `\#12\catcode `\^12\catcode `\_12\catcode `\%12\relax}%
\providecommand \@@startlink[1]{}%
\providecommand \@@endlink[0]{}%
\providecommand \url  [0]{\begingroup\@sanitize@url \@url }%
\providecommand \@url [1]{\endgroup\@href {#1}{\urlprefix }}%
\providecommand \urlprefix  [0]{URL }%
\providecommand \Eprint [0]{\href }%
\providecommand \doibase [0]{https://doi.org/}%
\providecommand \selectlanguage [0]{\@gobble}%
\providecommand \bibinfo  [0]{\@secondoftwo}%
\providecommand \bibfield  [0]{\@secondoftwo}%
\providecommand \translation [1]{[#1]}%
\providecommand \BibitemOpen [0]{}%
\providecommand \bibitemStop [0]{}%
\providecommand \bibitemNoStop [0]{.\EOS\space}%
\providecommand \EOS [0]{\spacefactor3000\relax}%
\providecommand \BibitemShut  [1]{\csname bibitem#1\endcsname}%
\let\auto@bib@innerbib\@empty
\bibitem [{\citenamefont {Deutscher}\ and\ \citenamefont
  {de~Gennes}(1969)}]{deutcher1969proximity}%
  \BibitemOpen
  \bibfield  {author} {\bibinfo {author} {\bibfnamefont {G.}~\bibnamefont
  {Deutscher}}\ and\ \bibinfo {author} {\bibfnamefont {P.~G.}\ \bibnamefont
  {de~Gennes}},\ }\href@noop {} {\emph {\bibinfo {title} {Proximity Effects}}}\
  (\bibinfo  {publisher} {Routledge},\ \bibinfo {year} {1969})\BibitemShut
  {NoStop}%
\bibitem [{\citenamefont
  {Tinkham}(2004{\natexlab{a}})}]{tinkham2004introduction}%
  \BibitemOpen
  \bibfield  {author} {\bibinfo {author} {\bibfnamefont {M.}~\bibnamefont
  {Tinkham}},\ }\href {http://www.worldcat.org/isbn/0486435032} {\emph
  {\bibinfo {title} {Introduction to Superconductivity}}},\ \bibinfo {edition}
  {2nd}\ ed.\ (\bibinfo  {publisher} {Dover Publications},\ \bibinfo {year}
  {2004})\BibitemShut {NoStop}%
\bibitem [{\citenamefont {Cohen}(1964)}]{cohen1964existence}%
  \BibitemOpen
  \bibfield  {author} {\bibinfo {author} {\bibfnamefont {M.~L.}\ \bibnamefont
  {Cohen}},\ }\bibfield  {title} {\bibinfo {title} {The existence of a
  superconducting state in semiconductors},\ }\href@noop {} {\bibfield
  {journal} {\bibinfo  {journal} {Reviews of Modern Physics}\ }\textbf
  {\bibinfo {volume} {36}},\ \bibinfo {pages} {240} (\bibinfo {year}
  {1964})}\BibitemShut {NoStop}%
\bibitem [{\citenamefont {Burkard}\ \emph {et~al.}(2020)\citenamefont
  {Burkard}, \citenamefont {Gullans}, \citenamefont {Mi},\ and\ \citenamefont
  {Petta}}]{burkard2020superconductor}%
  \BibitemOpen
  \bibfield  {author} {\bibinfo {author} {\bibfnamefont {G.}~\bibnamefont
  {Burkard}}, \bibinfo {author} {\bibfnamefont {M.~J.}\ \bibnamefont
  {Gullans}}, \bibinfo {author} {\bibfnamefont {X.}~\bibnamefont {Mi}},\ and\
  \bibinfo {author} {\bibfnamefont {J.~R.}\ \bibnamefont {Petta}},\ }\bibfield
  {title} {\bibinfo {title} {Superconductor--semiconductor hybrid-circuit
  quantum electrodynamics},\ }\href@noop {} {\bibfield  {journal} {\bibinfo
  {journal} {Nature Reviews Physics}\ }\textbf {\bibinfo {volume} {2}},\
  \bibinfo {pages} {129} (\bibinfo {year} {2020})}\BibitemShut {NoStop}%
\bibitem [{\citenamefont {Van~Duzer}(1988)}]{van1988superconductor}%
  \BibitemOpen
  \bibfield  {author} {\bibinfo {author} {\bibfnamefont {T.}~\bibnamefont
  {Van~Duzer}},\ }\bibfield  {title} {\bibinfo {title}
  {Superconductor—semiconductor hybrid devices, circuits and systems},\
  }\href@noop {} {\bibfield  {journal} {\bibinfo  {journal} {Cryogenics}\
  }\textbf {\bibinfo {volume} {28}},\ \bibinfo {pages} {527} (\bibinfo {year}
  {1988})}\BibitemShut {NoStop}%
\bibitem [{\citenamefont {Lu}\ and\ \citenamefont
  {Lieber}(2006)}]{lu2006semiconductor}%
  \BibitemOpen
  \bibfield  {author} {\bibinfo {author} {\bibfnamefont {W.}~\bibnamefont
  {Lu}}\ and\ \bibinfo {author} {\bibfnamefont {C.~M.}\ \bibnamefont
  {Lieber}},\ }\bibfield  {title} {\bibinfo {title} {Semiconductor nanowires},\
  }\href@noop {} {\bibfield  {journal} {\bibinfo  {journal} {Journal of Physics
  D: Applied Physics}\ }\textbf {\bibinfo {volume} {39}},\ \bibinfo {pages}
  {R387} (\bibinfo {year} {2006})}\BibitemShut {NoStop}%
\bibitem [{\citenamefont {P\"oschl}\ \emph {et~al.}(2022)\citenamefont
  {P\"oschl}, \citenamefont {Danilenko}, \citenamefont {Sabonis}, \citenamefont
  {Kristjuhan}, \citenamefont {Lindemann}, \citenamefont {Thomas},
  \citenamefont {Manfra},\ and\ \citenamefont {Marcus}}]{PhysRevB.106.L241301}%
  \BibitemOpen
  \bibfield  {author} {\bibinfo {author} {\bibfnamefont {A.}~\bibnamefont
  {P\"oschl}}, \bibinfo {author} {\bibfnamefont {A.}~\bibnamefont {Danilenko}},
  \bibinfo {author} {\bibfnamefont {D.}~\bibnamefont {Sabonis}}, \bibinfo
  {author} {\bibfnamefont {K.}~\bibnamefont {Kristjuhan}}, \bibinfo {author}
  {\bibfnamefont {T.}~\bibnamefont {Lindemann}}, \bibinfo {author}
  {\bibfnamefont {C.}~\bibnamefont {Thomas}}, \bibinfo {author} {\bibfnamefont
  {M.~J.}\ \bibnamefont {Manfra}},\ and\ \bibinfo {author} {\bibfnamefont
  {C.~M.}\ \bibnamefont {Marcus}},\ }\bibfield  {title} {\bibinfo {title}
  {Nonlocal conductance spectroscopy of andreev bound states in gate-defined
  inas/al nanowires},\ }\href {https://doi.org/10.1103/PhysRevB.106.L241301}
  {\bibfield  {journal} {\bibinfo  {journal} {Phys. Rev. B}\ }\textbf {\bibinfo
  {volume} {106}},\ \bibinfo {pages} {L241301} (\bibinfo {year}
  {2022})}\BibitemShut {NoStop}%
\bibitem [{\citenamefont {Zavaritskii}(1952)}]{zavaritskii1952properties}%
  \BibitemOpen
  \bibfield  {author} {\bibinfo {author} {\bibfnamefont {N.}~\bibnamefont
  {Zavaritskii}},\ }\bibfield  {title} {\bibinfo {title} {Properties of
  superconducting films of thallium and indium},\ }in\ \href@noop {} {\emph
  {\bibinfo {booktitle} {Doklady Akad Nauk SSSR}}},\ Vol.~\bibinfo {volume}
  {85}\ (\bibinfo {year} {1952})\ p.\ \bibinfo {pages} {749}\BibitemShut
  {NoStop}%
\bibitem [{\citenamefont {Toxen}(1961{\natexlab{a}})}]{toxen1961size}%
  \BibitemOpen
  \bibfield  {author} {\bibinfo {author} {\bibfnamefont {A.}~\bibnamefont
  {Toxen}},\ }\bibfield  {title} {\bibinfo {title} {Size effects in thin
  superconducting indium films},\ }\href@noop {} {\bibfield  {journal}
  {\bibinfo  {journal} {Physical Review}\ }\textbf {\bibinfo {volume} {123}},\
  \bibinfo {pages} {442} (\bibinfo {year} {1961}{\natexlab{a}})}\BibitemShut
  {NoStop}%
\bibitem [{\citenamefont {Khukhareva}(1963)}]{khukhareva1963superconducting}%
  \BibitemOpen
  \bibfield  {author} {\bibinfo {author} {\bibfnamefont {I.}~\bibnamefont
  {Khukhareva}},\ }\bibfield  {title} {\bibinfo {title} {The superconducting
  properties of thin aluminum films},\ }\href@noop {} {\bibfield  {journal}
  {\bibinfo  {journal} {Soviet Physics JETP}\ }\textbf {\bibinfo {volume}
  {16}},\ \bibinfo {pages} {828} (\bibinfo {year} {1963})}\BibitemShut
  {NoStop}%
\bibitem [{\citenamefont {Vogel}\ and\ \citenamefont
  {Garland}(1967)}]{vogel1967superconductivity}%
  \BibitemOpen
  \bibfield  {author} {\bibinfo {author} {\bibfnamefont {H.}~\bibnamefont
  {Vogel}}\ and\ \bibinfo {author} {\bibfnamefont {M.}~\bibnamefont
  {Garland}},\ }\bibfield  {title} {\bibinfo {title} {Superconductivity in thin
  indium films},\ }\href@noop {} {\bibfield  {journal} {\bibinfo  {journal}
  {Journal of Applied Physics}\ }\textbf {\bibinfo {volume} {38}},\ \bibinfo
  {pages} {5116} (\bibinfo {year} {1967})}\BibitemShut {NoStop}%
\bibitem [{\citenamefont {Linder}\ and\ \citenamefont
  {Robinson}(2015)}]{linder2015superconducting}%
  \BibitemOpen
  \bibfield  {author} {\bibinfo {author} {\bibfnamefont {J.}~\bibnamefont
  {Linder}}\ and\ \bibinfo {author} {\bibfnamefont {J.~W.}\ \bibnamefont
  {Robinson}},\ }\bibfield  {title} {\bibinfo {title} {Superconducting
  spintronics},\ }\href@noop {} {\bibfield  {journal} {\bibinfo  {journal}
  {Nature Physics}\ }\textbf {\bibinfo {volume} {11}},\ \bibinfo {pages} {307}
  (\bibinfo {year} {2015})}\BibitemShut {NoStop}%
\bibitem [{\citenamefont {Mart{\'\i}nez}\ \emph {et~al.}(2020)\citenamefont
  {Mart{\'\i}nez}, \citenamefont {H{\"o}gl}, \citenamefont
  {Gonz{\'a}lez-Ruano}, \citenamefont {Cascales}, \citenamefont {Tiusan},
  \citenamefont {Lu}, \citenamefont {Hehn}, \citenamefont {Matos-Abiague},
  \citenamefont {Fabian}, \citenamefont {{\v{Z}}uti{\'c}} \emph
  {et~al.}}]{martinez2020interfacial}%
  \BibitemOpen
  \bibfield  {author} {\bibinfo {author} {\bibfnamefont {I.}~\bibnamefont
  {Mart{\'\i}nez}}, \bibinfo {author} {\bibfnamefont {P.}~\bibnamefont
  {H{\"o}gl}}, \bibinfo {author} {\bibfnamefont {C.}~\bibnamefont
  {Gonz{\'a}lez-Ruano}}, \bibinfo {author} {\bibfnamefont {J.~P.}\ \bibnamefont
  {Cascales}}, \bibinfo {author} {\bibfnamefont {C.}~\bibnamefont {Tiusan}},
  \bibinfo {author} {\bibfnamefont {Y.}~\bibnamefont {Lu}}, \bibinfo {author}
  {\bibfnamefont {M.}~\bibnamefont {Hehn}}, \bibinfo {author} {\bibfnamefont
  {A.}~\bibnamefont {Matos-Abiague}}, \bibinfo {author} {\bibfnamefont
  {J.}~\bibnamefont {Fabian}}, \bibinfo {author} {\bibfnamefont
  {I.}~\bibnamefont {{\v{Z}}uti{\'c}}}, \emph {et~al.},\ }\bibfield  {title}
  {\bibinfo {title} {Interfacial spin-orbit coupling: A platform for
  superconducting spintronics},\ }\href@noop {} {\bibfield  {journal} {\bibinfo
   {journal} {Physical Review Applied}\ }\textbf {\bibinfo {volume} {13}},\
  \bibinfo {pages} {014030} (\bibinfo {year} {2020})}\BibitemShut {NoStop}%
\bibitem [{\citenamefont {Eschrig}(2015)}]{eschrig2015spin}%
  \BibitemOpen
  \bibfield  {author} {\bibinfo {author} {\bibfnamefont {M.}~\bibnamefont
  {Eschrig}},\ }\bibfield  {title} {\bibinfo {title} {Spin-polarized
  supercurrents for spintronics: a review of current progress},\ }\href@noop {}
  {\bibfield  {journal} {\bibinfo  {journal} {Reports on Progress in Physics}\
  }\textbf {\bibinfo {volume} {78}},\ \bibinfo {pages} {104501} (\bibinfo
  {year} {2015})}\BibitemShut {NoStop}%
\bibitem [{\citenamefont {Nadeem}\ \emph {et~al.}(2023)\citenamefont {Nadeem},
  \citenamefont {Fuhrer},\ and\ \citenamefont
  {Wang}}]{nadeem2023superconducting}%
  \BibitemOpen
  \bibfield  {author} {\bibinfo {author} {\bibfnamefont {M.}~\bibnamefont
  {Nadeem}}, \bibinfo {author} {\bibfnamefont {M.~S.}\ \bibnamefont {Fuhrer}},\
  and\ \bibinfo {author} {\bibfnamefont {X.}~\bibnamefont {Wang}},\ }\bibfield
  {title} {\bibinfo {title} {The superconducting diode effect},\ }\href@noop {}
  {\bibfield  {journal} {\bibinfo  {journal} {Nature Reviews Physics}\ }\textbf
  {\bibinfo {volume} {5}},\ \bibinfo {pages} {558} (\bibinfo {year}
  {2023})}\BibitemShut {NoStop}%
\bibitem [{\citenamefont {Kitaev}(2003)}]{KITAEV20032}%
  \BibitemOpen
  \bibfield  {author} {\bibinfo {author} {\bibfnamefont {A.}~\bibnamefont
  {Kitaev}},\ }\bibfield  {title} {\bibinfo {title} {Fault-tolerant quantum
  computation by anyons},\ }\href
  {https://doi.org/https://doi.org/10.1016/S0003-4916(02)00018-0} {\bibfield
  {journal} {\bibinfo  {journal} {Annals of Physics}\ }\textbf {\bibinfo
  {volume} {303}},\ \bibinfo {pages} {2} (\bibinfo {year} {2003})}\BibitemShut
  {NoStop}%
\bibitem [{\citenamefont {Nayak}\ \emph {et~al.}(2008)\citenamefont {Nayak},
  \citenamefont {Simon}, \citenamefont {Stern}, \citenamefont {Freedman},\ and\
  \citenamefont {Das~Sarma}}]{nayak2008non}%
  \BibitemOpen
  \bibfield  {author} {\bibinfo {author} {\bibfnamefont {C.}~\bibnamefont
  {Nayak}}, \bibinfo {author} {\bibfnamefont {S.~H.}\ \bibnamefont {Simon}},
  \bibinfo {author} {\bibfnamefont {A.}~\bibnamefont {Stern}}, \bibinfo
  {author} {\bibfnamefont {M.}~\bibnamefont {Freedman}},\ and\ \bibinfo
  {author} {\bibfnamefont {S.}~\bibnamefont {Das~Sarma}},\ }\bibfield  {title}
  {\bibinfo {title} {Non-abelian anyons and topological quantum computation},\
  }\href@noop {} {\bibfield  {journal} {\bibinfo  {journal} {Reviews of Modern
  Physics}\ }\textbf {\bibinfo {volume} {80}},\ \bibinfo {pages} {1083}
  (\bibinfo {year} {2008})}\BibitemShut {NoStop}%
\bibitem [{\citenamefont {Oreg}\ \emph {et~al.}(2010)\citenamefont {Oreg},
  \citenamefont {Refael},\ and\ \citenamefont {von Oppen}}]{oreg2010helical}%
  \BibitemOpen
  \bibfield  {author} {\bibinfo {author} {\bibfnamefont {Y.}~\bibnamefont
  {Oreg}}, \bibinfo {author} {\bibfnamefont {G.}~\bibnamefont {Refael}},\ and\
  \bibinfo {author} {\bibfnamefont {F.}~\bibnamefont {von Oppen}},\ }\bibfield
  {title} {\bibinfo {title} {{Helical liquids and Majorana bound states in
  quantum wires}},\ }\href@noop {} {\bibfield  {journal} {\bibinfo  {journal}
  {Physical review letters}\ }\textbf {\bibinfo {volume} {105}},\ \bibinfo
  {pages} {177002} (\bibinfo {year} {2010})}\BibitemShut {NoStop}%
\bibitem [{\citenamefont {Lutchyn}\ \emph {et~al.}(2010)\citenamefont
  {Lutchyn}, \citenamefont {Sau},\ and\ \citenamefont
  {Das~Sarma}}]{Lutchyn2010Aug}%
  \BibitemOpen
  \bibfield  {author} {\bibinfo {author} {\bibfnamefont {R.~M.}\ \bibnamefont
  {Lutchyn}}, \bibinfo {author} {\bibfnamefont {J.~D.}\ \bibnamefont {Sau}},\
  and\ \bibinfo {author} {\bibfnamefont {S.}~\bibnamefont {Das~Sarma}},\
  }\bibfield  {title} {\bibinfo {title} {{Majorana Fermions and a Topological
  Phase Transition in Semiconductor-Superconductor Heterostructures}},\ }\href
  {https://doi.org/10.1103/PhysRevLett.105.077001} {\bibfield  {journal}
  {\bibinfo  {journal} {Phys. Rev. Lett.}\ }\textbf {\bibinfo {volume} {105}},\
  \bibinfo {pages} {077001} (\bibinfo {year} {2010})}\BibitemShut {NoStop}%
\bibitem [{\citenamefont {Leijnse}\ and\ \citenamefont
  {Flensberg}(2012)}]{leijnse2012introduction}%
  \BibitemOpen
  \bibfield  {author} {\bibinfo {author} {\bibfnamefont {M.}~\bibnamefont
  {Leijnse}}\ and\ \bibinfo {author} {\bibfnamefont {K.}~\bibnamefont
  {Flensberg}},\ }\bibfield  {title} {\bibinfo {title} {Introduction to
  topological superconductivity and majorana fermions},\ }\href@noop {}
  {\bibfield  {journal} {\bibinfo  {journal} {Semiconductor Science and
  Technology}\ }\textbf {\bibinfo {volume} {27}},\ \bibinfo {pages} {124003}
  (\bibinfo {year} {2012})}\BibitemShut {NoStop}%
\bibitem [{\citenamefont {Lutchyn}\ \emph {et~al.}(2018)\citenamefont
  {Lutchyn}, \citenamefont {Bakkers}, \citenamefont {Kouwenhoven},
  \citenamefont {Krogstrup}, \citenamefont {Marcus},\ and\ \citenamefont
  {Oreg}}]{Lutchyn2018May}%
  \BibitemOpen
  \bibfield  {author} {\bibinfo {author} {\bibfnamefont {R.~M.}\ \bibnamefont
  {Lutchyn}}, \bibinfo {author} {\bibfnamefont {E.~P. A.~M.}\ \bibnamefont
  {Bakkers}}, \bibinfo {author} {\bibfnamefont {L.~P.}\ \bibnamefont
  {Kouwenhoven}}, \bibinfo {author} {\bibfnamefont {P.}~\bibnamefont
  {Krogstrup}}, \bibinfo {author} {\bibfnamefont {C.~M.}\ \bibnamefont
  {Marcus}},\ and\ \bibinfo {author} {\bibfnamefont {Y.}~\bibnamefont {Oreg}},\
  }\bibfield  {title} {\bibinfo {title} {{Majorana zero modes in
  superconductor{\textendash}semiconductor heterostructures}},\ }\href
  {https://doi.org/10.1038/s41578-018-0003-1} {\bibfield  {journal} {\bibinfo
  {journal} {Nat. Rev. Mater.}\ }\textbf {\bibinfo {volume} {3}},\ \bibinfo
  {pages} {52} (\bibinfo {year} {2018})}\BibitemShut {NoStop}%
\bibitem [{\citenamefont {Gao}\ \emph {et~al.}(2024)\citenamefont {Gao},
  \citenamefont {Song}, \citenamefont {Yang}, \citenamefont {Yu}, \citenamefont
  {Li}, \citenamefont {Miao}, \citenamefont {Wang}, \citenamefont {Chen},
  \citenamefont {Geng}, \citenamefont {Yang}, \citenamefont {Xia},
  \citenamefont {Feng}, \citenamefont {Zang}, \citenamefont {Li}, \citenamefont
  {Shang}, \citenamefont {Xue}, \citenamefont {He},\ and\ \citenamefont
  {Zhang}}]{gao2024hard}%
  \BibitemOpen
  \bibfield  {author} {\bibinfo {author} {\bibfnamefont {Y.}~\bibnamefont
  {Gao}}, \bibinfo {author} {\bibfnamefont {W.}~\bibnamefont {Song}}, \bibinfo
  {author} {\bibfnamefont {S.}~\bibnamefont {Yang}}, \bibinfo {author}
  {\bibfnamefont {Z.}~\bibnamefont {Yu}}, \bibinfo {author} {\bibfnamefont
  {R.}~\bibnamefont {Li}}, \bibinfo {author} {\bibfnamefont {W.}~\bibnamefont
  {Miao}}, \bibinfo {author} {\bibfnamefont {Y.}~\bibnamefont {Wang}}, \bibinfo
  {author} {\bibfnamefont {F.}~\bibnamefont {Chen}}, \bibinfo {author}
  {\bibfnamefont {Z.}~\bibnamefont {Geng}}, \bibinfo {author} {\bibfnamefont
  {L.}~\bibnamefont {Yang}}, \bibinfo {author} {\bibfnamefont {Z.}~\bibnamefont
  {Xia}}, \bibinfo {author} {\bibfnamefont {X.}~\bibnamefont {Feng}}, \bibinfo
  {author} {\bibfnamefont {Y.}~\bibnamefont {Zang}}, \bibinfo {author}
  {\bibfnamefont {L.}~\bibnamefont {Li}}, \bibinfo {author} {\bibfnamefont
  {R.}~\bibnamefont {Shang}}, \bibinfo {author} {\bibfnamefont {Q.-K.}\
  \bibnamefont {Xue}}, \bibinfo {author} {\bibfnamefont {K.}~\bibnamefont
  {He}},\ and\ \bibinfo {author} {\bibfnamefont {H.}~\bibnamefont {Zhang}},\
  }\bibfield  {title} {\bibinfo {title} {Hard superconducting gap in {PbTe}
  nanowires},\ }\href {https://doi.org/10.1088/0256-307X/41/3/038502}
  {\bibfield  {journal} {\bibinfo  {journal} {Chinese Physics Letters}\
  }\textbf {\bibinfo {volume} {41}},\ \bibinfo {pages} {038502} (\bibinfo
  {year} {2024})}\BibitemShut {NoStop}%
\bibitem [{\citenamefont {Queisser}\ and\ \citenamefont
  {Haller}(1998)}]{queisser1998defects}%
  \BibitemOpen
  \bibfield  {author} {\bibinfo {author} {\bibfnamefont {H.~J.}\ \bibnamefont
  {Queisser}}\ and\ \bibinfo {author} {\bibfnamefont {E.~E.}\ \bibnamefont
  {Haller}},\ }\bibfield  {title} {\bibinfo {title} {Defects in semiconductors:
  some fatal, some vital},\ }\href@noop {} {\bibfield  {journal} {\bibinfo
  {journal} {Science}\ }\textbf {\bibinfo {volume} {281}},\ \bibinfo {pages}
  {945} (\bibinfo {year} {1998})}\BibitemShut {NoStop}%
\bibitem [{\citenamefont {Zhang}\ and\ \citenamefont
  {Yates~Jr}(2012)}]{zhang2012band}%
  \BibitemOpen
  \bibfield  {author} {\bibinfo {author} {\bibfnamefont {Z.}~\bibnamefont
  {Zhang}}\ and\ \bibinfo {author} {\bibfnamefont {J.~T.}\ \bibnamefont
  {Yates~Jr}},\ }\bibfield  {title} {\bibinfo {title} {Band bending in
  semiconductors: chemical and physical consequences at surfaces and
  interfaces},\ }\href@noop {} {\bibfield  {journal} {\bibinfo  {journal}
  {Chemical reviews}\ }\textbf {\bibinfo {volume} {112}},\ \bibinfo {pages}
  {5520} (\bibinfo {year} {2012})}\BibitemShut {NoStop}%
\bibitem [{\citenamefont {Mourik}\ \emph {et~al.}(2012)\citenamefont {Mourik},
  \citenamefont {Zuo}, \citenamefont {Frolov}, \citenamefont {Plissard},
  \citenamefont {Bakkers},\ and\ \citenamefont
  {Kouwenhoven}}]{mourik2012signatures}%
  \BibitemOpen
  \bibfield  {author} {\bibinfo {author} {\bibfnamefont {V.}~\bibnamefont
  {Mourik}}, \bibinfo {author} {\bibfnamefont {K.}~\bibnamefont {Zuo}},
  \bibinfo {author} {\bibfnamefont {S.~M.}\ \bibnamefont {Frolov}}, \bibinfo
  {author} {\bibfnamefont {S.}~\bibnamefont {Plissard}}, \bibinfo {author}
  {\bibfnamefont {E.~P.}\ \bibnamefont {Bakkers}},\ and\ \bibinfo {author}
  {\bibfnamefont {L.~P.}\ \bibnamefont {Kouwenhoven}},\ }\bibfield  {title}
  {\bibinfo {title} {Signatures of majorana fermions in hybrid
  superconductor-semiconductor nanowire devices},\ }\href@noop {} {\bibfield
  {journal} {\bibinfo  {journal} {Science}\ }\textbf {\bibinfo {volume}
  {336}},\ \bibinfo {pages} {1003} (\bibinfo {year} {2012})}\BibitemShut
  {NoStop}%
\bibitem [{\citenamefont {Das}\ \emph {et~al.}(2012)\citenamefont {Das},
  \citenamefont {Ronen}, \citenamefont {Most}, \citenamefont {Oreg},
  \citenamefont {Heiblum},\ and\ \citenamefont {Shtrikman}}]{das2012zero}%
  \BibitemOpen
  \bibfield  {author} {\bibinfo {author} {\bibfnamefont {A.}~\bibnamefont
  {Das}}, \bibinfo {author} {\bibfnamefont {Y.}~\bibnamefont {Ronen}}, \bibinfo
  {author} {\bibfnamefont {Y.}~\bibnamefont {Most}}, \bibinfo {author}
  {\bibfnamefont {Y.}~\bibnamefont {Oreg}}, \bibinfo {author} {\bibfnamefont
  {M.}~\bibnamefont {Heiblum}},\ and\ \bibinfo {author} {\bibfnamefont
  {H.}~\bibnamefont {Shtrikman}},\ }\bibfield  {title} {\bibinfo {title}
  {{Zero-bias peaks and splitting in an Al--InAs nanowire topological
  superconductor as a signature of Majorana fermions}},\ }\href@noop {}
  {\bibfield  {journal} {\bibinfo  {journal} {Nature Physics}\ }\textbf
  {\bibinfo {volume} {8}},\ \bibinfo {pages} {887} (\bibinfo {year}
  {2012})}\BibitemShut {NoStop}%
\bibitem [{\citenamefont {Deng}\ \emph {et~al.}(2012)\citenamefont {Deng},
  \citenamefont {Yu}, \citenamefont {Huang}, \citenamefont {Larsson},
  \citenamefont {Caroff},\ and\ \citenamefont {Xu}}]{deng2012anomalous}%
  \BibitemOpen
  \bibfield  {author} {\bibinfo {author} {\bibfnamefont {M.}~\bibnamefont
  {Deng}}, \bibinfo {author} {\bibfnamefont {C.}~\bibnamefont {Yu}}, \bibinfo
  {author} {\bibfnamefont {G.}~\bibnamefont {Huang}}, \bibinfo {author}
  {\bibfnamefont {M.}~\bibnamefont {Larsson}}, \bibinfo {author} {\bibfnamefont
  {P.}~\bibnamefont {Caroff}},\ and\ \bibinfo {author} {\bibfnamefont
  {H.}~\bibnamefont {Xu}},\ }\bibfield  {title} {\bibinfo {title} {Anomalous
  zero-bias conductance peak in a {Nb}--{InSb} nanowire--{Nb} hybrid device},\
  }\href@noop {} {\bibfield  {journal} {\bibinfo  {journal} {Nano letters}\
  }\textbf {\bibinfo {volume} {12}},\ \bibinfo {pages} {6414} (\bibinfo {year}
  {2012})}\BibitemShut {NoStop}%
\bibitem [{\citenamefont {Takei}\ \emph {et~al.}(2013)\citenamefont {Takei},
  \citenamefont {Fregoso}, \citenamefont {Hui}, \citenamefont {Lobos},\ and\
  \citenamefont {Das~Sarma}}]{PhysRevLett.110.186803}%
  \BibitemOpen
  \bibfield  {author} {\bibinfo {author} {\bibfnamefont {S.}~\bibnamefont
  {Takei}}, \bibinfo {author} {\bibfnamefont {B.~M.}\ \bibnamefont {Fregoso}},
  \bibinfo {author} {\bibfnamefont {H.-Y.}\ \bibnamefont {Hui}}, \bibinfo
  {author} {\bibfnamefont {A.~M.}\ \bibnamefont {Lobos}},\ and\ \bibinfo
  {author} {\bibfnamefont {S.}~\bibnamefont {Das~Sarma}},\ }\bibfield  {title}
  {\bibinfo {title} {Soft superconducting gap in semiconductor majorana
  nanowires},\ }\href {https://doi.org/10.1103/PhysRevLett.110.186803}
  {\bibfield  {journal} {\bibinfo  {journal} {Phys. Rev. Lett.}\ }\textbf
  {\bibinfo {volume} {110}},\ \bibinfo {pages} {186803} (\bibinfo {year}
  {2013})}\BibitemShut {NoStop}%
\bibitem [{\citenamefont {G\"ul}\ \emph {et~al.}(2017)\citenamefont {G\"ul},
  \citenamefont {Zhang}, \citenamefont {de~Vries}, \citenamefont {van Veen},
  \citenamefont {Zuo}, \citenamefont {Mourik}, \citenamefont {Conesa-Boj},
  \citenamefont {Nowak}, \citenamefont {van Woerkom}, \citenamefont
  {Quintero-P\'erez}, \citenamefont {Cassidy}, \citenamefont {Geresdi},
  \citenamefont {Koelling}, \citenamefont {Car}, \citenamefont {Plissard},
  \citenamefont {Bakkers},\ and\ \citenamefont {Kouwenhoven}}]{gül2017hard}%
  \BibitemOpen
  \bibfield  {author} {\bibinfo {author} {\bibfnamefont {O.}~\bibnamefont
  {G\"ul}}, \bibinfo {author} {\bibfnamefont {H.}~\bibnamefont {Zhang}},
  \bibinfo {author} {\bibfnamefont {F.~K.}\ \bibnamefont {de~Vries}}, \bibinfo
  {author} {\bibfnamefont {J.}~\bibnamefont {van Veen}}, \bibinfo {author}
  {\bibfnamefont {K.}~\bibnamefont {Zuo}}, \bibinfo {author} {\bibfnamefont
  {V.}~\bibnamefont {Mourik}}, \bibinfo {author} {\bibfnamefont
  {S.}~\bibnamefont {Conesa-Boj}}, \bibinfo {author} {\bibfnamefont {M.~P.}\
  \bibnamefont {Nowak}}, \bibinfo {author} {\bibfnamefont {D.~J.}\ \bibnamefont
  {van Woerkom}}, \bibinfo {author} {\bibfnamefont {M.}~\bibnamefont
  {Quintero-P\'erez}}, \bibinfo {author} {\bibfnamefont {M.~C.}\ \bibnamefont
  {Cassidy}}, \bibinfo {author} {\bibfnamefont {A.}~\bibnamefont {Geresdi}},
  \bibinfo {author} {\bibfnamefont {S.}~\bibnamefont {Koelling}}, \bibinfo
  {author} {\bibfnamefont {D.}~\bibnamefont {Car}}, \bibinfo {author}
  {\bibfnamefont {S.~R.}\ \bibnamefont {Plissard}}, \bibinfo {author}
  {\bibfnamefont {E.~P. A.~M.}\ \bibnamefont {Bakkers}},\ and\ \bibinfo
  {author} {\bibfnamefont {L.~P.}\ \bibnamefont {Kouwenhoven}},\ }\bibfield
  {title} {\bibinfo {title} {Hard superconducting gap in {InSb} nanowires},\
  }\href {https://doi.org/10.1021/acs.nanolett.7b00540} {\bibfield  {journal}
  {\bibinfo  {journal} {Nano Letters}\ }\textbf {\bibinfo {volume} {17}},\
  \bibinfo {pages} {2690} (\bibinfo {year} {2017})},\ \bibinfo {note} {pMID:
  28355877},\ \Eprint
  {https://arxiv.org/abs/https://doi.org/10.1021/acs.nanolett.7b00540}
  {https://doi.org/10.1021/acs.nanolett.7b00540} \BibitemShut {NoStop}%
\bibitem [{\citenamefont {Chang}\ \emph {et~al.}(2015)\citenamefont {Chang},
  \citenamefont {Albrecht}, \citenamefont {Jespersen}, \citenamefont
  {Kuemmeth}, \citenamefont {Krogstrup}, \citenamefont {Nyg{\aa}rd},\ and\
  \citenamefont {Marcus}}]{chang2015hard}%
  \BibitemOpen
  \bibfield  {author} {\bibinfo {author} {\bibfnamefont {W.}~\bibnamefont
  {Chang}}, \bibinfo {author} {\bibfnamefont {S.}~\bibnamefont {Albrecht}},
  \bibinfo {author} {\bibfnamefont {T.}~\bibnamefont {Jespersen}}, \bibinfo
  {author} {\bibfnamefont {F.}~\bibnamefont {Kuemmeth}}, \bibinfo {author}
  {\bibfnamefont {P.}~\bibnamefont {Krogstrup}}, \bibinfo {author}
  {\bibfnamefont {J.}~\bibnamefont {Nyg{\aa}rd}},\ and\ \bibinfo {author}
  {\bibfnamefont {C.~M.}\ \bibnamefont {Marcus}},\ }\bibfield  {title}
  {\bibinfo {title} {Hard gap in epitaxial semiconductor--superconductor
  nanowires},\ }\href@noop {} {\bibfield  {journal} {\bibinfo  {journal}
  {Nature nanotechnology}\ }\textbf {\bibinfo {volume} {10}},\ \bibinfo {pages}
  {232} (\bibinfo {year} {2015})}\BibitemShut {NoStop}%
\bibitem [{\citenamefont {Kang}\ \emph {et~al.}(2017)\citenamefont {Kang},
  \citenamefont {Grivnin}, \citenamefont {Bor}, \citenamefont {Reiner},
  \citenamefont {Avraham}, \citenamefont {Ronen}, \citenamefont {Cohen},
  \citenamefont {Kacman}, \citenamefont {Shtrikman},\ and\ \citenamefont
  {Beidenkopf}}]{kang2017robust}%
  \BibitemOpen
  \bibfield  {author} {\bibinfo {author} {\bibfnamefont {J.-H.}\ \bibnamefont
  {Kang}}, \bibinfo {author} {\bibfnamefont {A.}~\bibnamefont {Grivnin}},
  \bibinfo {author} {\bibfnamefont {E.}~\bibnamefont {Bor}}, \bibinfo {author}
  {\bibfnamefont {J.}~\bibnamefont {Reiner}}, \bibinfo {author} {\bibfnamefont
  {N.}~\bibnamefont {Avraham}}, \bibinfo {author} {\bibfnamefont
  {Y.}~\bibnamefont {Ronen}}, \bibinfo {author} {\bibfnamefont
  {Y.}~\bibnamefont {Cohen}}, \bibinfo {author} {\bibfnamefont
  {P.}~\bibnamefont {Kacman}}, \bibinfo {author} {\bibfnamefont
  {H.}~\bibnamefont {Shtrikman}},\ and\ \bibinfo {author} {\bibfnamefont
  {H.}~\bibnamefont {Beidenkopf}},\ }\bibfield  {title} {\bibinfo {title}
  {Robust epitaxial al coating of reclined inas nanowires},\ }\href@noop {}
  {\bibfield  {journal} {\bibinfo  {journal} {Nano letters}\ }\textbf {\bibinfo
  {volume} {17}},\ \bibinfo {pages} {7520} (\bibinfo {year}
  {2017})}\BibitemShut {NoStop}%
\bibitem [{\citenamefont {Geng}\ \emph {et~al.}(2024)\citenamefont {Geng},
  \citenamefont {Chen}, \citenamefont {Gao}, \citenamefont {Yang},
  \citenamefont {Wang}, \citenamefont {Yang}, \citenamefont {Zhang},
  \citenamefont {Li}, \citenamefont {Song}, \citenamefont {Xu} \emph
  {et~al.}}]{geng2024epitaxial}%
  \BibitemOpen
  \bibfield  {author} {\bibinfo {author} {\bibfnamefont {Z.}~\bibnamefont
  {Geng}}, \bibinfo {author} {\bibfnamefont {F.}~\bibnamefont {Chen}}, \bibinfo
  {author} {\bibfnamefont {Y.}~\bibnamefont {Gao}}, \bibinfo {author}
  {\bibfnamefont {L.}~\bibnamefont {Yang}}, \bibinfo {author} {\bibfnamefont
  {Y.}~\bibnamefont {Wang}}, \bibinfo {author} {\bibfnamefont {S.}~\bibnamefont
  {Yang}}, \bibinfo {author} {\bibfnamefont {S.}~\bibnamefont {Zhang}},
  \bibinfo {author} {\bibfnamefont {Z.}~\bibnamefont {Li}}, \bibinfo {author}
  {\bibfnamefont {W.}~\bibnamefont {Song}}, \bibinfo {author} {\bibfnamefont
  {J.}~\bibnamefont {Xu}}, \emph {et~al.},\ }\bibfield  {title} {\bibinfo
  {title} {Epitaxial {Indium} on {PbTe} nanowires for quantum devices},\
  }\href@noop {} {\bibfield  {journal} {\bibinfo  {journal} {arXiv preprint
  arXiv:2402.04024}\ } (\bibinfo {year} {2024})}\BibitemShut {NoStop}%
\bibitem [{\citenamefont {Cole}\ \emph {et~al.}(2015)\citenamefont {Cole},
  \citenamefont {Das~Sarma},\ and\ \citenamefont {Stanescu}}]{cole2015effects}%
  \BibitemOpen
  \bibfield  {author} {\bibinfo {author} {\bibfnamefont {W.~S.}\ \bibnamefont
  {Cole}}, \bibinfo {author} {\bibfnamefont {S.}~\bibnamefont {Das~Sarma}},\
  and\ \bibinfo {author} {\bibfnamefont {T.~D.}\ \bibnamefont {Stanescu}},\
  }\bibfield  {title} {\bibinfo {title} {Effects of large induced
  superconducting gap on semiconductor majorana nanowires},\ }\href@noop {}
  {\bibfield  {journal} {\bibinfo  {journal} {Physical Review B}\ }\textbf
  {\bibinfo {volume} {92}},\ \bibinfo {pages} {174511} (\bibinfo {year}
  {2015})}\BibitemShut {NoStop}%
\bibitem [{\citenamefont {Pan}\ \emph {et~al.}(1998)\citenamefont {Pan},
  \citenamefont {Hudson},\ and\ \citenamefont {Davis}}]{pan1998vacuum}%
  \BibitemOpen
  \bibfield  {author} {\bibinfo {author} {\bibfnamefont {S.}~\bibnamefont
  {Pan}}, \bibinfo {author} {\bibfnamefont {E.}~\bibnamefont {Hudson}},\ and\
  \bibinfo {author} {\bibfnamefont {J.}~\bibnamefont {Davis}},\ }\bibfield
  {title} {\bibinfo {title} {Vacuum tunneling of superconducting quasiparticles
  from atomically sharp scanning tunneling microscope tips},\ }\href@noop {}
  {\bibfield  {journal} {\bibinfo  {journal} {Applied Physics Letters}\
  }\textbf {\bibinfo {volume} {73}},\ \bibinfo {pages} {2992} (\bibinfo {year}
  {1998})}\BibitemShut {NoStop}%
\bibitem [{\citenamefont {Bjergfelt}\ \emph {et~al.}(2021)\citenamefont
  {Bjergfelt}, \citenamefont {Carrad}, \citenamefont {Kanne}, \citenamefont
  {Johnson}, \citenamefont {Fiordaliso}, \citenamefont {Jespersen},\ and\
  \citenamefont {Nyg{\aa}rd}}]{bjergfelt2021superconductivity}%
  \BibitemOpen
  \bibfield  {author} {\bibinfo {author} {\bibfnamefont {M.~S.}\ \bibnamefont
  {Bjergfelt}}, \bibinfo {author} {\bibfnamefont {D.~J.}\ \bibnamefont
  {Carrad}}, \bibinfo {author} {\bibfnamefont {T.}~\bibnamefont {Kanne}},
  \bibinfo {author} {\bibfnamefont {E.}~\bibnamefont {Johnson}}, \bibinfo
  {author} {\bibfnamefont {E.~M.}\ \bibnamefont {Fiordaliso}}, \bibinfo
  {author} {\bibfnamefont {T.~S.}\ \bibnamefont {Jespersen}},\ and\ \bibinfo
  {author} {\bibfnamefont {J.}~\bibnamefont {Nyg{\aa}rd}},\ }\bibfield  {title}
  {\bibinfo {title} {Superconductivity and parity preservation in as-grown in
  islands on {InAs} nanowires},\ }\href@noop {} {\bibfield  {journal} {\bibinfo
   {journal} {Nano letters}\ }\textbf {\bibinfo {volume} {21}},\ \bibinfo
  {pages} {9875} (\bibinfo {year} {2021})}\BibitemShut {NoStop}%
\bibitem [{\citenamefont {Winkler}\ \emph {et~al.}(2016)\citenamefont
  {Winkler}, \citenamefont {Wu}, \citenamefont {Troyer}, \citenamefont
  {Krogstrup},\ and\ \citenamefont {Soluyanov}}]{PhysRevLett.117.076403}%
  \BibitemOpen
  \bibfield  {author} {\bibinfo {author} {\bibfnamefont {G.~W.}\ \bibnamefont
  {Winkler}}, \bibinfo {author} {\bibfnamefont {Q.}~\bibnamefont {Wu}},
  \bibinfo {author} {\bibfnamefont {M.}~\bibnamefont {Troyer}}, \bibinfo
  {author} {\bibfnamefont {P.}~\bibnamefont {Krogstrup}},\ and\ \bibinfo
  {author} {\bibfnamefont {A.~A.}\ \bibnamefont {Soluyanov}},\ }\bibfield
  {title} {\bibinfo {title} {Topological phases in
  {${\mathrm{InAs}}_{1\ensuremath{-}x}{\mathrm{Sb}}_{x}$}: From novel
  topological semimetal to majorana wire},\ }\href
  {https://doi.org/10.1103/PhysRevLett.117.076403} {\bibfield  {journal}
  {\bibinfo  {journal} {Phys. Rev. Lett.}\ }\textbf {\bibinfo {volume} {117}},\
  \bibinfo {pages} {076403} (\bibinfo {year} {2016})}\BibitemShut {NoStop}%
\bibitem [{\citenamefont {Sestoft}\ \emph {et~al.}(2018)\citenamefont
  {Sestoft}, \citenamefont {Kanne}, \citenamefont {Gejl}, \citenamefont {von
  Soosten}, \citenamefont {Yodh}, \citenamefont {Sherman}, \citenamefont
  {Tarasinski}, \citenamefont {Wimmer}, \citenamefont {Johnson}, \citenamefont
  {Deng}, \citenamefont {Nyg\aa{}rd}, \citenamefont {Jespersen}, \citenamefont
  {Marcus},\ and\ \citenamefont {Krogstrup}}]{PhysRevMaterials.2.044202}%
  \BibitemOpen
  \bibfield  {author} {\bibinfo {author} {\bibfnamefont {J.~E.}\ \bibnamefont
  {Sestoft}}, \bibinfo {author} {\bibfnamefont {T.}~\bibnamefont {Kanne}},
  \bibinfo {author} {\bibfnamefont {A.~N.}\ \bibnamefont {Gejl}}, \bibinfo
  {author} {\bibfnamefont {M.}~\bibnamefont {von Soosten}}, \bibinfo {author}
  {\bibfnamefont {J.~S.}\ \bibnamefont {Yodh}}, \bibinfo {author}
  {\bibfnamefont {D.}~\bibnamefont {Sherman}}, \bibinfo {author} {\bibfnamefont
  {B.}~\bibnamefont {Tarasinski}}, \bibinfo {author} {\bibfnamefont
  {M.}~\bibnamefont {Wimmer}}, \bibinfo {author} {\bibfnamefont
  {E.}~\bibnamefont {Johnson}}, \bibinfo {author} {\bibfnamefont
  {M.}~\bibnamefont {Deng}}, \bibinfo {author} {\bibfnamefont {J.}~\bibnamefont
  {Nyg\aa{}rd}}, \bibinfo {author} {\bibfnamefont {T.~S.}\ \bibnamefont
  {Jespersen}}, \bibinfo {author} {\bibfnamefont {C.~M.}\ \bibnamefont
  {Marcus}},\ and\ \bibinfo {author} {\bibfnamefont {P.}~\bibnamefont
  {Krogstrup}},\ }\bibfield  {title} {\bibinfo {title} {Engineering hybrid
  epitaxial {InAsSb/Al} nanowires for stronger topological protection},\ }\href
  {https://doi.org/10.1103/PhysRevMaterials.2.044202} {\bibfield  {journal}
  {\bibinfo  {journal} {Phys. Rev. Mater.}\ }\textbf {\bibinfo {volume} {2}},\
  \bibinfo {pages} {044202} (\bibinfo {year} {2018})}\BibitemShut {NoStop}%
\bibitem [{\citenamefont {Ercolani}\ \emph {et~al.}(2012)\citenamefont
  {Ercolani}, \citenamefont {Gemmi}, \citenamefont {Nasi}, \citenamefont
  {Rossi}, \citenamefont {Pea}, \citenamefont {Li}, \citenamefont {Salviati},
  \citenamefont {Beltram},\ and\ \citenamefont {Sorba}}]{ercolani2012growth}%
  \BibitemOpen
  \bibfield  {author} {\bibinfo {author} {\bibfnamefont {D.}~\bibnamefont
  {Ercolani}}, \bibinfo {author} {\bibfnamefont {M.}~\bibnamefont {Gemmi}},
  \bibinfo {author} {\bibfnamefont {L.}~\bibnamefont {Nasi}}, \bibinfo {author}
  {\bibfnamefont {F.}~\bibnamefont {Rossi}}, \bibinfo {author} {\bibfnamefont
  {M.}~\bibnamefont {Pea}}, \bibinfo {author} {\bibfnamefont {A.}~\bibnamefont
  {Li}}, \bibinfo {author} {\bibfnamefont {G.}~\bibnamefont {Salviati}},
  \bibinfo {author} {\bibfnamefont {F.}~\bibnamefont {Beltram}},\ and\ \bibinfo
  {author} {\bibfnamefont {L.}~\bibnamefont {Sorba}},\ }\bibfield  {title}
  {\bibinfo {title} {Growth of {InAs/InAsSb} heterostructured nanowires},\
  }\href@noop {} {\bibfield  {journal} {\bibinfo  {journal} {Nanotechnology}\
  }\textbf {\bibinfo {volume} {23}},\ \bibinfo {pages} {115606} (\bibinfo
  {year} {2012})}\BibitemShut {NoStop}%
\bibitem [{\citenamefont {Reiner}\ \emph {et~al.}(2017)\citenamefont {Reiner},
  \citenamefont {Nayak}, \citenamefont {Avraham}, \citenamefont {Norris},
  \citenamefont {Yan}, \citenamefont {Fulga}, \citenamefont {Kang},
  \citenamefont {Karzig}, \citenamefont {Shtrikman},\ and\ \citenamefont
  {Beidenkopf}}]{reiner2017hot}%
  \BibitemOpen
  \bibfield  {author} {\bibinfo {author} {\bibfnamefont {J.}~\bibnamefont
  {Reiner}}, \bibinfo {author} {\bibfnamefont {A.~K.}\ \bibnamefont {Nayak}},
  \bibinfo {author} {\bibfnamefont {N.}~\bibnamefont {Avraham}}, \bibinfo
  {author} {\bibfnamefont {A.}~\bibnamefont {Norris}}, \bibinfo {author}
  {\bibfnamefont {B.}~\bibnamefont {Yan}}, \bibinfo {author} {\bibfnamefont
  {I.~C.}\ \bibnamefont {Fulga}}, \bibinfo {author} {\bibfnamefont {J.-H.}\
  \bibnamefont {Kang}}, \bibinfo {author} {\bibfnamefont {T.}~\bibnamefont
  {Karzig}}, \bibinfo {author} {\bibfnamefont {H.}~\bibnamefont {Shtrikman}},\
  and\ \bibinfo {author} {\bibfnamefont {H.}~\bibnamefont {Beidenkopf}},\
  }\bibfield  {title} {\bibinfo {title} {Hot electrons regain coherence in
  semiconducting nanowires},\ }\href@noop {} {\bibfield  {journal} {\bibinfo
  {journal} {Physical Review X}\ }\textbf {\bibinfo {volume} {7}},\ \bibinfo
  {pages} {021016} (\bibinfo {year} {2017})}\BibitemShut {NoStop}%
\bibitem [{\citenamefont {Dynes}\ \emph {et~al.}(1978)\citenamefont {Dynes},
  \citenamefont {Narayanamurti},\ and\ \citenamefont
  {Garno}}]{dynes1978direct}%
  \BibitemOpen
  \bibfield  {author} {\bibinfo {author} {\bibfnamefont {R.~C.}\ \bibnamefont
  {Dynes}}, \bibinfo {author} {\bibfnamefont {V.}~\bibnamefont
  {Narayanamurti}},\ and\ \bibinfo {author} {\bibfnamefont {J.~P.}\
  \bibnamefont {Garno}},\ }\bibfield  {title} {\bibinfo {title} {Direct
  measurement of quasiparticle-lifetime broadening in a strong-coupled
  superconductor},\ }\href@noop {} {\bibfield  {journal} {\bibinfo  {journal}
  {Physical Review Letters}\ }\textbf {\bibinfo {volume} {41}},\ \bibinfo
  {pages} {1509} (\bibinfo {year} {1978})}\BibitemShut {NoStop}%
\bibitem [{\citenamefont {Herman}\ and\ \citenamefont
  {Hlubina}(2016)}]{PhysRevB.94.144508}%
  \BibitemOpen
  \bibfield  {author} {\bibinfo {author} {\bibfnamefont {F.~c.~v.}\
  \bibnamefont {Herman}}\ and\ \bibinfo {author} {\bibfnamefont
  {R.}~\bibnamefont {Hlubina}},\ }\bibfield  {title} {\bibinfo {title}
  {Microscopic interpretation of the dynes formula for the tunneling density of
  states},\ }\href {https://doi.org/10.1103/PhysRevB.94.144508} {\bibfield
  {journal} {\bibinfo  {journal} {Phys. Rev. B}\ }\textbf {\bibinfo {volume}
  {94}},\ \bibinfo {pages} {144508} (\bibinfo {year} {2016})}\BibitemShut
  {NoStop}%
\bibitem [{\citenamefont {Matthias}\ \emph {et~al.}(1963)\citenamefont
  {Matthias}, \citenamefont {Geballe},\ and\ \citenamefont
  {Compton}}]{Matthias1963Jan}%
  \BibitemOpen
  \bibfield  {author} {\bibinfo {author} {\bibfnamefont {B.~T.}\ \bibnamefont
  {Matthias}}, \bibinfo {author} {\bibfnamefont {T.~H.}\ \bibnamefont
  {Geballe}},\ and\ \bibinfo {author} {\bibfnamefont {V.~B.}\ \bibnamefont
  {Compton}},\ }\bibfield  {title} {\bibinfo {title} {{Superconductivity}},\
  }\href {https://doi.org/10.1103/RevModPhys.35.1} {\bibfield  {journal}
  {\bibinfo  {journal} {Rev. Mod. Phys.}\ }\textbf {\bibinfo {volume} {35}},\
  \bibinfo {pages} {1} (\bibinfo {year} {1963})}\BibitemShut {NoStop}%
\bibitem [{\citenamefont {Eisenstein}(1954)}]{Eisenstein1954Jul}%
  \BibitemOpen
  \bibfield  {author} {\bibinfo {author} {\bibfnamefont {J.}~\bibnamefont
  {Eisenstein}},\ }\bibfield  {title} {\bibinfo {title} {{Superconducting
  Elements}},\ }\href {https://doi.org/10.1103/RevModPhys.26.277} {\bibfield
  {journal} {\bibinfo  {journal} {Rev. Mod. Phys.}\ }\textbf {\bibinfo {volume}
  {26}},\ \bibinfo {pages} {277} (\bibinfo {year} {1954})}\BibitemShut
  {NoStop}%
\bibitem [{\citenamefont {Antipov}\ \emph
  {et~al.}(2018{\natexlab{a}})\citenamefont {Antipov}, \citenamefont
  {Bargerbos}, \citenamefont {Winkler}, \citenamefont {Bauer}, \citenamefont
  {Rossi},\ and\ \citenamefont {Lutchyn}}]{antipov2018effects}%
  \BibitemOpen
  \bibfield  {author} {\bibinfo {author} {\bibfnamefont {A.~E.}\ \bibnamefont
  {Antipov}}, \bibinfo {author} {\bibfnamefont {A.}~\bibnamefont {Bargerbos}},
  \bibinfo {author} {\bibfnamefont {G.~W.}\ \bibnamefont {Winkler}}, \bibinfo
  {author} {\bibfnamefont {B.}~\bibnamefont {Bauer}}, \bibinfo {author}
  {\bibfnamefont {E.}~\bibnamefont {Rossi}},\ and\ \bibinfo {author}
  {\bibfnamefont {R.~M.}\ \bibnamefont {Lutchyn}},\ }\bibfield  {title}
  {\bibinfo {title} {Effects of gate-induced electric fields on semiconductor
  majorana nanowires},\ }\href@noop {} {\bibfield  {journal} {\bibinfo
  {journal} {Physical Review X}\ }\textbf {\bibinfo {volume} {8}},\ \bibinfo
  {pages} {031041} (\bibinfo {year} {2018}{\natexlab{a}})}\BibitemShut
  {NoStop}%
\bibitem [{\citenamefont {Escribano}(2022)}]{Thesis_sam}%
  \BibitemOpen
  \bibfield  {author} {\bibinfo {author} {\bibfnamefont {S.~D.}\ \bibnamefont
  {Escribano}},\ }\emph {\bibinfo {title} {Towards a realistic description of
  topological hybrid semiconductor, superconductor and ferromagnetic-insulator
  systems}},\ \href@noop {} {\bibinfo {type} {Phd thesis}},\ \bibinfo  {school}
  {Universidad Autonoma de Madrid} (\bibinfo {year} {2022}),\ \bibinfo {note}
  {available at \url{http://hdl.handle.net/10486/706437}}\BibitemShut {NoStop}%
\bibitem [{\citenamefont {Antipov}\ \emph
  {et~al.}(2018{\natexlab{b}})\citenamefont {Antipov}, \citenamefont
  {Bargerbos}, \citenamefont {Winkler}, \citenamefont {Bauer}, \citenamefont
  {Rossi},\ and\ \citenamefont {Lutchyn}}]{Antipov2018EffectsNanowires}%
  \BibitemOpen
  \bibfield  {author} {\bibinfo {author} {\bibfnamefont {A.~E.}\ \bibnamefont
  {Antipov}}, \bibinfo {author} {\bibfnamefont {A.}~\bibnamefont {Bargerbos}},
  \bibinfo {author} {\bibfnamefont {G.~W.}\ \bibnamefont {Winkler}}, \bibinfo
  {author} {\bibfnamefont {B.}~\bibnamefont {Bauer}}, \bibinfo {author}
  {\bibfnamefont {E.}~\bibnamefont {Rossi}},\ and\ \bibinfo {author}
  {\bibfnamefont {R.~M.}\ \bibnamefont {Lutchyn}},\ }\bibfield  {title}
  {\bibinfo {title} {Effects of gate-induced electric fields on semiconductor
  majorana nanowires},\ }\href {https://doi.org/10.1103/PhysRevX.8.031041}
  {\bibfield  {journal} {\bibinfo  {journal} {Phys. Rev. X}\ }\textbf {\bibinfo
  {volume} {8}},\ \bibinfo {pages} {031041} (\bibinfo {year}
  {2018}{\natexlab{b}})}\BibitemShut {NoStop}%
\bibitem [{\citenamefont {Tinkham}(1996)}]{Tinkham1996}%
  \BibitemOpen
  \bibfield  {author} {\bibinfo {author} {\bibfnamefont {M.}~\bibnamefont
  {Tinkham}},\ }\href
  {https://books.google.co.il/books/about/Introduction_to_Superconductivity.html?id=XP_uAAAAMAAJ&redir_esc=y}
  {\emph {\bibinfo {title} {{Introduction to Superconductivity}}}}\ (\bibinfo
  {publisher} {McGraw Hill},\ \bibinfo {address} {Maidenhead, England, UK},\
  \bibinfo {year} {1996})\BibitemShut {NoStop}%
\bibitem [{\citenamefont {Toxen}(1961{\natexlab{b}})}]{Toxen:PR61}%
  \BibitemOpen
  \bibfield  {author} {\bibinfo {author} {\bibfnamefont {A.~M.}\ \bibnamefont
  {Toxen}},\ }\bibfield  {title} {\bibinfo {title} {Size effects in thin
  superconducting indium films},\ }\href
  {https://doi.org/10.1103/PhysRev.124.2052.2} {\bibfield  {journal} {\bibinfo
  {journal} {Phys. Rev.}\ }\textbf {\bibinfo {volume} {124}},\ \bibinfo {pages}
  {2052} (\bibinfo {year} {1961}{\natexlab{b}})}\BibitemShut {NoStop}%
\bibitem [{\citenamefont {Toxen}(1962)}]{Toxen:PR62}%
  \BibitemOpen
  \bibfield  {author} {\bibinfo {author} {\bibfnamefont {A.~M.}\ \bibnamefont
  {Toxen}},\ }\bibfield  {title} {\bibinfo {title} {Critical fields of thin
  superconducting films. i. thickness effects},\ }\href
  {https://doi.org/10.1103/PhysRev.127.382} {\bibfield  {journal} {\bibinfo
  {journal} {Phys. Rev.}\ }\textbf {\bibinfo {volume} {127}},\ \bibinfo {pages}
  {382} (\bibinfo {year} {1962})}\BibitemShut {NoStop}%
\bibitem [{\citenamefont {Chaudhari}(1966)}]{Chaudhari:PR66}%
  \BibitemOpen
  \bibfield  {author} {\bibinfo {author} {\bibfnamefont {R.~D.}\ \bibnamefont
  {Chaudhari}},\ }\bibfield  {title} {\bibinfo {title} {Critical magnetic
  fields in superconducting films of indium},\ }\href
  {https://doi.org/10.1103/PhysRev.151.96} {\bibfield  {journal} {\bibinfo
  {journal} {Phys. Rev.}\ }\textbf {\bibinfo {volume} {151}},\ \bibinfo {pages}
  {96} (\bibinfo {year} {1966})}\BibitemShut {NoStop}%
\bibitem [{\citenamefont {Tinkham}(2004{\natexlab{b}})}]{tinkham:book04}%
  \BibitemOpen
  \bibfield  {author} {\bibinfo {author} {\bibfnamefont {M.}~\bibnamefont
  {Tinkham}},\ }\href {http://www.worldcat.org/isbn/0486435032} {\emph
  {\bibinfo {title} {Introduction to Superconductivity}}},\ \bibinfo {edition}
  {2nd}\ ed.\ (\bibinfo  {publisher} {Dover Publications},\ \bibinfo {year}
  {2004})\BibitemShut {NoStop}%
\bibitem [{\citenamefont {Kang}\ \emph {et~al.}(2013)\citenamefont {Kang},
  \citenamefont {Cohen}, \citenamefont {Ronen}, \citenamefont {Heiblum},
  \citenamefont {Buczko}, \citenamefont {Kacman}, \citenamefont
  {Popovitz-Biro},\ and\ \citenamefont {Shtrikman}}]{Kang2013Nov}%
  \BibitemOpen
  \bibfield  {author} {\bibinfo {author} {\bibfnamefont {J.-H.}\ \bibnamefont
  {Kang}}, \bibinfo {author} {\bibfnamefont {Y.}~\bibnamefont {Cohen}},
  \bibinfo {author} {\bibfnamefont {Y.}~\bibnamefont {Ronen}}, \bibinfo
  {author} {\bibfnamefont {M.}~\bibnamefont {Heiblum}}, \bibinfo {author}
  {\bibfnamefont {R.}~\bibnamefont {Buczko}}, \bibinfo {author} {\bibfnamefont
  {P.}~\bibnamefont {Kacman}}, \bibinfo {author} {\bibfnamefont
  {R.}~\bibnamefont {Popovitz-Biro}},\ and\ \bibinfo {author} {\bibfnamefont
  {H.}~\bibnamefont {Shtrikman}},\ }\bibfield  {title} {\bibinfo {title}
  {{Crystal Structure and Transport in Merged {InAs} Nanowires {MBE} Grown on
  (001) {InAs}}},\ }\href {https://doi.org/10.1021/nl402571s} {\bibfield
  {journal} {\bibinfo  {journal} {Nano Lett.}\ }\textbf {\bibinfo {volume}
  {13}},\ \bibinfo {pages} {5190} (\bibinfo {year} {2013})}\BibitemShut
  {NoStop}%
\bibitem [{\citenamefont {Kang}\ \emph {et~al.}(2018)\citenamefont {Kang},
  \citenamefont {Krizek}, \citenamefont {Zaluska-Kotur}, \citenamefont
  {Krogstrup}, \citenamefont {Kacman}, \citenamefont {Beidenkopf},\ and\
  \citenamefont {Shtrikman}}]{Kang2018Jul}%
  \BibitemOpen
  \bibfield  {author} {\bibinfo {author} {\bibfnamefont {J.-H.}\ \bibnamefont
  {Kang}}, \bibinfo {author} {\bibfnamefont {F.}~\bibnamefont {Krizek}},
  \bibinfo {author} {\bibfnamefont {M.}~\bibnamefont {Zaluska-Kotur}}, \bibinfo
  {author} {\bibfnamefont {P.}~\bibnamefont {Krogstrup}}, \bibinfo {author}
  {\bibfnamefont {P.}~\bibnamefont {Kacman}}, \bibinfo {author} {\bibfnamefont
  {H.}~\bibnamefont {Beidenkopf}},\ and\ \bibinfo {author} {\bibfnamefont
  {H.}~\bibnamefont {Shtrikman}},\ }\bibfield  {title} {\bibinfo {title}
  {{Au-Assisted Substrate-Faceting for Inclined Nanowire Growth}},\ }\href
  {https://doi.org/10.1021/acs.nanolett.8b00853} {\bibfield  {journal}
  {\bibinfo  {journal} {Nano Lett.}\ }\textbf {\bibinfo {volume} {18}},\
  \bibinfo {pages} {4115} (\bibinfo {year} {2018})}\BibitemShut {NoStop}%
\bibitem [{Note1()}]{Note1}%
  \BibitemOpen
  \bibinfo {note} {In principle, the penetration depth also depends on the
  thickness of the SupC in thin films as $\lambda \simeq \lambda ^{\protect \rm
  (bulk)}\coth {\left (d/2\lambda _0\right )}$ (see general discussion on \ch
  {In} films in Refs.~\cite {Toxen:PR61, Toxen:PR62, Chaudhari:PR66,
  tinkham:book04}). However, this equation provides values for $\lambda $
  orders of magnitude larger than $30$~nm, meaning that it may work for a dirty
  SC. For a clean one, if $\lambda \ll l_e$ (being $l_e$ the electron mean free
  path), the equation that holds is $\lambda \simeq 0.64\lambda ^{\protect \rm
  (bulk)}\protect \sqrt {\xi _0/l_e}$ that we approximate to
  $30$~nm.}\BibitemShut {Stop}%
\bibitem [{\citenamefont {Webster}\ \emph {et~al.}(2015)\citenamefont
  {Webster}, \citenamefont {Riordan}, \citenamefont {Liu}, \citenamefont
  {Steenbergen}, \citenamefont {Synowicki}, \citenamefont {Zhang},\ and\
  \citenamefont {Johnson}}]{Webster:JOP15}%
  \BibitemOpen
  \bibfield  {author} {\bibinfo {author} {\bibfnamefont {P.~T.}\ \bibnamefont
  {Webster}}, \bibinfo {author} {\bibfnamefont {N.~A.}\ \bibnamefont
  {Riordan}}, \bibinfo {author} {\bibfnamefont {S.}~\bibnamefont {Liu}},
  \bibinfo {author} {\bibfnamefont {E.~H.}\ \bibnamefont {Steenbergen}},
  \bibinfo {author} {\bibfnamefont {R.~A.}\ \bibnamefont {Synowicki}}, \bibinfo
  {author} {\bibfnamefont {Y.-H.}\ \bibnamefont {Zhang}},\ and\ \bibinfo
  {author} {\bibfnamefont {S.~R.}\ \bibnamefont {Johnson}},\ }\bibfield
  {title} {\bibinfo {title} {Measurement of {InAsSb} bandgap energy and
  {InAs/InAsSb} band edge positions using spectroscopic ellipsometry and
  photoluminescence spectroscopy},\ }\href {https://doi.org/10.1063/1.4939293}
  {\bibfield  {journal} {\bibinfo  {journal} {Journal of Applied Physics}\
  }\textbf {\bibinfo {volume} {118}},\ \bibinfo {pages} {245706} (\bibinfo
  {year} {2015})}\BibitemShut {NoStop}%
\bibitem [{\citenamefont {Suchalkin}\ \emph {et~al.}(2016)\citenamefont
  {Suchalkin}, \citenamefont {Ludwig}, \citenamefont {Belenky}, \citenamefont
  {Laikhtman}, \citenamefont {Kipshidze}, \citenamefont {Lin}, \citenamefont
  {Shterengas}, \citenamefont {Smirnov}, \citenamefont {Luryi}, \citenamefont
  {Sarney},\ and\ \citenamefont {Svensson}}]{Suchalkin:IOP16}%
  \BibitemOpen
  \bibfield  {author} {\bibinfo {author} {\bibfnamefont {S.}~\bibnamefont
  {Suchalkin}}, \bibinfo {author} {\bibfnamefont {J.}~\bibnamefont {Ludwig}},
  \bibinfo {author} {\bibfnamefont {G.}~\bibnamefont {Belenky}}, \bibinfo
  {author} {\bibfnamefont {B.}~\bibnamefont {Laikhtman}}, \bibinfo {author}
  {\bibfnamefont {G.}~\bibnamefont {Kipshidze}}, \bibinfo {author}
  {\bibfnamefont {Y.}~\bibnamefont {Lin}}, \bibinfo {author} {\bibfnamefont
  {L.}~\bibnamefont {Shterengas}}, \bibinfo {author} {\bibfnamefont
  {D.}~\bibnamefont {Smirnov}}, \bibinfo {author} {\bibfnamefont
  {S.}~\bibnamefont {Luryi}}, \bibinfo {author} {\bibfnamefont {W.~L.}\
  \bibnamefont {Sarney}},\ and\ \bibinfo {author} {\bibfnamefont {S.~P.}\
  \bibnamefont {Svensson}},\ }\bibfield  {title} {\bibinfo {title} {Electronic
  properties of unstrained unrelaxed narrow gap inassb alloys},\ }\href
  {https://doi.org/10.1088/0022-3727/49/10/105101} {\bibfield  {journal}
  {\bibinfo  {journal} {Journal of Physics D: Applied Physics}\ }\textbf
  {\bibinfo {volume} {49}},\ \bibinfo {pages} {105101} (\bibinfo {year}
  {2016})}\BibitemShut {NoStop}%
\bibitem [{\citenamefont {Manago}\ \emph {et~al.}(2021)\citenamefont {Manago},
  \citenamefont {Kasahara},\ and\ \citenamefont {Shibasaki}}]{Manago:AIP21}%
  \BibitemOpen
  \bibfield  {author} {\bibinfo {author} {\bibfnamefont {T.}~\bibnamefont
  {Manago}}, \bibinfo {author} {\bibfnamefont {K.}~\bibnamefont {Kasahara}},\
  and\ \bibinfo {author} {\bibfnamefont {I.}~\bibnamefont {Shibasaki}},\
  }\bibfield  {title} {\bibinfo {title} {Composition optimization of
  {InAsSb/AlInSb} quantum wells for hall sensors with high sensitivity and high
  thermal stability},\ }\href {https://doi.org/10.1063/5.0039809} {\bibfield
  {journal} {\bibinfo  {journal} {AIP Advances}\ }\textbf {\bibinfo {volume}
  {11}},\ \bibinfo {pages} {035213} (\bibinfo {year} {2021})}\BibitemShut
  {NoStop}%
\bibitem [{\citenamefont {Escribano}\ \emph {et~al.}(2020)\citenamefont
  {Escribano}, \citenamefont {Yeyati},\ and\ \citenamefont
  {Prada}}]{Escribano2020Jul}%
  \BibitemOpen
  \bibfield  {author} {\bibinfo {author} {\bibfnamefont {S.~D.}\ \bibnamefont
  {Escribano}}, \bibinfo {author} {\bibfnamefont {A.~L.}\ \bibnamefont
  {Yeyati}},\ and\ \bibinfo {author} {\bibfnamefont {E.}~\bibnamefont
  {Prada}},\ }\bibfield  {title} {\bibinfo {title} {Improved effective equation
  for the rashba spin-orbit coupling in semiconductor nanowires},\ }\href
  {https://doi.org/10.1103/PhysRevResearch.2.033264} {\bibfield  {journal}
  {\bibinfo  {journal} {Phys. Rev. Res.}\ }\textbf {\bibinfo {volume} {2}},\
  \bibinfo {pages} {033264} (\bibinfo {year} {2020})}\BibitemShut {NoStop}%
\bibitem [{\citenamefont {Escribano}(2020)}]{MajoranaNanowiresQSP_v1}%
  \BibitemOpen
  \bibfield  {author} {\bibinfo {author} {\bibfnamefont {S.~D.}\ \bibnamefont
  {Escribano}},\ }\href {https://doi.org/10.5281/zenodo.3974287} {\bibinfo
  {title} {{MajoranaNanowires: Quantum Simulation Package}}} (\bibinfo {year}
  {2020})\BibitemShut {NoStop}%
\end{thebibliography}
\end{document}